\documentclass[12pt]{article}

\RequirePackage[T1]{fontenc}
\usepackage{natbib}
\RequirePackage[colorlinks,citecolor=blue,urlcolor=blue]{hyperref}
\usepackage[english]{babel}
\usepackage{amsthm}
\usepackage{amsmath}
\usepackage{amsfonts}
\usepackage{amssymb}
\usepackage{graphicx}
\usepackage{enumerate}
\usepackage{color}
\usepackage{array}
\usepackage{bbm}
\usepackage{multirow}
\usepackage{rotating}
\usepackage{xr}
\usepackage{setspace}
\usepackage{algorithm,algpseudocode}
\usepackage{booktabs}

\def\frac#1#2{{\textstyle{#1\over#2}}}

\DeclareSymbolFont{AMSb}{U}{msb}{m}{n}
\DeclareMathSymbol{\Natural}{\mathbin}{AMSb}{"4E}
\DeclareMathSymbol{\Integer}{\mathbin}{AMSb}{"5A}
\DeclareMathSymbol{\Real}{\mathbin}{AMSb}{"52}
\DeclareMathSymbol{\Rational}{\mathbin}{AMSb}{"51}
\DeclareMathSymbol{\Imaginary}{\mathbin}{AMSb}{"49}
\DeclareMathSymbol{\Complex}{\mathbin}{AMSb}{"43} 
\DeclareMathSymbol{\Disk}{\mathbin}{AMSb}{"44} 
\def\bi{\begin{itemize}}
\def\ei{\end{itemize}}
\def\bd{\begin{description}}
\def\ed{\end{description}}
\def\ben{\begin{enumerate}}
\def\een{\end{enumerate}}
\def\bv{\begin{verbatim}}
\def\ev{\end{verbatim}}

\def\bar#1{{\overline{#1}}}
\def\hat#1{{\widehat{#1}}}

\def\pr{{\rm Pr}}
\def\Pr{\pr}

\def\2to{{\ {\buildrel 2\over \longrightarrow}\ }}

\def\I1ton{{$I_1,\ldots,I_n$}}
\def\X1ton{{$X_1,\ldots,X_n$}}
\def\Y1ton{{$Y_1,\ldots,Y_n$}}
\def\Z1ton{{$Z_1,\ldots,Z_n$}}
\def\R1ton{{$R_1,\ldots,R_n$}}
\def\e1ton{{$e_1,\ldots,e_n$}}
\def\t1ton{{$t_1,\ldots,t_n$}}
\def\x1ton{{$x_1,\ldots,x_n$}}
\def\y1ton{{$y_1,\ldots,y_n$}}
\def\z1ton{{$z_1,\ldots,z_n$}}
\begin{document}
\thispagestyle{empty}
\baselineskip=28pt
\vskip 5mm
\begin{center} {\Large{\bf Joint Estimation of Extreme Spatially Aggregated Precipitation at Different Scales through Mixture Modelling}}
\end{center}

\baselineskip=12pt
\vskip 5mm

\begin{center}
\large
Jordan Richards$^{12*}$ and Jonathan A. Tawn$^{2}$ and Simon Brown$^{3}$
\end{center}

\footnotetext[1]{
\baselineskip=10pt Statistics Program, Computer, Electrical and Mathematical Sciences and Engineering (CEMSE) Division, King Abdullah University of Science and Technology (KAUST), Saudi Arabia. \\$^*$E-mail: jordan.richards@kaust.edu.sa}
\footnotetext[2]{
\baselineskip=10pt Department of Mathematics and Statistics, Lancaster University, UK}
\footnotetext[3]{
\baselineskip=10pt Hadley Centre for Climate Prediction and Research, Met Office, UK}

\baselineskip=17pt
\vskip 4mm
\centerline{\today}
\vskip 6mm

%%%%%%%%%%%%%%%%%%%%%%%%%%%%%%%%%%%%%%%%%%%%%%%%%%%%%%%%%%%%%%%%%%%%%%%%
\begin{center}
{\large{\bf Abstract}}
\end{center}
Although most models for rainfall extremes focus on pointwise values, it is aggregated precipitation over areas up to river catchment scale that is of the most interest. To capture the joint behaviour of precipitation aggregates evaluated at different spatial scales, parsimonious and effective models must be built with knowledge of the underlying spatial process. Precipitation is driven by a mixture of processes acting at different scales and intensities, e.g., convective and frontal, with extremes of aggregates for typical catchment sizes arising from extremes of only one of these processes, rather than a combination of them. 
High-intensity convective events cause extreme spatial aggregates at small scales but the contribution of lower-intensity large-scale fronts is likely to increase as the area aggregated increases. Thus, to capture small to large scale spatial aggregates within a single approach requires a model that can accurately capture the extremal properties of both convective and frontal events.  Previous extreme value methods have ignored this mixture structure; we propose a spatial extreme value model which is a mixture of two components with different marginal and dependence models that are able to capture the extremal behaviour of convective and frontal rainfall and more  faithfully reproduces spatial aggregates for a wide range of scales. Modelling extremes of the frontal component raises new challenges due to it exhibiting strong long-range extremal spatial dependence. Our modelling approach is applied to fine-scale, high-dimensional, gridded precipitation data. We show that accounting for the mixture structure improves the joint inference on extremes of spatial aggregates over regions of different sizes.
\baselineskip=16pt

\par\vfill\noindent

\pagenumbering{arabic}
\baselineskip=24pt

\newpage

%%%%%%%%%%%%%%%%%%%%%%%%%%%%%%%%%
%%%%%%%%%%%%%%%%%%%%%%%%%%%%%%%%%
\section{Introduction}
\subsection{Motivation}
In light of climate change, there is a real need to assess the resilience of national infrastructure to extreme rainfall, both now and under future warming scenarios. This requires knowledge of extreme rainfall behaviour at a range of scales and durations, from the very small and short (about 1km and $10$ minutes) for building design and urban drainage, up to thousands of square kilometres and days for river catchments and flooding defence.  Acquiring such knowledge is very difficult as observations tend to be sparse spatially, with time series too short in length to reliably inform practitioners about future risk. Climate models can provide spatially complete data for both the present and the future, but when run at resolutions that are too coarse to resolve convection they struggle to reproduce realistic extreme rainfall \citep{weller2013two}.  With the advent of convection permitting models (CPMs) run at resolutions that can explicitly represent convection, albeit only at the larger scales, the realism of extreme rainfall is much improved  particularly in summer when convection is the dominant mechanism \citep{kendon2012realism,chan2014value, kendon2021}. However, such CPMs are very expensive to run and thus currently produce data records which are of insufficient length to characterise extreme rainfall with high precision particularly if considering the area affected by extreme rainfall.\par
\begin{figure}[h!]
\centering
\begin{minipage}{0.32\linewidth}
\centering
\includegraphics[width=\linewidth]{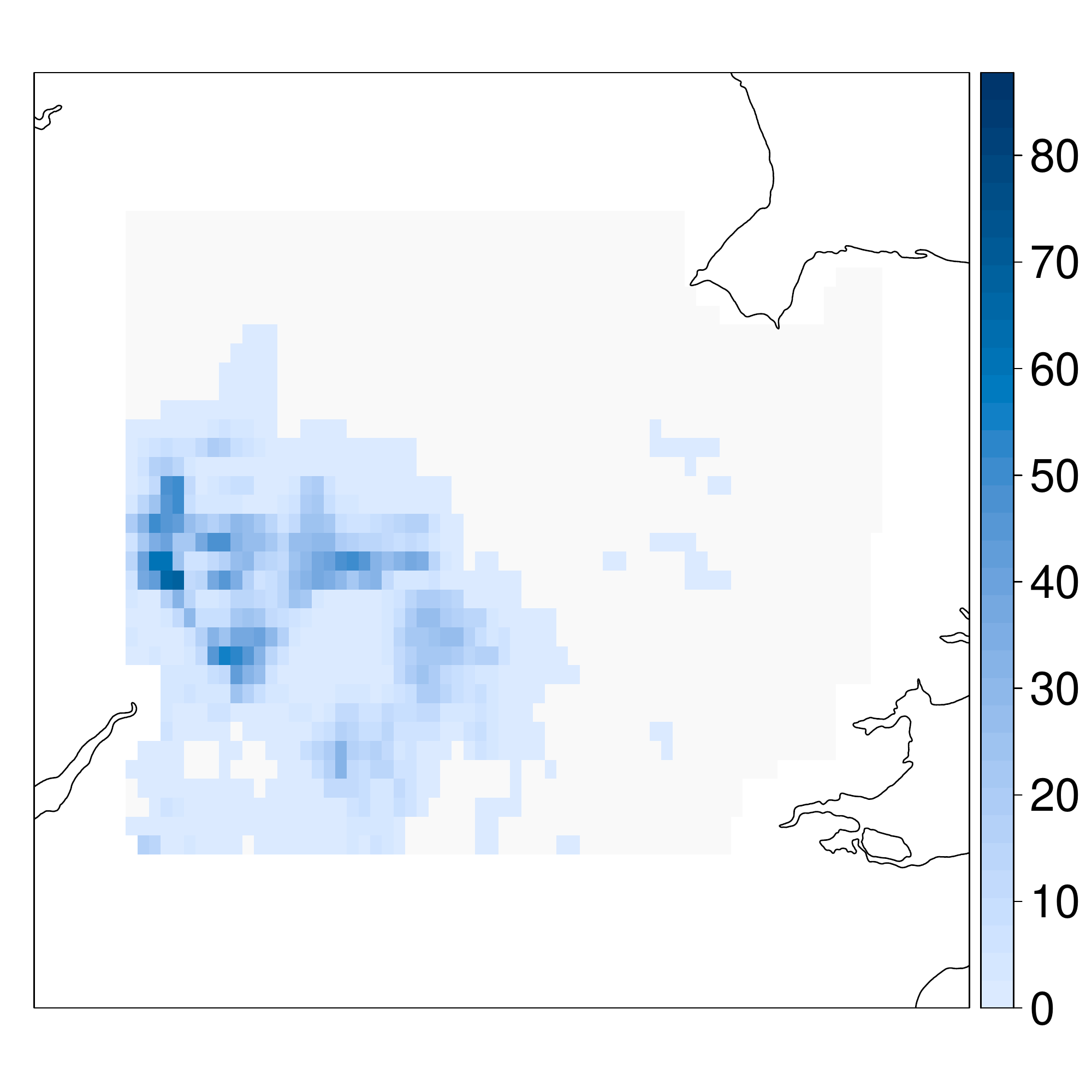} 
\end{minipage}
\begin{minipage}{0.32\linewidth}
\centering
\includegraphics[width=\linewidth]{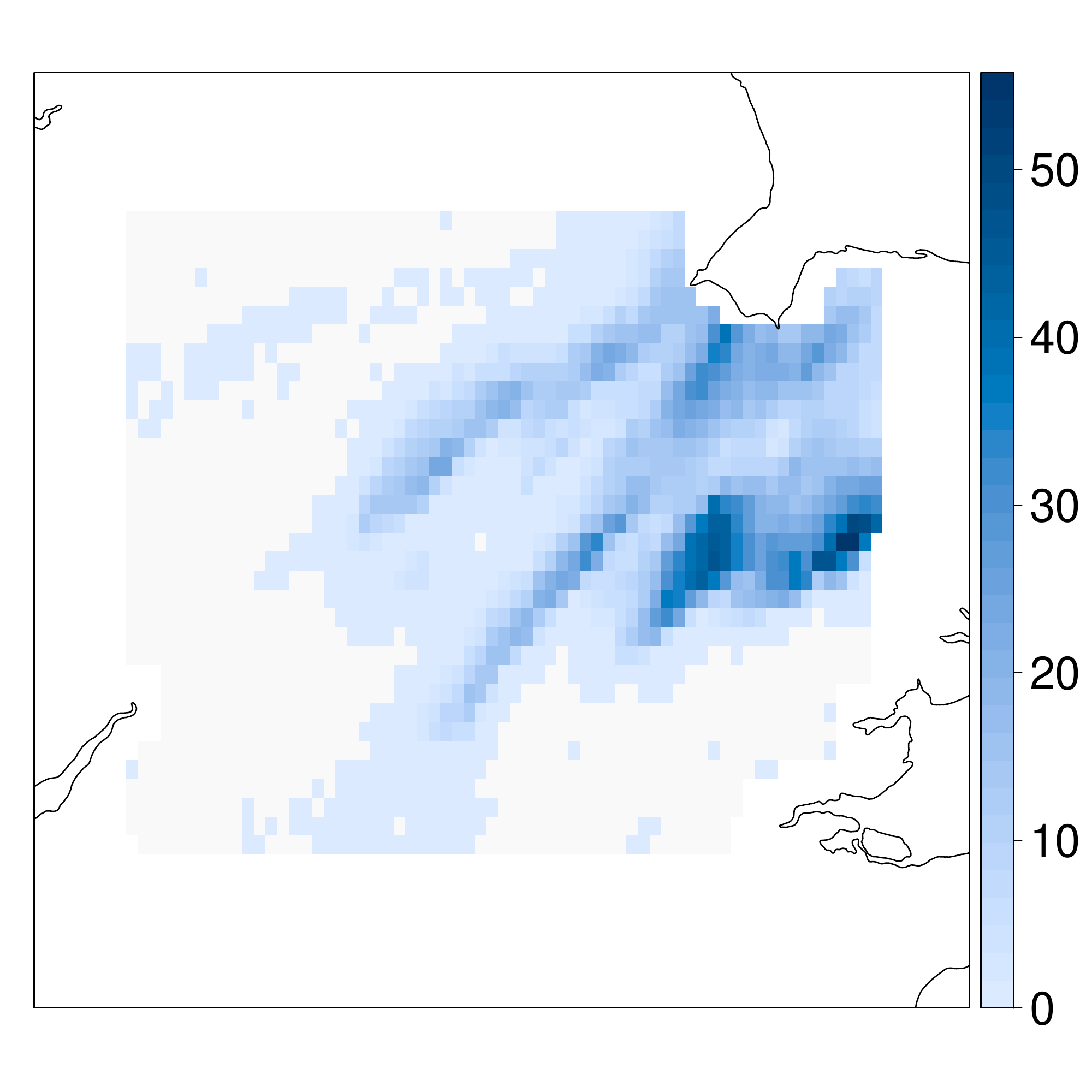} 
\end{minipage}
\begin{minipage}{0.32\linewidth}
\centering
\includegraphics[width=\linewidth]{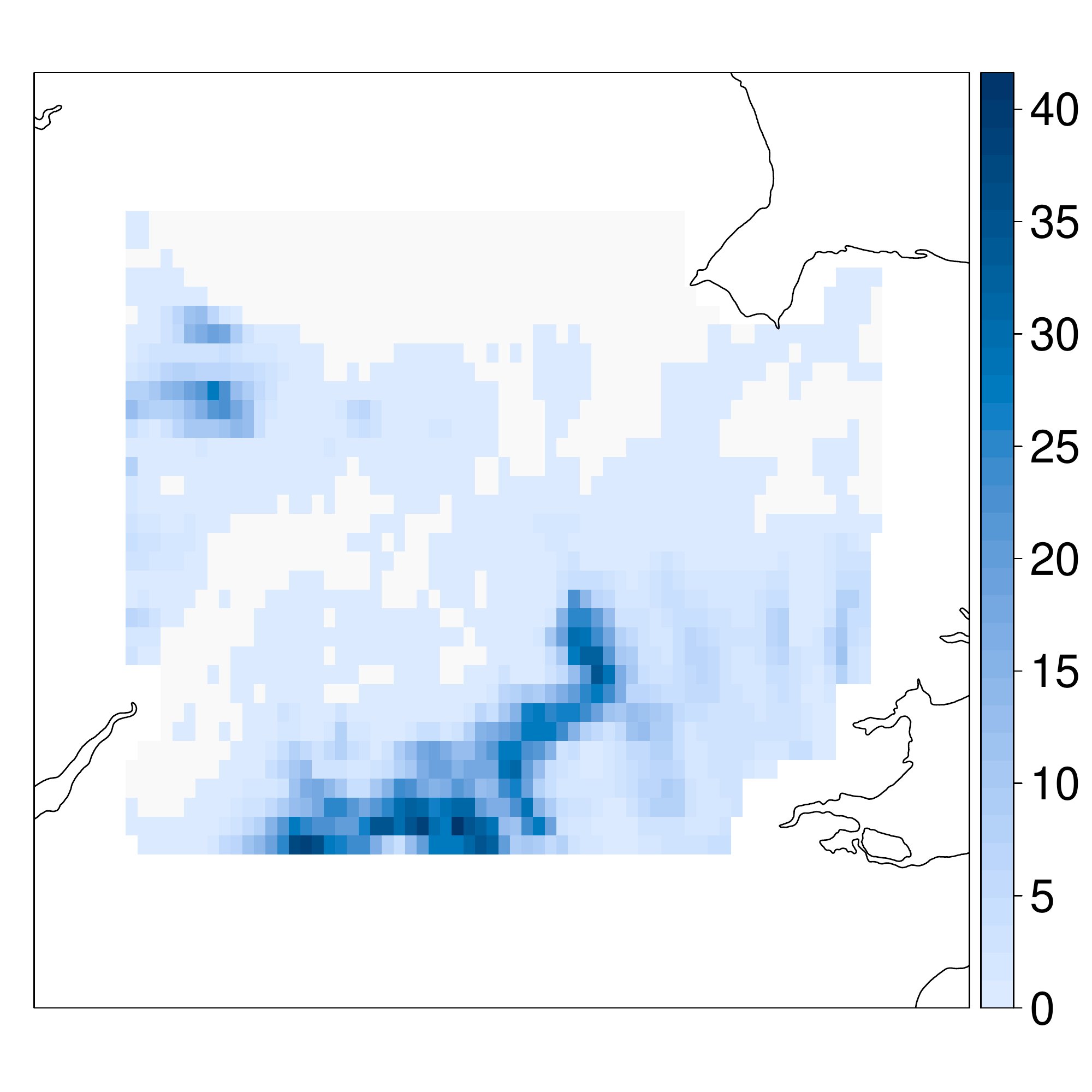} 
\end{minipage}
\begin{minipage}{0.32\linewidth}
\centering
\includegraphics[width=\linewidth]{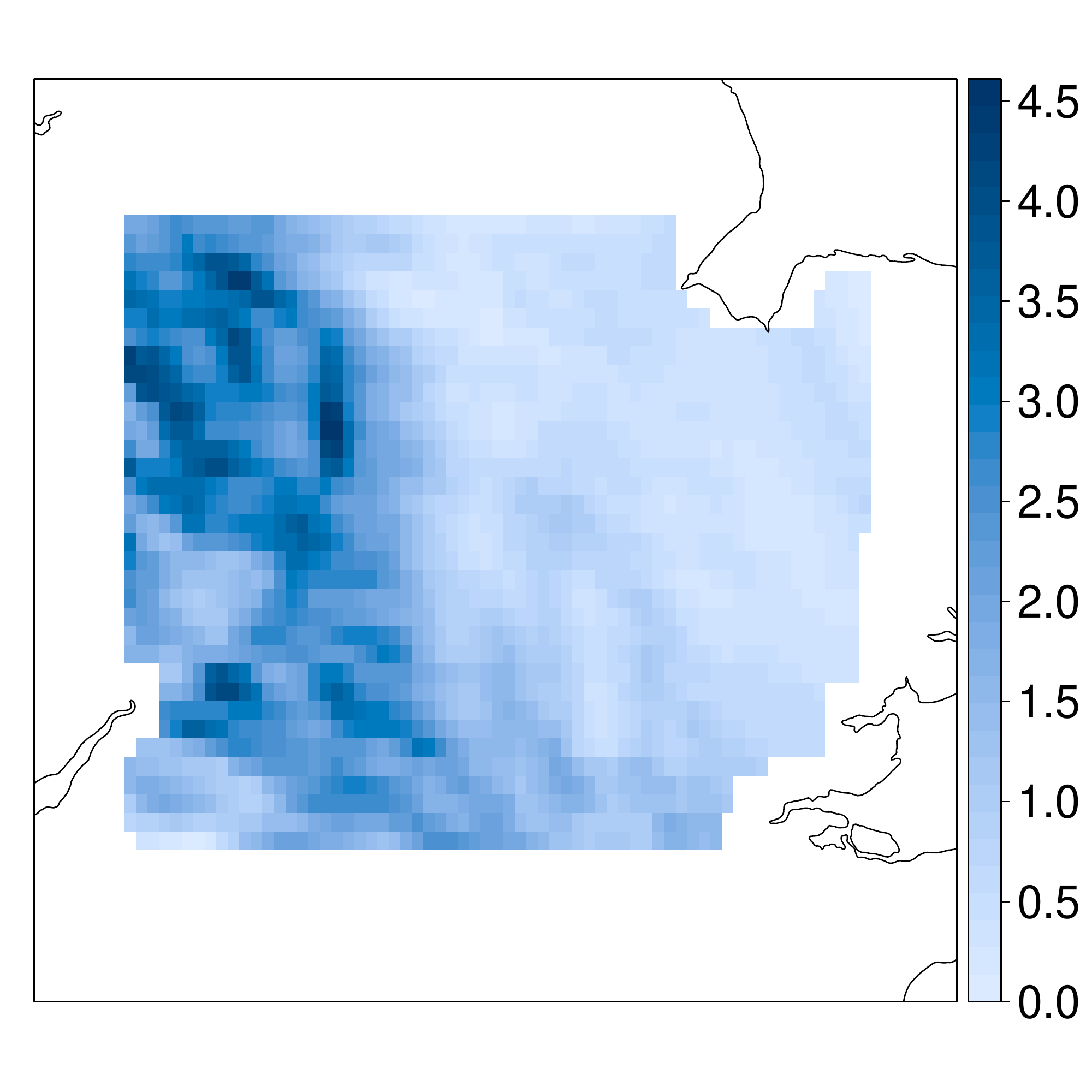} 
\end{minipage}
\begin{minipage}{0.32\linewidth}
\centering
\includegraphics[width=\linewidth]{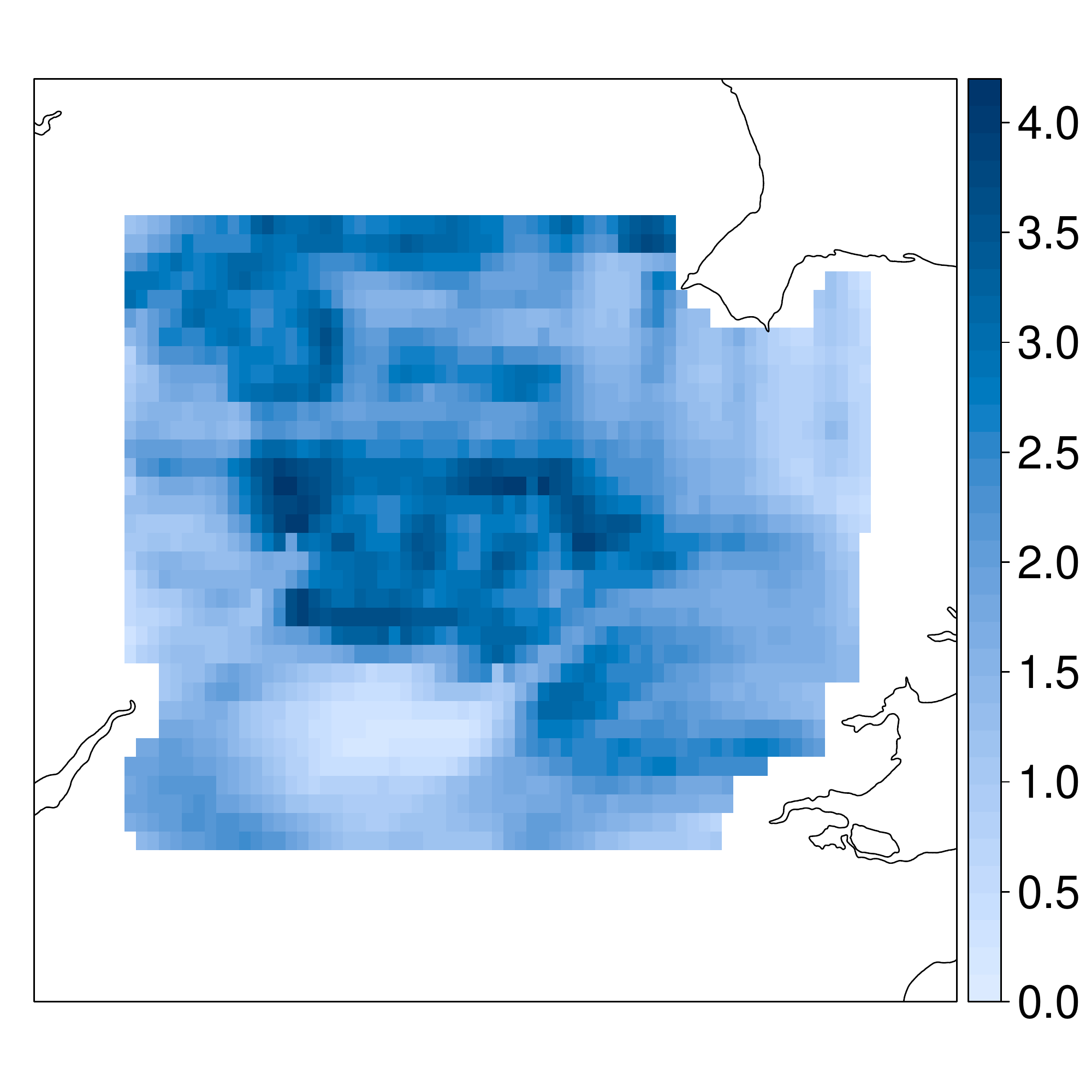} 
\end{minipage}
\begin{minipage}{0.32\linewidth}
\centering
\includegraphics[width=\linewidth]{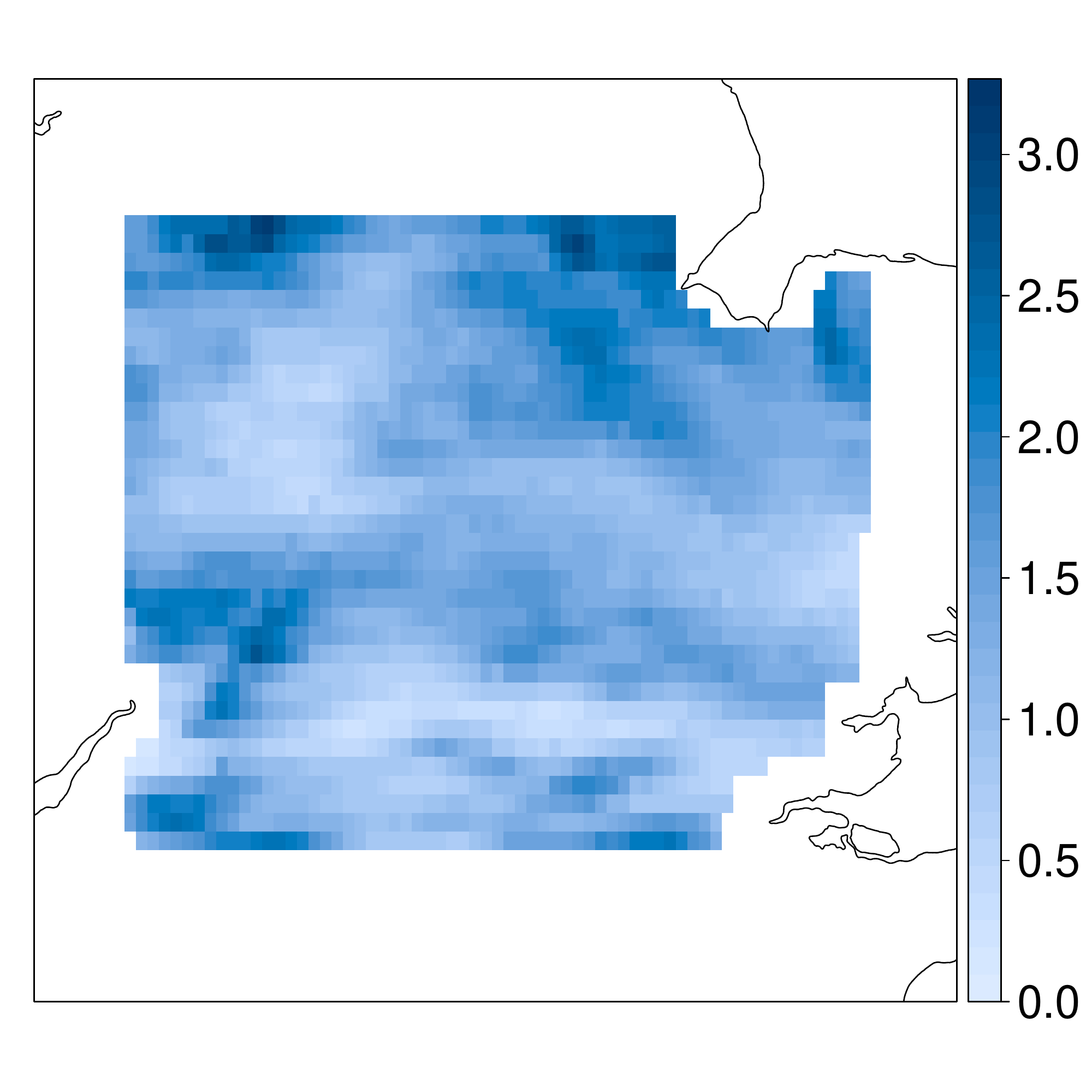} 
\end{minipage}
\caption{Observed extreme fields identified as convective (top) and non-convective (bottom) (mm/hr).}
\label{observations}
\end{figure}

 %River flooding is caused by prolonged heavy rainfall over a designated catchment area, with the size of such areas varying dramatically between rivers in the UK. For example, the approximate catchment areas of the UK river Lune, that flows through Lancaster, and the Severn, the largest river in the UK, are $1,300 km^2$ and $11,000 km^2$, respectively. Hence, a key aspect of river flood risk management is understanding the extremal behaviour of precipitation accumulations at different spatial scales. Our focus will be on catchments of appropriate size for the UK, and we will discuss the complexities of working with larger regions in Section~\ref{sec_discuss}. Whilst inference on precipitation extremes can be conducted at different spatial scales separately, and many studies have done so (see \cite{feh1999}), ideally we require an approach that can be used for joint estimation of these variables at multiple spatial scales simultaneously. The benefits of such an approach include a framework for modelling dependence between aggregates over different regions and physically consistent inference, i.e., a natural ordering for quantiles of aggregates over nested regions \citep{richards2021modelling}. Moreover, this approach offers an increase in the reliability of inference on aggregates over small regions as more data will be used for modelling.\par
Building a parsimonious model for extreme precipitation at multiple spatial scales is difficult, as the important aspects of the underlying processes that drive the extremal behaviour of precipitation change with spatial resolution. Extreme precipitation is caused by a mixture of known and well-understood processes.  The primary drivers are convection, which produces spatially localised, high-intensity precipitation events over short time scales \citep{Schroeer2018}, and processes that cause large scale extreme events, e.g., frontal systems producing stratiform precipitation of lower intensity but affecting a much larger area for a longer duration \citep{berg2013strong, catto2013importance, gregersen2013assessing}; Figure~\ref{observations} shows observations of these two classes of process. The distinction between these two types of events and their space-time characteristics is an important consideration when modelling the upper-tail behaviour of aggregates of rain over different spatial scales; we highlight major differences in the marginal, and extremal dependence, structures of both processes in our application, and show that the relative contribution of frontal events to extreme spatial aggregates of precipitation increases with the size of the region over which aggregates are made. In addition, future changes due to global warming may not affect these two mechanisms equally.
\par Whilst inference on precipitation extremes can be conducted at different spatial resolutions separately \citep{feh1999}, ideally we would adopt an approach that can be used for joint estimation of these variables at multiple spatial scales simultaneously; the benefits include a framework for modelling dependence between aggregates over different regions and physically consistent inference, i.e., a natural ordering for quantiles of aggregates over nested regions \citep{richards2021modelling}. Moreover, this approach has the potential to increase the reliability of inference on aggregates at lower-resolution spatial scales as more data will be used for modelling. \par
 In this work we focus on the spatial aggregation of short duration (hourly) extreme rainfall and the modelling of the mixture between convective and non-convective rainfall, showing how that mixture depends on aggregation scale.  We restrict the spatial scales considered to the native resolution of the CPM used ($2.2$km) up to that of UK river catchments, the largest being the Severn at $11,000$km$^2$, in anticipation of future work that will consider the issues of aggregation over time and how future extreme rainfall might change in the future.
\subsection{Data}
\label{data-sec}
\begin{figure}[h!]
\centering
\includegraphics[width=0.4\linewidth]{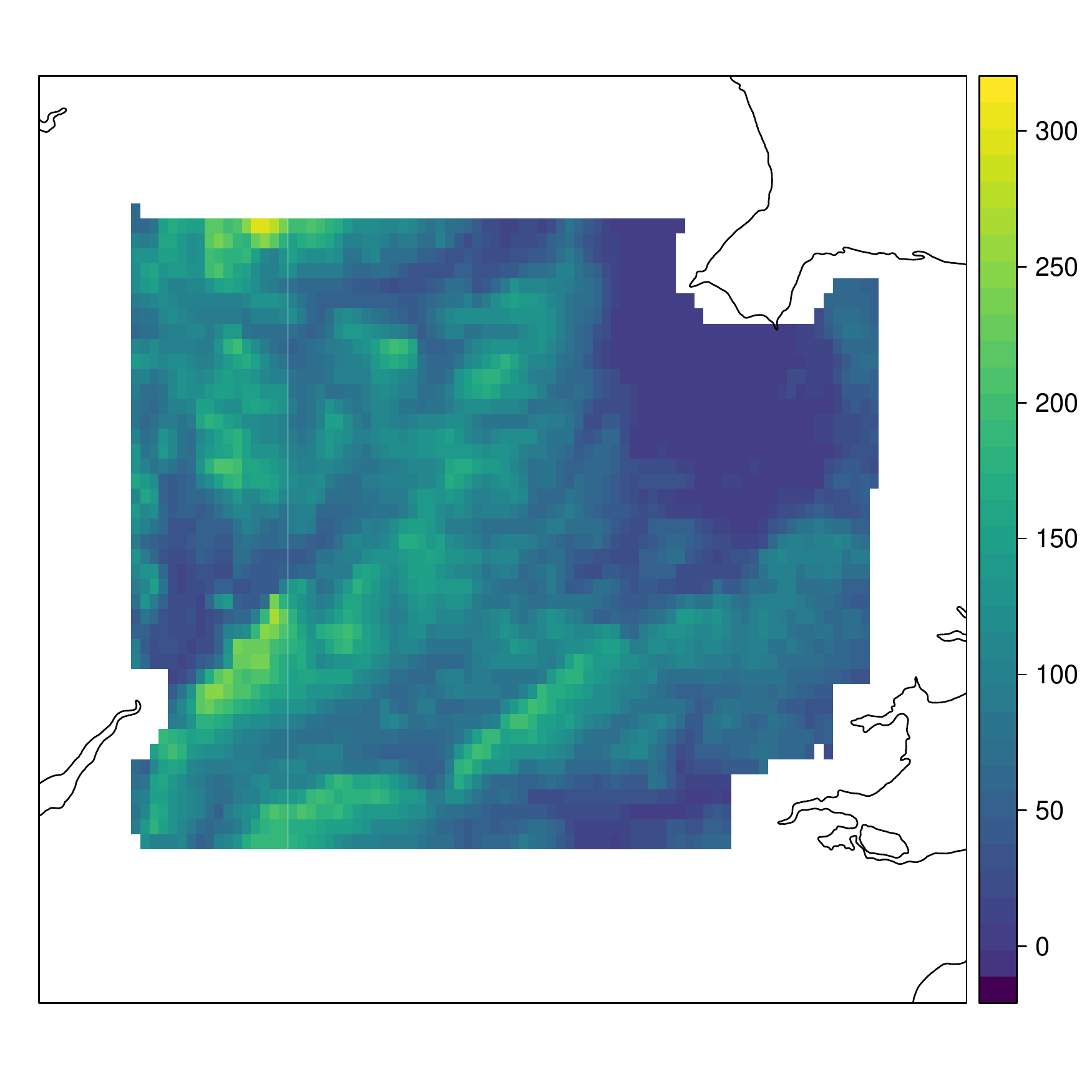} 
\caption{A map of elevation (m) for the spatial domain $\mathcal{S}$ of interest.}
\label{elev}
\end{figure}

We analyse average hourly precipitation rates (mm/hour) taken from the UKCP18 CPM projections \citep{ukcp18CPMscience, kendon2021}, 1981--2000. \citet{richards2021modelling} consider the same data on the coarser resolution 5km$\;\times\;$5km grid, but data we use are on the finer resolution native climate model grid over a much larger spatial domain and the number of sampling locations increases from $934$ to $7526$. % We analysed the first and fourth ensemble members from the UKCP18; however, we present the latter analysis only, as we identified abnormally large and physically unrealistic values in the former.
The sampling locations are $2.2 \times 2.2\; $km$^2$ grid boxes and the spatial domain $\mathcal{S}$ of interest is the southern U.K., approximately centred at Northampton and covering an area of $198 \times 231\; $km$^2$ with data only sampled over land. The study domain $\mathcal{S}$ is illustrated in Figure~\ref{elev}.\par
 Each observation corresponds to the average precipitation over a grid-box. As extreme hourly precipitation is more likely to occur in summer, we use June--August observations only, yielding $43200$ fields and eliminating the need to include seasonality effects; Figure~\ref{observations} illustrates six such fields with the top, and bottom, rows of panels displaying events that produce extreme spatial aggregates over small, and large, scales respectively. We follow \cite{richards2021modelling} and treat the centre of each grid box as a sampling location, and use the great-circle distance metric. We also set all values of the data less than $1 \times 10^{-5}$mm/hour to zero, since this level of precipitation would be recorded as zero by a rain gauge.
\subsection{Overview}
\cite{richards2021modelling} develop methodology for joint inference on precipitation extremes at different spatial scales; they consider a spatial process $\{Y(s): s \in \mathcal{S}\}$ for some spatial domain $\mathcal{S} \subset \mathbb{R}^2$, and
let data used for modelling be observations $\mathbf{y}_{t}=(y_t(s_1),\dots,y_t(s_d))^T$ at times $t\in\mathbf{T}=\{1,\dots, n\}$ and sampling locations $\boldsymbol{S}=(s_1,\dots,s_d)\subset \mathcal{S}$. Specifically, they perform inference on the upper-tail of the aggregate variable
\begin{equation}
\label{agg_eq}
\bar{R}_{\mathcal{A}}=\frac{1}{|\mathcal{A}|}\int_{\mathcal{A}}Y(s)\mathrm{d}s,
\end{equation}
where $\mathcal{A}\subset \mathcal{S}$ is a sub-region of interest. To achieve this, they propose a model for the extremal behaviour of $\{Y(s):s \in \mathcal{S}\}$ by adapting the spatial conditional extremes approach, first proposed by \cite{wadsworth2018spatial} and with extensions by \cite{Tawn2018,shooter2019, shooter2021spatial,shooterinpress,huser2020advances,simpson2020conditional} and \cite{simpson2020highdimensional}. Realisations taken from this model are used to conduct inference on $\bar{R}_\mathcal{A}$ for numerous regions $\mathcal{A} \subset \mathcal{S}$ simultaneously. Whilst the approach of \cite{richards2021modelling} improved on previous approaches by ensuring self-consistent inference across different sub-regions of interest and flexible extremal dependence modelling, they make the restrictive assumption that both the marginal and dependence behaviour of the data generating process is constant over time which, as we illustrate in Section~\ref{sec-classify}, is a poor assumption for precipitation; we adapt their approach and propose separate models for convective and non-convective precipitation, and illustrate throughout the paper that this distinction is necessary due to the disparity between both the marginal and dependence behaviour of extreme events arising from the two classes of process. Alongside new parametric forms for the dependence parameters of the \cite{wadsworth2018spatial} model, further extensions of the approach of \cite{richards2021modelling} include an approach for simulating events from processes which suffer particularly badly from boundary effects due to their long-range dependence, e.g., non-convective events.\par
Although we employ a conditional approach to model extremal dependence within the process $Y(s)$, other models have been applied in this context, e.g., max-stable processes \citep{coles1993,padoan2010likelihood,westra2011detection,reich2012hierarchical}, Pareto processes \citep{palacios2020generalized,deFondeville2020functional} and censored Gaussian copulas \citep{sang2010continuous,thibaud2013threshold}. These models make restrictive assumptions about extremal dependence which are not satisfied by our data, described in Section~\ref{EXPLORE_SEC}; moreover, \cite{richards2021modelling} illustrate that modelling $Y(s)$ with the same extremal dependence as a max-stable or Pareto process leads to severe bias in estimates of return-levels for $\bar{R}_\mathcal{A}$.
\par
The structure of the paper is as follows. Section~\ref{sec-classify} describes an algorithm for classifying observed fields into convective or non-convective precipitation events, and gives an overview of our mixture model setup. Details for modelling the extremes of mixture components are provided in Section~\ref{model_Extend}, and the inference framework is described in Section~\ref{infer_sec}. We simulate from the two fitted models for both classes of data using a procedure described in Section~\ref{sec-sim}, and then combine realisations from both models to estimate the upper-tail behaviour of $\bar{R}_\mathcal{A}$. To illustrate the improvements to modelling $\bar{R}_\mathcal{A}$ that our approach provides to that detailed by \cite{richards2021modelling}, we apply both methods to the precipitation data described in Section~\ref{data-sec}.
\pagebreak
 \section{Mixture process}
 \label{sec-classify}
 \subsection{Mixture component classification}
We classify observation times  $ t \in \mathbf{T}$ for the entire observed fields $\{y_t(s): s \in \mathcal{S}, t = 1,\dots, n\}$ into two sets, denoted $\mathcal{C}$ and $\mathcal{N}$, which correspond to ``convective'' and ``non-convective'' times, respectively; that is, if $t\in\mathcal{C}$, then the observed field $\{y_t(s):s \in \mathcal{S}\}$ is a convective event, and similarly for non-convective events. For the classification, we use Algorithm~\ref{convect-ident-algo}, developed at the Met Office Hadley Centre, U.K., for summer precipitation data with gridded sampling locations \citep{kendon2012realism}. At time $t$, the algorithm makes its classification based on the gradient of the surface of $y_t(s)$ in a neighbourhood around $s_i$ for all sites $s_i \in \boldsymbol{S}$, as convective rainfall are highly localised and so large gradients are expected if convection is present. Gradients are calculated using the difference between precipitation rates at adjacent grid-boxes in a $n_g \times n_g$ grid centred at $s_i$, denoted $\mathcal{L}_i$, as the data are observed on a regular grid. If it is not possible to create $\mathcal{L}_i$, e.g., at the edges of $\mathcal{S}$, then $s_i$ is removed. Any field not identified as convective is automatically classified as non-convective; this includes any fields with no rain at any $s \in \mathcal{S}$.\par
\begin{algorithm}[h]
\caption{Identify convective fields}
\label{convect-ident-algo}
\textbf{[Hyperparameters]}: gradient thresholds $g_l > 0, g_u > g_l$, proportion of large rainfall $p^*\in[0,1]$, neighbourhood size $n_g \in \{2d^*-1,d^* \in \mathbb{N}\}$\\
\textbf{[Data]}: Observations $\{y_t(s):s \in \boldsymbol{S},t \in \mathbf{T}\}$\\
For all $t=1,\dots n$:
\begin{enumerate}
\item For all $i=1,\dots,d$:
\begin{enumerate}
\item Identify the $n_g \times n_g$ local neighbourhood $\mathcal{L}_i$ for $s_i$, defined in Section~\ref{sec-classify}.
\item Evaluate all differences $\mathcal{G}_i=\{y_t(s_j)-y_t(s_k): s_j,s_k \in \mathcal{L}_i\}$.
\item Calculate the proportion $p_{g,i}=|\{g \in \mathcal{G}_i:g \geq g_u\}|/ |\{g \in \mathcal{G}_i: g \geq g_l\}|$.
\item \label{algo-step} If $p_{g,i} \geq p^*$, then label $y_t(s_i)$ as convective.
\end{enumerate}
\item \label{algo-step2}  If any elements of $\{y_t(s): s \in \boldsymbol{S}\}$ are labelled as convective, then the entire field is classified as convective, i.e., $t\in\mathcal{C}$. Otherwise, $t \in \mathcal{N}$.
\end{enumerate}
\end{algorithm}
   Algorithm~\ref{convect-ident-algo} involves four hyperparameters, which we take to be $g_l = 0.01, g_u=1$ and $p^* = 0.2$ and $n_g = 9$; these were tuned by the Met Office through visual inspection, and specifically for the climate model we consider (Roberts, N. and Kendon, E., 2020, personal communication). The most critical hyperparameters are $g_u$ and $p^*$; if $g_u$ is set too low or $p^*$ too high, then stratiform, or frontal, precipitation may be classified as convective; the lower threshold $g_l$ simply removes the effect of any grid-cells with very little rainfall, i.e., drizzle, and $n_g$ determines the size of the region around site $s_i$ for which the surface of $y_t(s)$ is considered. With these hyperparameters, the algorithm gives $|\mathcal{C}|=17724$ convective, and $|\mathcal{N}|=25976$ non-convective, fields of our data, with examples of both classes presented in Figure~\ref{observations}; these selected fields were randomly sampled from fields which gave site-wise maxima for their respective processes. Note that the 1481 fully-dry fields, where there is zero rainfall at all sites, are classified as non-convective. Fields identified as being non-convective appear smoother over space and with much lower marginal magnitude than the convective fields; note the difference in the scales of Figure~\ref{observations}.
 \subsection{Mixture model}
We assume that, for each $t \in \mathbf{T}$, the entirety of the process $\{Y_t(s):s \in \mathcal{S}\}$ can be described by one of two mixture components, denoted by $\{Y_{\mathcal{C},t}(s): t \in \mathcal{C} \}$ and $\{Y_{\mathcal{N},t}(s):t \in \mathcal{N}\}$, that describe convective and non-convective rainfall, respectively. If $t\in \mathcal{C}$, then $Y_t$ can be described solely by $Y_{\mathcal{C},t}$, rather than a mixture of both components, and similarly for $t\in\mathcal{N}$ and $Y_{\mathcal{N},t}$. Differences in the marginal and dependence behaviour of the two mixture components can be observed in Figures~\ref{observations} and \ref{chiempfig}, respectively, and so we propose different marginal and dependence models for $Y_{\mathcal{C},t}$ and $Y_{\mathcal{N},t}$. We assume that, for each $\mathcal{C}$ and $\mathcal{N}$ process, both the marginal behaviour and dependence structures are stationary with respect to time; that is, the process $\{Y_{\mathcal{C},k}(s)\}$ is identically distributed to $\{Y_{\mathcal{C},l}(s)\}$ for all $l,k \in \mathcal{C}$, and similarly for the non-convective process $Y_{\mathcal{N},t}$. We then define the mixture process $\{Y_{\mathcal{M},t}(s)\}$ by
\begin{equation}
\label{totalprob}
\{Y_{\mathcal{M},t}(s):s \in \mathcal{S}\}=\begin{cases}\{Y_{\mathcal{C},t}(s):s\in\mathcal{S}\},\;\;\;&\text{with probability }p_\mathcal{C},\\
\{Y_{\mathcal{N},t}(s):s\in\mathcal{S}\},\;\;\;&\text{with probability }1-p_\mathcal{C},
\end{cases}
\end{equation}
where $p_\mathcal{C}\in[0,1]$. We estimate $p_\mathcal{C}$ empirically using $|\mathcal{C}|/(|\mathcal{C}|+|\mathcal{N}|)$ and Algorithm~\ref{convect-ident-algo} gives $\hat{p}_\mathcal{C}\approx 39.9\%$. As our interest lies only in spatial aggregates, we do not model temporal dependence in either component of $Y_{\mathcal{M},t}$ but account for this dependence through use of a block bootstrap in Section~\ref{App_sec}; henceforth we drop the time index from notation and use the shorthand $Y_\mathcal{M}:=\{Y_\mathcal{M}(s):s\in\mathcal{S}\}$ (and similarly for $Y_\mathcal{C}$ and $Y_\mathcal{N}$). \par
We compare two modelling approaches for $\{Y(s)\}$: an approach using the existing model developed by \cite{richards2021modelling}, which we denote throughout by $\{Y_\mathcal{E}(s)\}$ or the shorthand $Y_\mathcal{E}$ , which ignores any mixture structure in $Y$; and our improved approach which uses model \eqref{totalprob}. Note that $Y_\mathcal{E}$ and $Y_\mathcal{M}$ are equivalent when $p_\mathcal{C}=0$ or $p_\mathcal{C}=1$, or if $Y_\mathcal{C}(s)=Y_\mathcal{N}(s)$ for all $s\in\mathcal{S}$. Although observations of the process $Y(s)$ are assumed to arise from some probabilistic mixture, we are able to pre-classify observations from each mixture component; in this regard, our approach differs from standard methods using mixture models. Whilst there has been some recent focus on incorporating mixture structures into multivariate extremal dependence models \citep{simpson2020determining,tendijck2021modeling}, very few methods has been proposed for mixture modelling of spatial extremes; the available models are typically restricted to a single extremal dependence class \citep{hazra2019estimating}, which, as we illustrate in Section~\ref{sec:dep_model}, make them inappropriate for our data. Although model \eqref{totalprob} is mathematically simple, from an applied perspective it represents an important and novel step towards flexible mixture modelling of spatial extremes. In Section~\ref{appl-inferaggs}, we illustrate that the inclusion of the ``rudimentary" mixture structure in eq.\ \eqref{totalprob} leads to a better model fit and improved inference on the extremes of $\bar{R}_\mathcal{A}$.
\subsection{Model justification}
Model \eqref{totalprob} introduces separate mixture components that describe both convective and non-convective events over the entire spatial domain $\mathcal{S}$. The assumption that an entire spatial field, i.e., any instantaneous precipitation event observed over $\mathcal{S}$, can be neatly classified into distinct categories is unlikely to be valid in general, particularly as $|\mathcal{S}|$ increases; for very large $|\mathcal{S}|$, we may observe convective and non-convective events occurring simultaneously within $\mathcal{S}$ or even multiple independent convective events. We aim to counter these concerns by focusing on only events leading to extreme precipitation aggregates and by restricting the scale of the domain $\mathcal{S}$ to that of river catchments, which in the UK has an area of less than $11,000$km$^2$. In Section~\ref{EXPLORE_SEC}, we highlight differences in the empirical extremal dependence structure of the two classes of data determined using Algorithm~\ref{convect-ident-algo}.
\par
For an observed field, Algorithm~\ref{convect-ident-algo} first identifies individual grid-boxes, i.e., sites, as convective (Step~\ref{algo-step} of Algorithm~\ref{convect-ident-algo}); it then classifies the entirety of the field as convective (Step~\ref{algo-step2} of Algorithm~\ref{convect-ident-algo}), if at least one grid-cell is convective. Theoretically, this can lead to situations where entire fields are classified as convective courtesy of a single site (although we did not find this in practice). We investigate the proportion of convective grid-cells within a field to determine if the assumption of model~\eqref{totalprob}, that a field at time $t$ is either wholly convective or non-convective but never a mixture of both, is appropriate for the data. As our interest lies in the extremes of $\bar{R}_\mathcal{A}$, we calculate the proportion of large values of empirical spatial aggregates (of different sizes) that are contributed to by convective sites, whenever any convective rainfall is present in the field; this is achieved via the heuristic described in Algorithm~\ref{heur_algo}. If the estimated proportion is significantly less than one, then this would be an indication of there being mixing of the two rainfall types within extreme spatial aggregates and evidence to suggest that the model~\eqref{totalprob} cannot be reasonably assumed for the data.

\begin{algorithm}[h]
\caption{Empirical proportion of convective rainfall leading to extreme spatial aggregates}
\label{heur_algo}
\textbf{[Hyperparameters]}: radius $r>0\;[r\in\{30,85\}$km], number of samples $[M=50]$\\
\textbf{[Data]}: All replications with $t\in\mathcal{C}$ temporal replications
\begin{enumerate}
\item For all $t\in\mathcal{C}$:
\begin{enumerate}
\item Repeat for $M$ samples:
\begin{enumerate}
\item Sample a spatial location uniformly over the domain $s_c\in\mathcal{S}$.
\item \label{heur_step} Compute sample total of hourly rainfall rate over all sites within a radius $r$.
\item \label{heur_step2} Compute sample total of hourly rainfall rate over all sites identified as convective (using Step~\ref{algo-step} of Algorithm~\ref{convect-ident-algo}) within a radius $r$.
\end{enumerate}
\end{enumerate}
\item Keep only values where samples from Step~\ref{heur_step} exceed the empirical $90\%$ quantile over all corresponding $M\times\mathcal{C}$ samples.
\item Compute proportions of extreme total hourly rainfall contributed to by convective sites, i.e., divide samples in Step~\ref{heur_step2} by those from Step~\ref{heur_step}. 
\end{enumerate}
\end{algorithm}
Ideally, for Algorithm~\ref{heur_algo} we would create samples of extreme aggregates over a specific aggregate region; however, this would leave very few samples available for estimation of the required proportion. We instead make an assumption of stationarity, i.e., that the distribution of the proportion of interest is constant over space and time; this allows us to pool information across sites and observations. Then to create a sample of spatial aggregates which is independent of the position of the aggregate region, we randomly sample a centre and evaluate the spatial average over a circular region with radius $r>0$ and repeat this process $M$ times for each convective field. We then consider only the largest values within the sample of the aggregates, i.e., those that exceed the sample $90\%$ quantile, and compute the proportion of their total value contributed to by observed values at convective grid-cells. We chose $M=50$, large enough that reasonably-sized samples  were created, with larger values not providing a significant change in the results. 
We found strong evidence that, for a range of radii ($r=30$km, $r=85$km and the whole study domain), the majority of extreme aggregates are driven by mostly convective rainfall, with the median estimate of the proportion for small and large aggregate regions being $0.92$, $0.81$ and $0.78$, respectively. Hence we proceed under the assumption that model~\eqref{totalprob} is appropriate for the data.
\subsection{Exploratory analysis for mixture components}
\label{EXPLORE_SEC}
\begin{figure}[h!]
\centering
\begin{minipage}{0.49\linewidth}
\includegraphics[width=\linewidth]{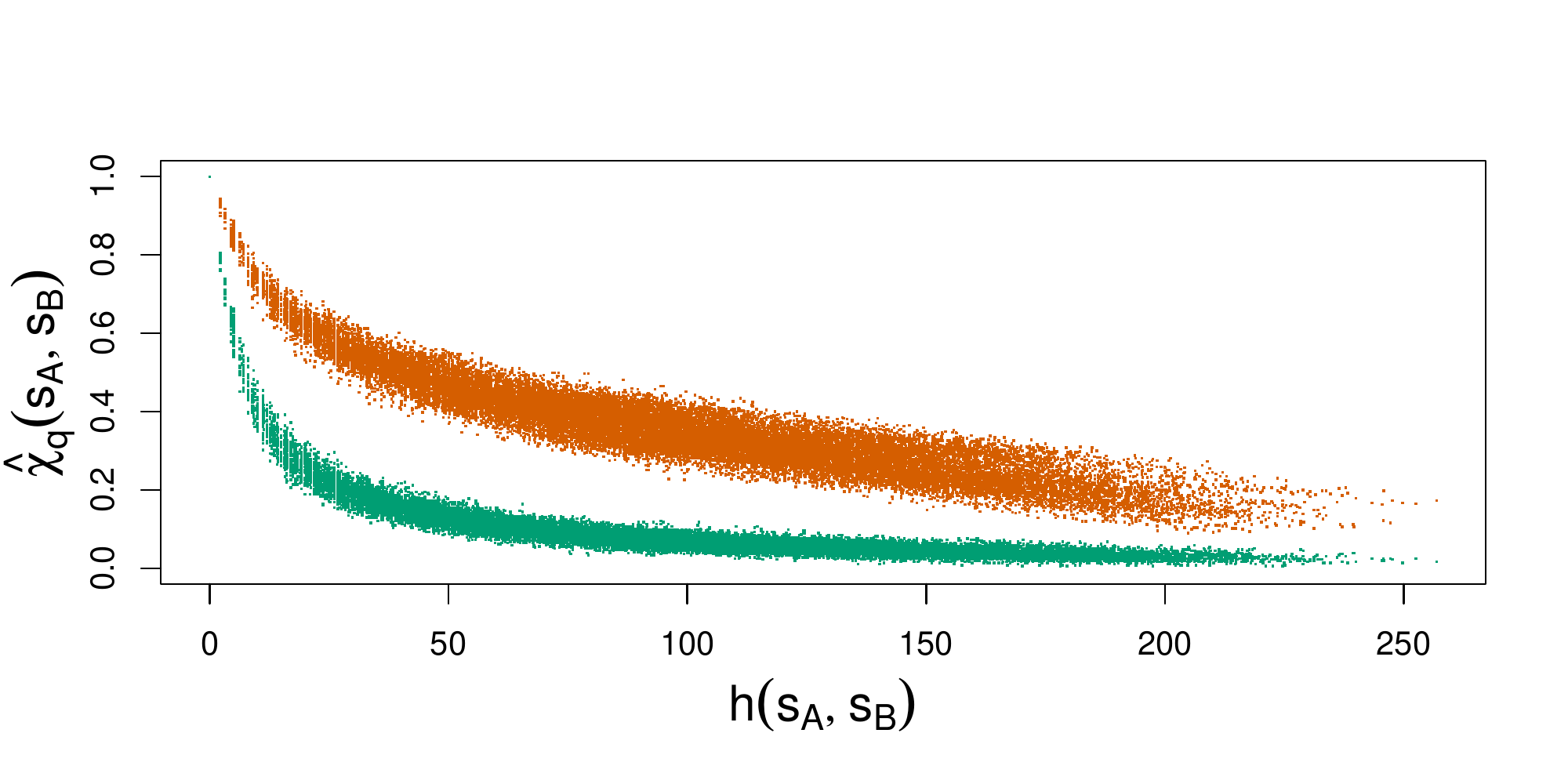} 
\end{minipage}
\begin{minipage}{0.49\linewidth}
\includegraphics[width=\linewidth]{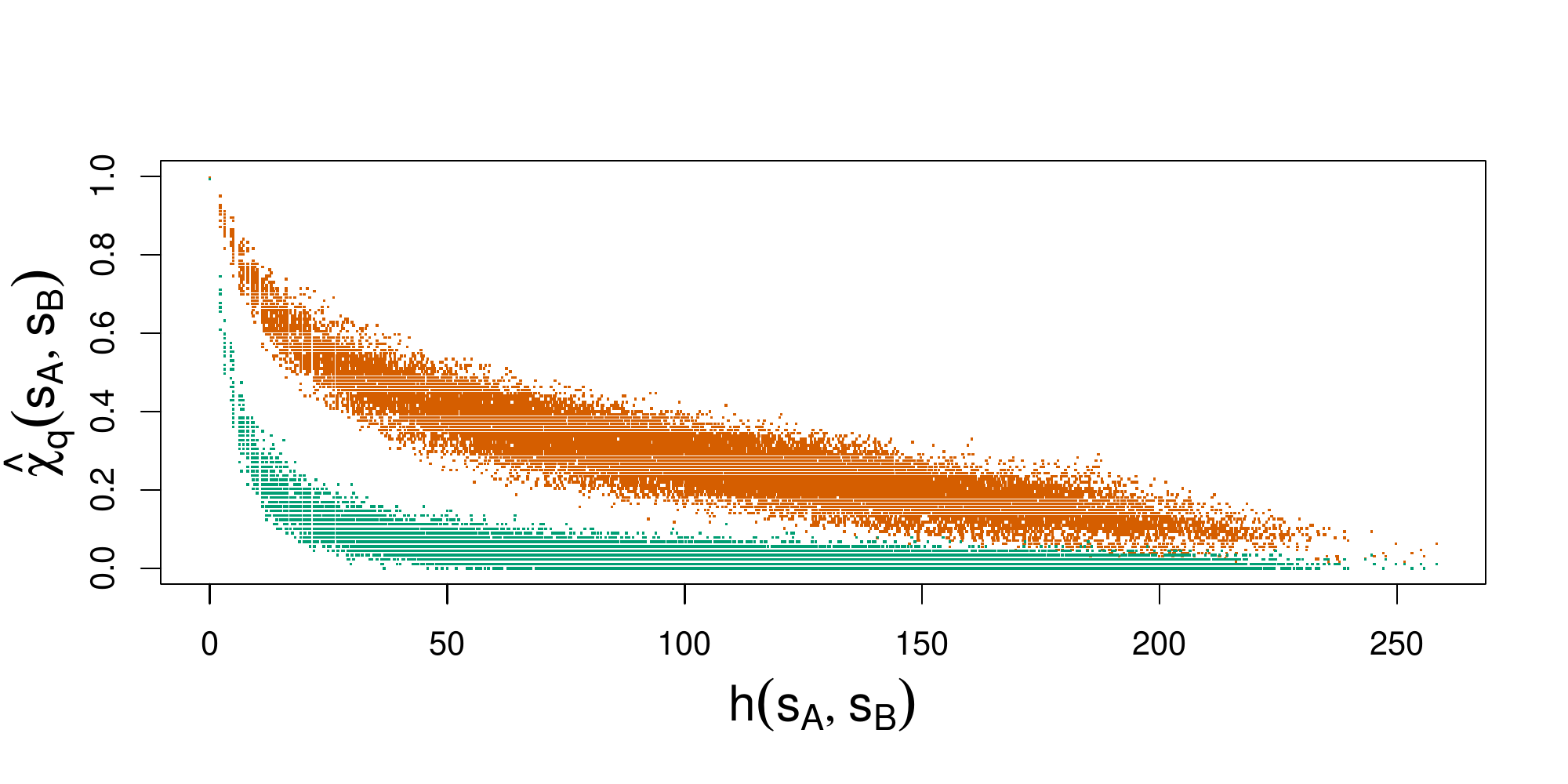} 
\end{minipage}
\caption{Empirical estimates of pairwise extremal dependence strength $\hat{\chi}_q(s_A,s_B)$ of eq.\ \eqref{chieq} for $q=0.98$ (left) and $q=0.995$ (right) $s_A,s_B\in\mathcal{S}$ against great-circle distances $h(s_A,s_B):=\|s_A-s_B\|$, which are given in km. The green and brown points correspond to estimates for convective and non-convective realisations, respectively.}
\label{chiempfig}
\end{figure}

We expect the two mixture components defined in eq.\ \eqref{totalprob} to exhibit different extremal dependence structures, as they describe very different rainfall processes: convective and non-convective rainfall. We quantify extremal dependence for a process $\{Y(s)\}$ using the upper tail index $\chi(s_A,s_B):=\lim_{q \uparrow 1}\chi_q(s_A,s_B)$ \citep{joe1997multivariate} for all $s_A,s_B \in \mathcal{S}$, defined by 
 \begin{equation}
\label{chieq}
 \chi_q(s_A,s_B)=\Pr\left\{Y(s_B) > F^{-1}_{Y(s_B)}(q)\mid Y(s_A) > F^{-1}_{Y(s_A)}(q)\right\},
 \end{equation}
where $F^{-1}_{Y(s_A)}$ denotes the quantile function of $Y(s_A)$. The measure $\chi(s_A,s_B)$ determines the extremal dependence class; if $\chi(s_A,s_B) > 0$ or $\chi(s_A,s_B)=0$, then $Y(s_A)$ and $Y(s_B)$ are asymptotically dependent or asymptotically independent, respectively \citep{Coles2001book}. \cite{richards2021modelling} found evidence that convective hourly rainfall exhibits asymptotic independence at the smallest observable spatial distances, i.e., $\chi(s_A,s_B) = 0$ for sites $s_A, s_B \in \mathcal{S}$ separated by at least $2.2$km, inferring that extreme events of $Y_\mathcal{C}$ become increasingly more localised as the magnitude of the events gets larger. However, we find that non-convective events exhibit asymptotic dependence at short-range and asymptotic independence at long-range. To illustrate the need for different extremal dependence structures for both classes of rainfall, we compute empirical estimates $\hat{\chi}_q(s_A,s_B)$ of eq.\ \eqref{chieq} for both classes of data; Figure~\ref{chiempfig} presents these for all pairs of a subset of 400 sites randomly sampled over $\mathcal{S}$ and with $q=0.98$ and $q=0.995$. We observe that non-convective rainfall (brown points) exhibits much stronger extremal dependence than convective rainfall (green points), as values of $\chi_q(s_A,s_B)$ decay much slower with distance for the brown points than the green points. We also observe that, as $q$ increases, $\chi_q(s_A,s_B)$ appears to tend to zero at a much faster rate for convective rainfall; our model fits in Section~\ref{sec:dep_model} suggest that the assumption of short-range asymptotic dependence, i.e., $\chi(s_A,s_B)>0$ for close $s_A$ and $s_B$, may be appropriate only for non-convective rainfall. We adapt the approach of \cite{richards2021modelling} and, in Section~\ref{model_Extend}, propose slightly different parametric sub-models for the two processes to accommodate the differences in their extremal dependence structure.
\pagebreak
\section{Modelling extremes of each mixture component}
\label{model_Extend}
\subsection{Overview}
\label{margsec_overview}
Following \cite{richards2021modelling}, we adopt a two-step modelling approach. We first standardise the data using the marginal model described in Section~\ref{marg_sec}; this model is presented for a generic stationary process $\{Y(s):s \in \mathcal{S}\}$, but is fitted separately for each of the processes $Y_\mathcal{C},Y_\mathcal{N}$ and $Y_\mathcal{E}$. The marginal fits are used to perform site-wise standardisation of data to standard Laplace margins; we denote these standardised processes by $X_\mathcal{C}$, $X_\mathcal{N}$ and $X_\mathcal{E}$.\par
To characterise extremal dependence, we use the spatial conditional extremes framework \citep{wadsworth2018spatial}; this is detailed for a generic standardised process $\{X(s): s \in \mathcal{S}\}$ in Section~\ref{dep_sec:generic}, but is applied, separately, to each of $X_\mathcal{C}$, $X_\mathcal{N}$ and $X_\mathcal{E}$. Slight differences in modelling choices are made for the three processes, i.e., choices of parametric forms for some dependence functions and hyper-parameters for simulation; discussions of these differences are provided in Sections~\ref{sec_differ} and \ref{choose_sec}, respectively.
\subsection{Marginal model}
\label{marg_sec}
We model the site-wise marginals of $\{Y(s):s \in \mathcal{S}\}$ using three components: generalised Pareto tails \citep[p.~75]{Coles2001book} above some high threshold $q_\lambda(s)$, the $(1-\lambda)-$quantile of $Y(s)$, the empirical distribution for the bulk and a discrete mass at the lower tail, denoted $p(s)$, which corresponds to the probability that no rain occurs at site $s\in\mathcal{S}$. The distribution function of $Y(s)$, for all $s \in \mathcal{S}$, is
\begin{equation}
\label{chap6-MargTransform}
F_{Y(s)}(y)=
\begin{cases}
p(s),&\text{if}\;\;y = 0,\\
\{1-\lambda-p(s)\}\frac{F_{Y_+(s)}(y)}{F_{Y_+(s)}\{q_\lambda(s)\}}+p(s),&\text{if}\;\;0 < y \leq q_\lambda(s),\\
1-\lambda\left[1+\frac{\xi\{y-q_\lambda(s)\}}{\upsilon(s)}\right]^{-1/\xi}_+, &\text{if}\;\;y > q_\lambda(s),\\
\end{cases}
\end{equation}
where $[x]_+=\max\{x,0\}$, and where $\xi\in \mathbb{R}$, $\upsilon(s) > 0$ and $p(s) \geq 0,\;0<\lambda < 1$ and $F_{Y_+(s)}(y)$ denotes the empirical distribution function of strictly positive values of $Y(s)$; we further require that $ {p(s) + \lambda < 1}$, suggesting that our marginal model is unlikely to perform well if $p(s)$ is close to $1$, as only a small number of observations would then be available to estimate the generalised Pareto distribution parameters. \par
The marginal parameter functions $p(s), q_\lambda(s)$ and $\upsilon(s)$ are represented through a basis of thin-plate splines. We estimate $p(s)$ using a logistic generalised additive model \citep{wood2006generalized} and estimate $q_\lambda(s)$ for all $s \in \mathcal{S}$ using additive quantile regression \citep{qgam}; the latter technique is computationally expensive and so we use only a subset of sites for estimation. \cite{richards2021modelling} propose a technique for estimating $q_\lambda(s)$ by fitting a thin-plate spline through point-wise estimates of $q_\lambda(s)$ for each $s \in (s_1,\dots,s_d)$; however, our approach better accounts for uncertainty in the spline estimates as more data are used in their estimation, i.e., all observations at a single site rather than a single quantile estimate. The scale parameter $\upsilon(s)$ is estimated by fitting a generalised Pareto, with parameters represented as additive functions, to the exceedances of $q_\lambda(s)$, i.e.,  $Y(s)-q_\lambda(s)$, with a fixed shape parameter $\xi$ for all $s\in \mathcal{S}$ \citep{youngman2019}. Our choice to fix $\xi$ over space was fully supported by our data and is a common approach taken when modelling spatial characteristics of extreme rainfall as it reduces the risk of having parameter identifiability issues \citep{thibaud2013threshold,Zheng2015,SAUNDERS201717, Brown2018}. \par Exploratory analysis of the data reveals that elevation is an important covariate in the marginal behaviour of precipitation, an observation supported by \cite{coles1996,cooley2007} and \cite{cooley2010spatial}. Hence, we allow $p(s), q_\lambda(s)$ and $\log \upsilon(s)$ to vary with both location and elevation; separate spline bases are used for both.
 Each spline consists of as few basis functions as possible, i.e., four, to ensure that the parameter surfaces are relatively smooth; this is to make the marginal fits physically interpretable and to reduce over-fitting. Our approach for inference differs from that proposed by \cite{richards2021modelling} in that  we allow the margins to change with elevation. 
\subsection{Extremal dependence model}
\label{dep_sec}
\subsubsection{Generic model}
\label{dep_sec:generic}
We model extremal dependence of the standardised process $\{X(s):s \in \mathcal{S}\}$, given that it is extreme for some $s \in \mathcal{S}$, by conditioning on the process being above some high threshold $u$ at a specified site $s_O \in \mathcal{S}$ for each $s_O \in \mathcal{S}$. The process $\{X(s)\}$ is assumed to be stationary with dependence a function of distance, i.e.,  $h(s_A,s_B)=\|s_A-s_B\|$ for $s_A, s_B \in \mathcal{S}$, and some distance metric $\|\cdot\|$. Following \cite{wadsworth2018spatial}, we assume that there exists normalising functions $\alpha:[0,\infty)\mapsto [0,1]$, with $\alpha(0) = 1$,  and $\beta:[0,\infty)\mapsto[0,1]$, such that for each $s_O \in \mathcal{S}$, as $ u\rightarrow \infty$,
\begin{equation}
\label{depmodeleq}
\left(\left\{\frac{X(s)-\alpha\{h(s,s_O)\}X(s_O)}{\{X(s_O)\}^{\beta\{h(s,s_O)\}}}:s \in \mathcal{S}\right\},X(s_O)-u\right)\;\Bigg|\;\bigg(X(s_O)>u\bigg)\xrightarrow{d} \Bigg(\bigg\{Z(s|s_O),s \in \mathcal{S}\bigg\},E\Bigg),
\end{equation}
where $E$ is a standard exponential variable, the process $\{Z(s|s_O)\}$ is non-degenerate for all $s \in \mathcal{S}$ where $s \neq s_O$ and it is independent of $E$. That is, there is convergence in distribution of the normalised process to the residual process $Z(s|s_O)$ which satisfies $Z(s_O|s_O) = 0$ almost surely.\par
We also follow \cite{wadsworth2018spatial} and take the parametric form for $\alpha(h)$ of 
\[
\alpha(h)=\begin{cases}1, &h \leq \Delta,\\
\exp(-\{(h - \Delta)/\kappa_{\alpha_1}\}^{\kappa_{\alpha_2}}), &h > \Delta,
\end{cases}
\]
where $\Delta \geq 0$ and $\kappa_{\alpha_1},\kappa_{\alpha_2} > 0$. This function determines the strength and class of extremal dependence within $\{X(s)\}$, see eq.\ \eqref{chieq}; allowing asymptotic dependence up to distance $\Delta$ from $s_O$, and asymptotic independence thereafter, with strength of dependence decreasing with $h$. Setting $\Delta=\infty$ and $\beta(h)=0$ for all $h\geq 0$ corresponds to $X$ having the same dependence structure as an $r-$Pareto process conditional on exceedances at a single site \citep{deFondeville2020functional}. \par \cite{richards2021modelling} impose that the residual process $Z(s|s_O)$ for $s \in \mathcal{S}$ has delta-Laplace margins with location, scale and shape parameters determined by the spatial functions $\mu:[0,\infty)\mapsto \mathbb{R}$, $\sigma:[0,\infty)\mapsto [0,\sqrt{2}]$ and $\delta:[0,\infty)\mapsto [1,\infty)$, respectively. We denote this by $Z(s|s_O)\sim\mbox{DL}(\mu\{h(s,s_O)\},\sigma\{h(s,s_O)\},\delta\{h(s,s_O)\})$, where the delta-Laplace density is \begin{equation}
\label{DL_dens}
f(z)=\delta\{2k\sigma\Gamma(1/\delta)\}^{-1}\exp\{-|(z-\mu)/(k\sigma)|^\delta\}
\end{equation} for $z \in \mathbb{R},\mu \in \mathbb{R},\sigma>0$ and $\delta >0$, and where $\Gamma$ denotes the standard gamma function and $k^2=\Gamma(1/\delta)/\Gamma(3/\delta)$. Dependence within $Z(s|s_O)$, for each $s_O\in\mathcal{S}$, is modelled using a standard Gaussian process with a stationary Mat\'ern correlation function, denoted $\rho$ (see Appendix~\ref{bivdens}), with $Z(s_O|s_O)=0$ almost surely. We adopt the same parametric forms for $\beta$, $\mu$, $\sigma$, and $\delta$ as \cite{richards2021modelling}, unless stated otherwise in Section~\ref{sec_differ}; full details are provided in Appendix~\ref{model_spec_sup}. These parametric forms allow the extremal dependence model for $\{X(s)\}$ to exhibit perfect independence at long-range as $\alpha(h)\rightarrow 0$,  $\beta(h)\rightarrow 0$, $\mu(h)\rightarrow 0$, $\sigma(h)\rightarrow \sqrt{2}$ and $\delta(h)\rightarrow 1$ when $h \rightarrow \infty$. \par
The function $h(s,s_O)$ accounts for potential geometric anisotropy in the extremal dependence structure of $\{X(s)\}$. As \cite{richards2021modelling}, we define the distance metric $\|s_A-s_B\|$ as an elliptical transformation of $(s_A,s_B) \in \mathcal{S} \times \mathcal{S}$ before finding the great-circle, or spherical, distance between $s_A$ and $s_B$ in eq. \eqref{anisoeq}; the metric is parametrised by $\theta \in [-\pi/2, 0]$ and $L > 0$ which control the rotation and coordinate stretching effect, respectively, with $L=1$ being isotropy.
\subsubsection{Differences for non-convective component}
\label{sec_differ}
We apply the model described in Section~\ref{dep_sec:generic} to each of $X_\mathcal{C}$, $X_\mathcal{N}$ and $X_\mathcal{E}$. For $X_\mathcal{C}$ and $X_\mathcal{E}$, we use exactly the same parametric forms for the dependence functions as proposed by \cite{richards2021modelling}, which are provided in Appendix~\ref{model_spec_sup}; for $X_\mathcal{N}$, we adapt the $\beta$ and $\sigma$ functions, denoting these new forms $\beta_\mathcal{N}$ and $\sigma_\mathcal{N}$.\par 
The $\beta$ function advocated by \cite{richards2021modelling} for $X_\mathcal{C}$, say $\beta_\mathcal{C}$, satisfies $\beta_\mathcal{C}(0)=1$ and allows the convective process to exhibit spatial roughness dependent on the magnitude of $X(s_O)$. Whilst this is a desirable property for a process generating convective precipitation events, it is not required for $X_\mathcal{N}$, which generates much smoother spatial events than $X_\mathcal{C}$; this claim is supported by observations of $Y_\mathcal{C}$ and $Y_\mathcal{N}$ in Figure~\ref{observations}. For $\beta_\mathcal{N}$, we adopt the approach of \cite{shooterinpress} and let
\begin{equation}
\label{chap6-betaeq}
\beta_{\mathcal{N}}(h)=\frac{\kappa_{\beta_1}h^{\kappa_{\beta_2}}\exp(-h/\kappa_{\beta_3})}{\max_{h_*>0}\{h_*^{\kappa_{\beta_2}}\exp(-h_*/\kappa_{\beta_3})\}},\qquad\kappa_{\beta_1}\in[0,1],\kappa_{\beta_2} >0, \kappa_{\beta_3} >0,
\end{equation}
so $0 \leq\beta_{\mathcal{N}}(h) \leq 1$ for all $h \geq 0$ and $\beta_{\mathcal{N}}(h)\rightarrow 0$ as $h\rightarrow \infty$ with $\beta_\mathcal{N}(0)=0$. In our application, we considered both forms of $\beta$ for $X_\mathcal{N}$, but found that eq.\ \eqref{chap6-betaeq} provided a better fitting model. \par
The scaling function $\sigma_\mathcal{C}$ gives long-range independence. However, non-convective events can have a very large spatial extent, e.g., up to $1000$km \citep{houze1997stratiform}. Long-range independence  is not required for $X_\mathcal{N}$ when the domain of interest $\mathcal{S}$ is small enough, as in our application in Section~\ref{App_sec}. To ensure a better fitting model for $X_\mathcal{N}$, we let the scale function $\sigma_\mathcal{N}$ be
\[
\sigma_{\mathcal{N}}(h)=\kappa_{\sigma_3}\left[1-\exp\{-(h/\kappa_{\sigma_1})^{\kappa_{\sigma_2}}\}\right],\qquad\kappa_{\sigma_1}>0, \kappa_{\sigma_2} >0,\kappa_{\sigma_3}>0,
\]
which is equivalent to $\sigma_\mathcal{C}$ when $\kappa_{\sigma_3}=\sqrt{2}$. 
\subsubsection{Inference}
\label{infer_sec}
Inference for each of the three extremal dependence models is conducted using the pseudo-likelihood procedure proposed by \cite{richards2021modelling}, which they illustrate works well for models of this type and for application to precipitation data. Full details of the inference procedure are given in Appendix~\ref{append-infer}.  This technique requires selection of a sub-sample of $d_s$ triples of sampling locations, subject to the maximum pairwise distances subceeding a value of $h_{max}$ km. Maximisation of a pseudo-likelihood is then conducted using observations of the process via a triple-wise censored likelihood; the latter is required to handle point masses in the marginals of $X(s)$ caused by zeroes in the marginals of $Y(s)$. The censoring threshold for $X(s)$, denoted $c(s)$, is found by transforming $p(s)$ to the Laplace scale, i.e., $c(s)=F^{-1}_L\{p(s)\}$, where $F_L$ denotes the standard Laplace distribution. We apply this approach and use different values of $h_{max}$ and $d_s$ for each of the three processes; further details are provided in Section~\ref{sec:dep_model}. Inference is conducted for model~\eqref{depmodeleq} by fixing a high threshold $u > 0$ and assuming that the limiting relation of eq.\ \eqref{depmodeleq} holds exactly; the value of $u$ differs between the three processes. Finding a threshold such that $Z(s|s_O)$ and $E$, defined in eq.\ \eqref{depmodeleq}, are independent may be infeasible with mixtures of different dependence present in $X(s)$, e.g., for $X_\mathcal{E}(s)$. This provides further support for our use of separate extremal dependence models for the mixture components, as the assumptions made in eq.\ \eqref{depmodeleq} are more likely to hold when applied to $X_\mathcal{N}$ and $X_\mathcal{C}$, rather than $X_\mathcal{E}$, and at lower thresholds $u$ (and thus allows us to use more data for inference); we find the latter to be the case in our analysis. 
\section{Simulating events}
\label{sec-sim}
\subsection{Efficient simulation}
\label{inferenceextend}
We now consider simulation of $\{Y(s):s \in \boldsymbol{S}\}$ on a dense discrete set of sites, denoted $\boldsymbol{S}\subseteq\mathcal{S}$;  we further distinguish between the continuous aggregate region $\mathcal{A}$ and a discretized version $\boldsymbol{A}\subseteq \mathcal{A}$, which must satisfy $\boldsymbol{A} \subset \boldsymbol{S}$. Typically, we would take $\boldsymbol{S}$ to be the set of sampling locations, such that $\boldsymbol{S}:=\mathbf{s}$, and we do so hereafter. When $|\boldsymbol{S}|$ is large, e.g., $|\boldsymbol{S}|>5,000$ as with our data, it may be too computationally expensive to simulate a large number of replicates of $\{Y(s): s\in \boldsymbol{S}\}$ in order to do reliable inference on the upper-tail behaviour of $\bar{R}_\mathcal{A}$. It may not be entirely necessary to simulate $Y$ over all of $\boldsymbol{S}$ if our interest lies in aggregates over sufficiently small regions; we can instead simulate fields of $Y$ on a sub-region $\boldsymbol{S}_\tau\subset \boldsymbol{S}$; we define this set for $\tau>0$  by \begin{equation}
\label{s_tau_def_eq}
\boldsymbol{S}_\tau=\left\{s \in \boldsymbol{S}: \min_{s_{\boldsymbol{A}}\in\boldsymbol{A}}\{\|s-s_{\boldsymbol{A}}\|\} \leq \tau\right\}, 
\end{equation}
 i.e., so $\boldsymbol{A}\subset \boldsymbol{S}_\tau$. \cite{richards2021modelling} chose $\tau$ so that there was a predetermined small probability of observing a large event at any $s \in \boldsymbol{A}$ given that there is an extreme event at any single site outside of $\boldsymbol{S}_\tau$. We also introduce a set of conditioning sites $\boldsymbol{A}_c=\boldsymbol{S}_\tau \cup \boldsymbol{S}^+$,  where $\boldsymbol{S}^+\cap \boldsymbol{S} = \emptyset$ for a set $\boldsymbol{S}^+$ to be defined, and create realisations of $\{Y(s):s \in \boldsymbol{S}_\tau\}$ given that the process is extreme at $s_O \in \boldsymbol{A}_c$. Illustrations of $\boldsymbol{A}$, $\boldsymbol{S}$, $\boldsymbol{S}_\tau$, $\boldsymbol{S}^+$ and $\boldsymbol{A}_c$ for our application are presented in Figure~\ref{agg_diags_locs} and a heuristic for choosing both $\boldsymbol{S}_\tau$ and $\boldsymbol{A}_c$ is given in Section~\ref{heuristic-sec-OD}; these sets vary over the three processes. 
\par
When considering aggregates over a fixed region $\boldsymbol{A}\subset \boldsymbol{S}_\tau$, the simulation procedure proposed by \cite{richards2021modelling} never generates events for which the conditioning site, $s_O$ in $\boldsymbol{A}_c$, lies outside of the boundaries of $\boldsymbol{S}_\tau$. When the process exhibits particularly strong extremal spatial dependence, such as for non-convective rainfall, we may bias the upper-quantiles of the distribution of $\bar{R}_\mathcal{A}$ if we set $\boldsymbol{A}_c=\boldsymbol{S}_\tau$, i.e., $\boldsymbol{S}^+=\emptyset$, since we are ignoring events which could be large somewhere outside of $\boldsymbol{S}_\tau$ but still impact the upper-tails of the aggregate over $\boldsymbol{A}$. So we need to account for the possibility of an extreme at conditioning sites $s_O \notin \boldsymbol{S}$ that are far from $\boldsymbol{A}$ to avoid this problem. When we require that $\boldsymbol{S}^+ \not=\emptyset$ we construct $\boldsymbol{S}^+$ to ensure that $\boldsymbol{A}_c$ contains additional, randomly selected, synthetic conditioning sites located outside of $\boldsymbol{S}$. \par
To simulate from the fitted models for $Y_{\mathcal{C}}$, $Y_{\mathcal{N}}$ and $Y_{\mathcal{E}}$, we need to extend the procedure proposed by \cite{richards2021modelling} due to long-range dependence features of frontal rain; we present this technique for a generic $Y$ and $X$, but this approach can be applied to any of the three processes considered. To simulate a realisation of $Y(s)$ at a site $s\in \boldsymbol{S}_\tau$, we draw a realisation of 
\begin{align}
\label{chap6-simexp}
Y(s)\;\bigg|\;\left\{ \max\limits_{s \in\boldsymbol{S}}\left( F^{-1}_L[F_{Y(s)}\{Y(s)\}]\right) > v\right\} \equiv F^{-1}_{Y(s)}[F_{L}\{X(s)\}]\;\bigg|\;\left[ \max\limits_{s \in\boldsymbol{S}}\left\{X(s)\right\} > v\right],
\end{align}
using models \eqref{chap6-MargTransform} and \eqref{depmodeleq} and for $v \geq u$, with probability
\[
\Pr\left\{ \max\limits_{s \in\boldsymbol{S}}\left(F^{-1}_L[F_{Y(s)}\{Y(s)\}]\right) > v\right\},
\]
and otherwise draw a realisation from the observed $Y(s)$ subject to $ \max\limits_{s \in\boldsymbol{S}}\left( F^{-1}_L[F_{Y(s)}\{Y(s)\}]\right) < v$.
To draw $b\in\mathbb{N}$ realisations of process~\eqref{chap6-simexp}, we use Algorithm~\ref{chap6-sim-algo}, which adjusts Algorithm~1 of \cite{richards2021modelling} to account for the introduction of $\boldsymbol{A}_c$ and $\boldsymbol{S}_\tau$. We use importance sampling to approximately draw realistions from the distribution conditional on  $\max\limits_{s \in\boldsymbol{S}}\left\{X(s)\right\}$ being large; we first simulate a population of $b'$ replicates and then sub-sample $b<b'$ replicates using an importance sampling scheme. \par
\begin{algorithm}[h!]
\caption{Simulating conditional spatial fields corresponding to formulation \eqref{chap6-simexp}}
\label{chap6-sim-algo}
\begin{enumerate}
\item For $i = 1,\dots, b'$ with $b' > b$ denoting the replicate population size: \begin{enumerate}
\item Draw a conditioning location $s^{(i)}_O$ from $\boldsymbol{A}_c$ with equal probability $1/|\boldsymbol{A}_c|$.
\item Simulate $E^{(i)} \sim \mbox{Exp}(1)$ and set $x_i(s^{(i)}_O) = v + E^{(i)}$.
\item Simulate a field $\{z_i(s|s_O^{(i)}): s \in \boldsymbol{S}_\tau\}$ from the residual process model.
\item Set $x_i(s)=x_i(s^{(i)}_O)\alpha\{h(s,s^{(i)}_O)\}+\{x_i(s^{(i)}_O)\}^{\beta\{h(s,s^{(i)}_O)\}} \{z_i(s|s_O^{(i)})\}$ for each $s \in \boldsymbol{S}_\tau$.
\end{enumerate}
\item Assign each simulated field  $\{x_i(s): s \in \boldsymbol{S}_\tau\}$ an importance weight of
\[
\begin{cases}
\left[\sum_{s \in\boldsymbol{S}_\tau}\mathbbm{1}\{x_i(s)>v\}\right]^{-1},\;&\text{if}\;\sum_{s \in \boldsymbol{S}_\tau}\mathbbm{1}\{x_i(s)>v\} >0,\\
0, &\text{otherwise},
\end{cases}
\]
for $i=1,\dots,b^{'}$. \label{agg_algo_step}
\item Sub-sample $b$ realisations from the collection of replicates with probabilities proportional to their assigned importance weights.
\item For all $s\in\boldsymbol{S}_\tau$, transform $x_i(s)$ to $y_i(s)$ using the marginal transformation in eq.\ \eqref{chap6-MargTransform}. Note that if $x_i(s) \leq c(s)$, we set $y_i(s)=0$; however, $y_i(s^{'})$ is above its $F_L(v)$-th quantile for some $s^{'} \in \boldsymbol{S}_\tau$.
\end{enumerate}
\end{algorithm}
Using Algorithm~\ref{chap6-sim-algo}, we can draw realisations of $\{Y_\mathcal{C}(s): s \in \boldsymbol{S}_\tau\}$ and $\{Y_\mathcal{N}(s): s \in \boldsymbol{S}_\tau\}$ and then use these to derive samples of $\{Y_\mathcal{M}(s):s \in \boldsymbol{S}_\tau\}$ by drawing from the samples of $\{Y_\mathcal{C}(s):s \in \boldsymbol{S}_\tau\}$ and $\{Y_\mathcal{N}(s):s \in \boldsymbol{S}_\tau\}$ with probabilities $p_\mathcal{C}$ and $1-p_\mathcal{C}$, respectively, see eq.\ \eqref{totalprob}, which are estimated empirically. Approximate realistions of $\bar{R}_\mathcal{A}$ are derived by taking replicates of $Y$ and computing the average of $\{Y(s):s \in \boldsymbol{A}\}$; we denote estimates of $\bar{R}_\mathcal{A}$ derived using $Y_\mathcal{M}$ and $Y_\mathcal{E}$ by $\bar{R}_{\mathcal{M},\mathcal{A}}$ and $\bar{R}_{\mathcal{E},\mathcal{A}}$, respectively. Although the approach we have described is tailored to a specific aggregate region $\boldsymbol{A} \subset \boldsymbol{S}_\tau$, our interest may lie in aggregates over multiple regions $\boldsymbol{A}_1,\dots,\boldsymbol{A}_m$ for $m \in \mathbb{N}$. To ensure that we can perform joint inference on aggregates at multiple scales, we can simulate replicates of $Y$ with $\boldsymbol{A}\supseteq \cup^m_{i=1}\boldsymbol{A}_i$; replicates of aggregates of $Y$ over the sub-regions $\boldsymbol{A}_1,\dots,\boldsymbol{A}_m$ are then derived from a common collection of replicates.
\subsection{Choosing $\boldsymbol{S}_\tau$ and $\boldsymbol{A}_c$}
\label{choose_sec}
\begin{figure}[t!]
\centering
\includegraphics[width=0.75\textwidth]{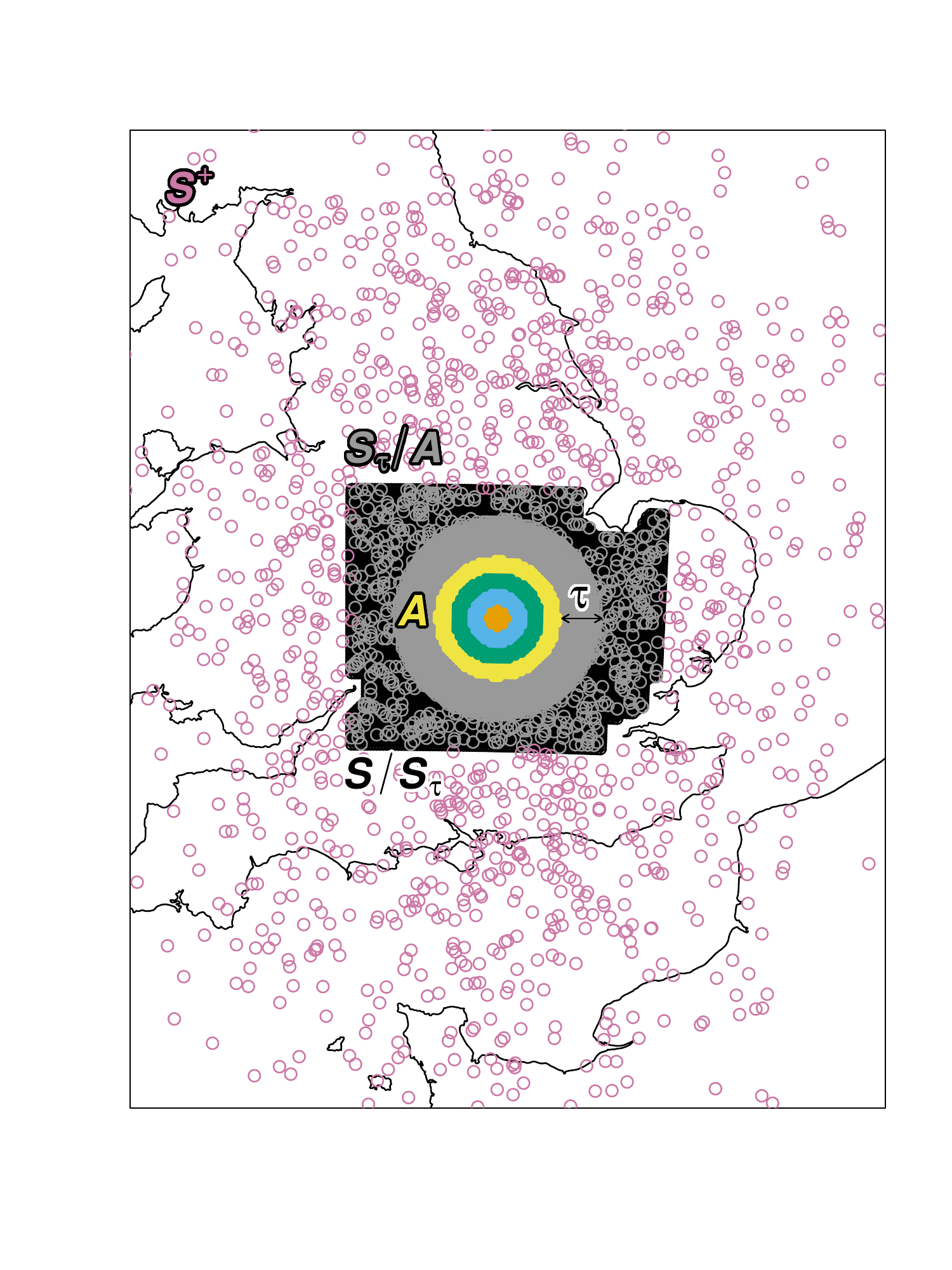} 
\caption{Regions $\boldsymbol{A}$ and $\boldsymbol{S}$ that correspond to discretized versions of the aggregate region $\mathcal{A}$ and study domain $\mathcal{S}$, respectively; regions $\boldsymbol{S}_\tau, \boldsymbol{A}_c$ and $\boldsymbol{S}^+=\boldsymbol{A}_c\setminus\boldsymbol{S}$ are used in Algorithm~\ref{chap6-sim-algo} for simulating model realisations. The aggregate region $\boldsymbol{A}$ encompasses four sub-regions, with corresponding areas $(179,1263,3257,6200)$ km$^2$, that are coloured orange, blue, green, yellow; sub-regions include both the coloured and interior points and are numbered 1 to 4 in Figures~\ref{agg_diags2} and \ref{agg_diags1}. The grey and black points denote $\boldsymbol{S}_\tau\setminus\boldsymbol{A}$ and $\boldsymbol{S}\setminus\boldsymbol{S}_\tau$, respectively. The $n_c=1250$ purple points, outside of the boundaries of $\boldsymbol{S}$, correspond to $\boldsymbol{S}^+$, and the region $\boldsymbol{S}_\tau$ encompasses all points within distance $\tau=27.5$km outside of $\boldsymbol{A}$ and $n_{s}=500$ additional randomly-sampled points within $\boldsymbol{S}$; these values mean that $|\boldsymbol{S}_\tau|=3625$ and $|\boldsymbol{A}_c|=4875$.}
\label{agg_diags_locs}
\end{figure}
We begin by considering the dense sub-region $\boldsymbol{S}_\tau$ of locations at which realisations of $Y(s)$ are to be simulated, ensuring that $\boldsymbol{A} \subset \boldsymbol{S}_\tau \subseteq \boldsymbol{S}$, and then subsequently choose $\boldsymbol{S}^+$ to build $\boldsymbol{A}_c\supseteq \boldsymbol{S}_\tau$; recall that $\boldsymbol{A}_c=\boldsymbol{S}_\tau \cup \boldsymbol{S}^+$ is the set of all possible conditioning sites used in simulation. Note that, in both cases, $\boldsymbol{S}_\tau$ and $\boldsymbol{A}_c$ should be chosen as large as is computationally feasible. For Step~\ref{agg_algo_step} of Algorithm~\ref{chap6-sim-algo}, we require that
\begin{equation}
\label{importanceweightseq}
\sum_{s \in \boldsymbol{S}_\tau}\mathbbm{1}\{x_i(s)>v\}\approx\sum_{s \in\boldsymbol{S}}\mathbbm{1}\{x_i(s)>v\},
\end{equation}
to account for the conditioning event in eq.\ \eqref{chap6-simexp}; we can improve the accuracy of this approximation by ensuring that sites in $\boldsymbol{S}_\tau$ are sufficiently spread out across $\boldsymbol{S}$. We first set $\boldsymbol{S}_\tau$ according to eq.\ \eqref{s_tau_def_eq}, as we expect extreme events within $\boldsymbol{A}$ to have the largest impact on the tail behaviour of $\bar{R}_{\mathcal{A}}$. If this choice of $\boldsymbol{S}_\tau$ does not give a good approximation in eq.\ \eqref{importanceweightseq}, then we additionally sample $n_s$ sites uniformly at random across $\boldsymbol{S}\setminus \boldsymbol{S}_\tau$ and add these to $\boldsymbol{S}_\tau$.\par
In our application, we found that setting $\boldsymbol{S}^+:=\emptyset \Rightarrow \boldsymbol{A}_c:=\boldsymbol{S}_\tau$ was a reasonable choice to make for $Y_{\mathcal{C}}$ and $Y_\mathcal{E}$, as no improvement in our inference for $\bar{R}_\mathcal{A}$ was achieved by using $\boldsymbol{A}_c\supset \boldsymbol{S}_\tau$. However, this was not true for $Y_\mathcal{N}$, which exhibits strong extremal dependence at even the greatest distances within $\boldsymbol{S}_\tau$. For such processes, we need to artificially increase the maximum possible distance between sites in $\boldsymbol{A}_c$ and $\boldsymbol{A}$, to ensure that any possible conditioning site $s_O$ that could contribute to an extreme event of $\bar{R}_\mathcal{A}$ is available in $\boldsymbol{A}_c$. To this end, we construct $\boldsymbol{A}_c$ by taking $\boldsymbol{S}^+$ to be a set of $n_c$ sites, sufficiently far from $\boldsymbol{A}$, where data are not observed. We create $\boldsymbol{S}^+$ via a brute-force approach; we add independent Gaussian noise to the coordinates of a site in the centre of $\boldsymbol{A}$, rejecting any new points that lie within the boundaries of $\mathcal{S}$, see Figure~\ref{agg_diags_locs}. 
\par
 Suitable values for the constants $n_s,\tau$ and $n_c$, and the variance of the noise term used to create $\boldsymbol{S}^+$, can be chosen through validation techniques, such as model fit diagnostics for the aggregate variable (Figures~\ref{agg_diags2} and \ref{agg_diags1}); if the fits look poor, one can repeatedly increase the value of these constants until no significant changes in fit are observed.
\label{heuristic-sec-OD}
\section{Application}
\label{App_sec}
\subsection{Marginal analysis}
\label{marganalysis}
Marginal analysis is conducted by fitting the model described in Section~\ref{marg_sec} to each of the three datasets, i.e., convective, non-convective, and pooled. We set $\lambda=0.005$ in eq.\ \eqref{chap6-MargTransform} for all three processes and use a subset of $500$ sites sampled randomly over $\mathcal{S}$ to estimate $q_\lambda(s)$; more sites were considered, but no significant difference was observed in the inference. The same sites are used for all processes and are illustrated in Figure~\ref{GAM_locs}; we ensure that sampled sites proportionately cover the distribution of elevation values by sampling sites with probability according to the approximate empirical distribution of elevation measurements. We use all sampling locations to estimate $p(s)$, $\upsilon(s)$ and $\xi$. Figures~\ref{convGAMfig}--\ref{BothGAMfig} give estimates for the parameters of the marginal models and $20$-year marginal return level estimates for convective, non-convective and all rainfall, respectively, with estimation of the return levels accounting for the differing length of the observation record for each process. We observe similar patterns in estimates of $p(s)$ and $q_\lambda(s)$ for each of the three processes, with $p(s)$ and $q_\lambda(s)$ decreasing and increasing with elevation, respectively. \par
There are differences between the estimates of $\upsilon(s)$ and the 20-year return levels for the three processes; whereas for $Y_\mathcal{N}$ both increase with elevation, for $Y_\mathcal{C}$ and $Y_\mathcal{E}$, we observe spatially smooth estimates of both, with larger values in the east of the domain (see Figures~\ref{convGAMfig} and \ref{BothGAMfig}), indicating that the physical processes driving convection are only weakly affected by orography, unlike frontal rain (see Figure~\ref{frontGAMfig}). The $20$-year return levels for $Y_\mathcal{C}$ and $Y_\mathcal{E}$ are much higher than for $Y_\mathcal{N}$, which is consistent with their different physical properties. This property also holds for all larger return levels since the median shape parameter estimates ($95\%$ confidence interval), across all bootstrap samples, for convective, non-convective and all rainfall are $\hat{\xi}_\mathcal{C}=0.226$ (0.201,0.245), $\hat{\xi}_\mathcal{N}=-0.074$ ($-$0.108, $-$0.03) and $\hat{\xi}_\mathcal{E}=0.286$ (0.266,0.307), respectively.  Although \cite{richards2021modelling} allow $\xi$ to vary with location, our inference for the marginal tails of all rainfall agree with theirs as our estimate, $\hat{\xi}_\mathcal{C}=0.226$, falls within the range of their estimates; similar agreement is found with results by \cite{chan2014value} and \cite{hosseinzadehtalaei2020climate}, and the estimate is not inconsistent with the range of feasible values for extreme precipitation proposed by \cite{martins2000generalized}. Note that the change in the shape estimate from 0.286 for all rainfall to 0.226 for convective rainfall supports our assumptions that, marginally, $Y_\mathcal{E}$ is a mixture of processes. \par
To assess the fit of the generalised Pareto additive models, we present Q-Q plots of the marginal fits for each of $Y_\mathcal{C}$, $Y_\mathcal{N}$ and $Y_\mathcal{E}$ at five randomly sampled locations in the Supplementary Material, Figures~\ref{conv_gamdiag}--\ref{both_gamdiag}. These figures show good individual fits for each of the processes at most locations; some slight biases in return level estimates are found at certain locations near the boundaries of the domain, but we found that this did not compromise the overall marginal fit or the inference on spatial aggregates. To evaluate the fit over all locations, we use a pooled Q-Q plot \citep{heffernan2001extreme}, transforming all data onto standard exponential margins using the fitted model; these plots are also given in these figures. Again, we observe good fits for each process. Confidence intervals for the Q-Q plots are estimated using the following bootstrap procedure: we create $250$ bootstrap samples of the data using the stationary bootstrap approach of \cite{politis1994stationary} with expected block size of 48 hours. With $q_\lambda(s)$ treated as fixed across all samples, we then estimate $\upsilon(s)$ and $\xi$ for each bootstrap sample. For the pooled diagnostic plot, we apply the marginal transformation to the original data using the $250$ estimated marginal parameter sets.
\subsection{Extremal dependence modelling}
\label{sec:dep_model}
We proceed by fitting the extremal dependence models described in Section~\ref{dep_sec} separately to each process. We use a different exceedance threshold, i.e., $u$ in eq.\ \eqref{depmodeleq}, for each process; the $96\%$ quantile for $X_\mathcal{C}$ and the $99\%$ quantile for $X_\mathcal{E}$ and $X_\mathcal{N}$. The thresholds were chosen so that the limiting assumptions in eq.\ \eqref{depmodeleq} were applicable; other choices were considered but these led to poorer inference on the extremal behaviour of spatial aggregates, which we quantified using the measures and diagnostics described in Section~\ref{appl-inferaggs}.
Ideally, for $X_\mathcal{E}$ we would set $u$ such that the number of observations used for inference was identical to the mixture approach to provide a fair comparison. However, we found that using such a threshold led to poorer inference for $X_\mathcal{E}$, with significant evidence that assumptions in eq.\ \eqref{depmodeleq} fail for any lower threshold choice.\par
Inference is conducted using the stratified sampling regime described in Section~\ref{dep_sec}. We use the same value of $d_s$ for all three fitted models; we follow \cite{richards2021modelling} and
set $d_s=5000$. Different values of $h_{max}$ were also taken: 35km for $X_\mathcal{E}$ and $X_\mathcal{C}$ and 250km for $X_\mathcal{N}$, with the latter reflecting the long-range extremal dependence of $X_\mathcal{N}$. Inference for all three extremal dependence models was conducted using different sub-samples of sampling locations. For $X_\mathcal{N}$, we tried both parametric forms of $\beta$, i.e., $\beta_\mathcal{N}$ in eq.\ \eqref{chap6-betaeq} and the form described in Appendix~\ref{model_spec_sup}; with $\beta_\mathcal{N}$ giving a better fit, we present findings for this form only. Parameter estimates for all models are given in Table~\ref{parEst1} of the Appendix. For $X_\mathcal{E}$ and $X_\mathcal{C}$, we found that fixing $\kappa_{\delta_4}=\kappa_{\beta_3}=1$ and $\Delta=0$ did not restrict the quality of either fit. \par
\begin{figure}[h!]
\centering
\begin{minipage}{0.32\linewidth}
\centering
\includegraphics[width=\linewidth]{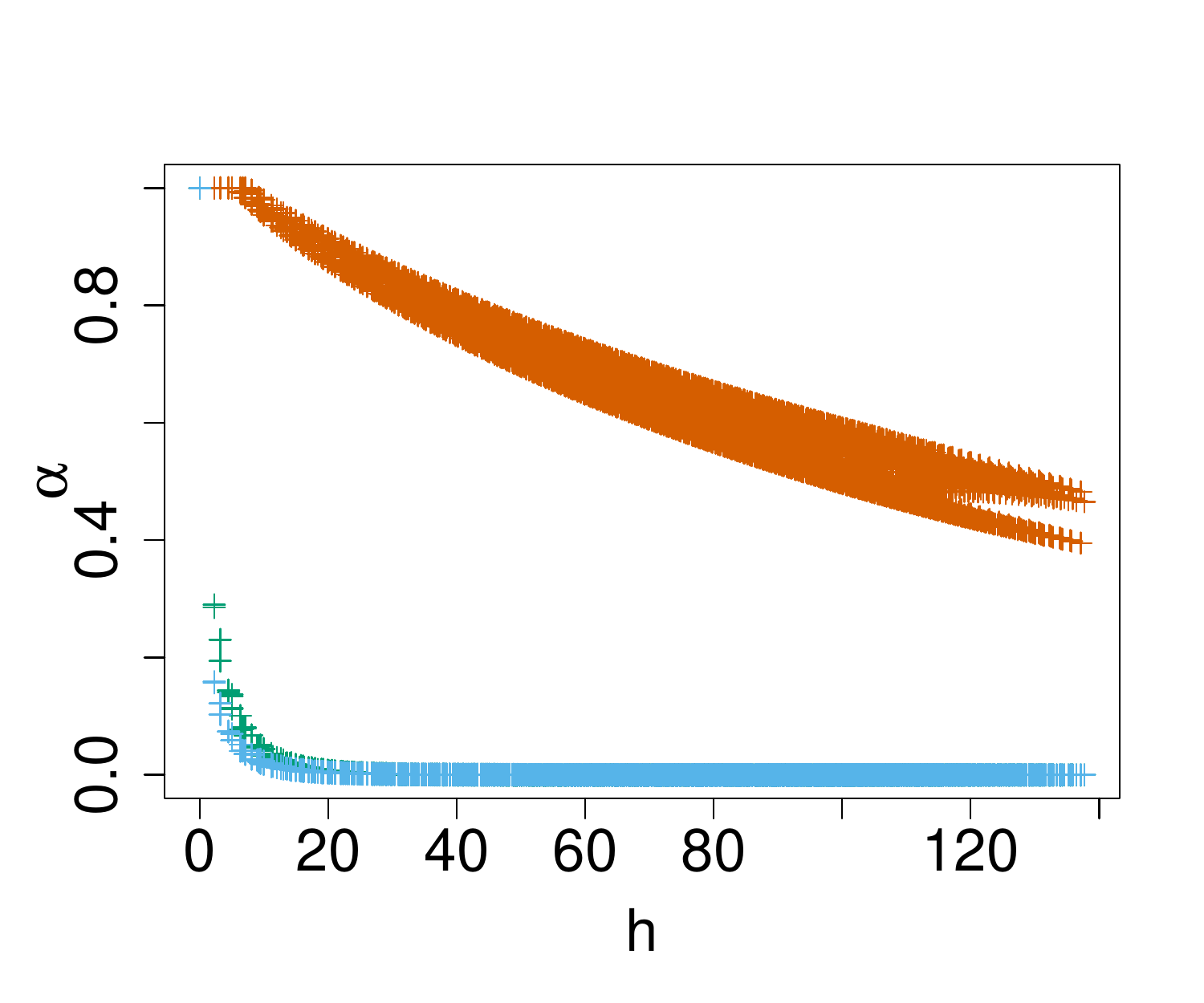} 
\end{minipage}
\begin{minipage}{0.32\linewidth}
\centering
\includegraphics[width=\linewidth]{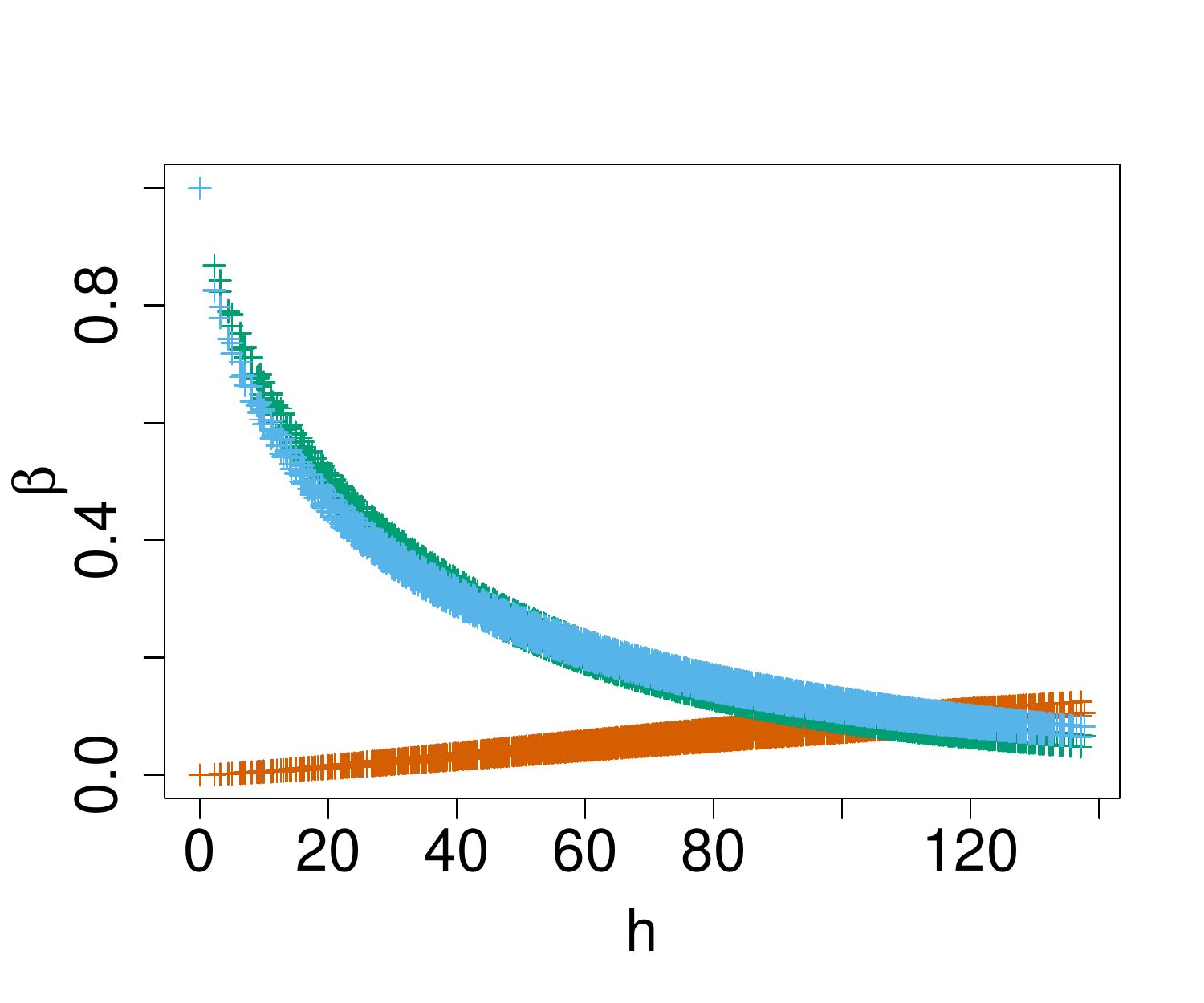} 
\end{minipage}
\vspace{-0.5cm}
\begin{minipage}{0.32\linewidth}
\centering
\includegraphics[width=\linewidth]{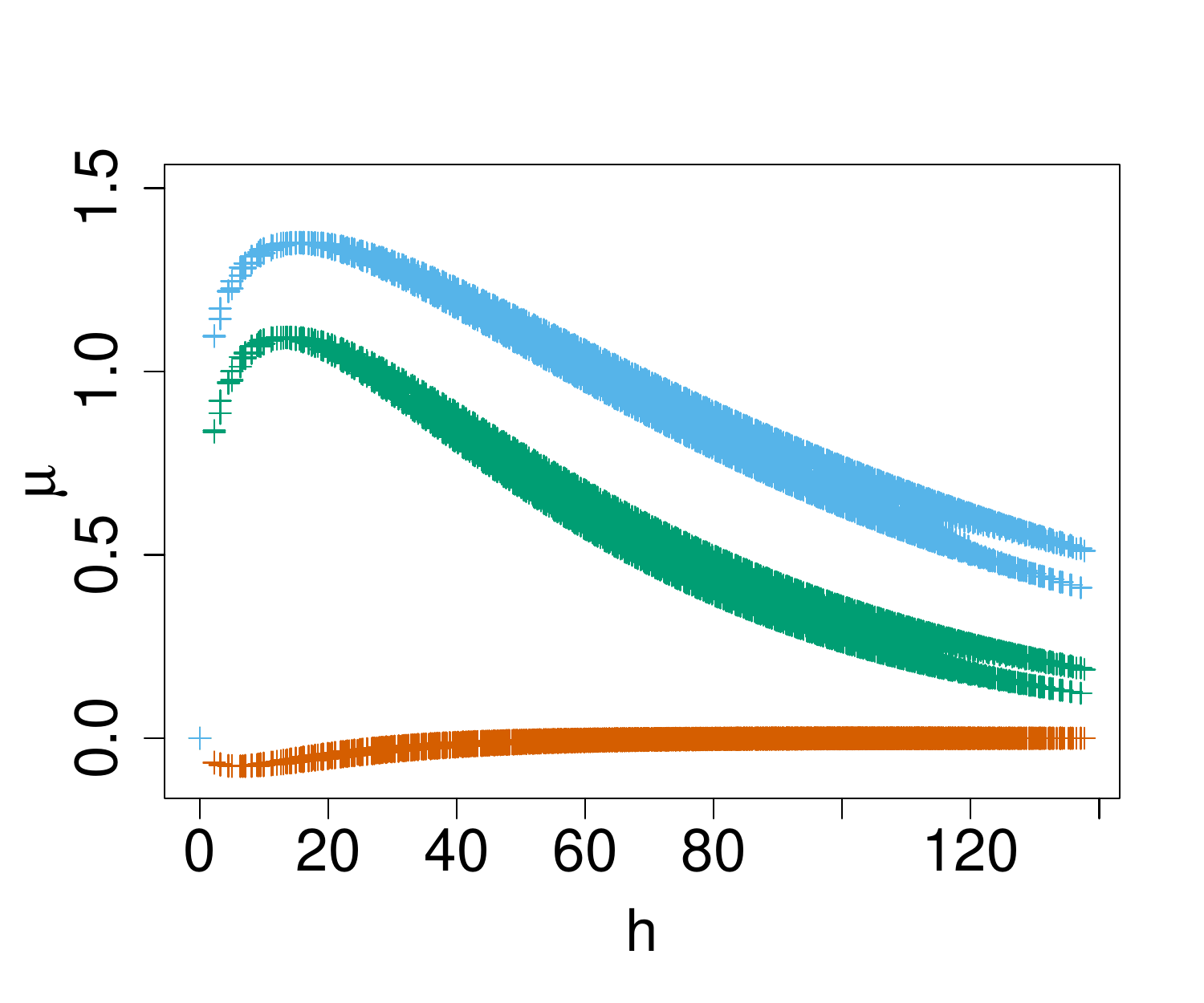} 
\end{minipage}
\begin{minipage}{0.32\linewidth}
\centering
\includegraphics[width=\linewidth]{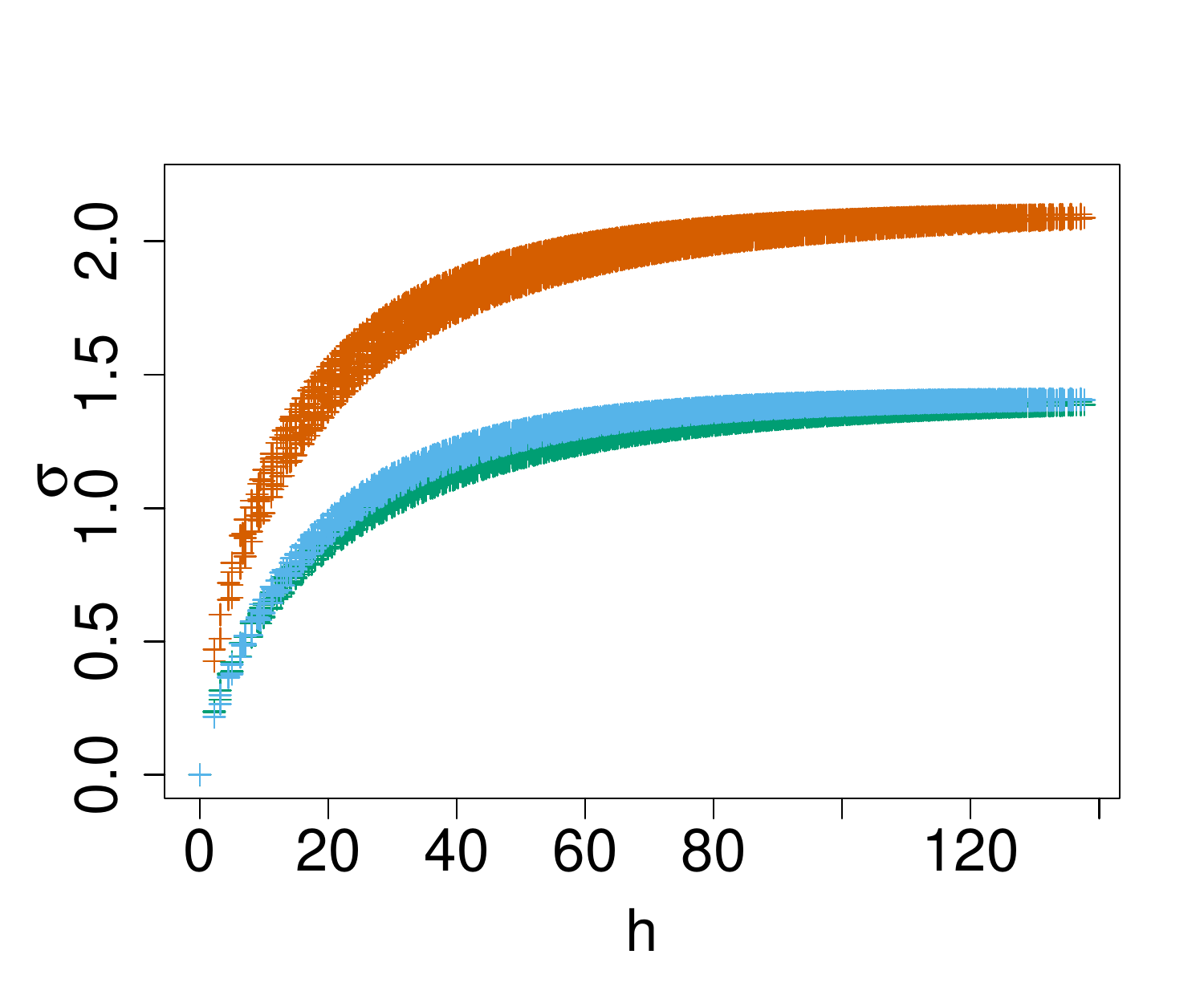} 
\end{minipage}
\begin{minipage}{0.32\linewidth}
\centering
\includegraphics[width=\linewidth]{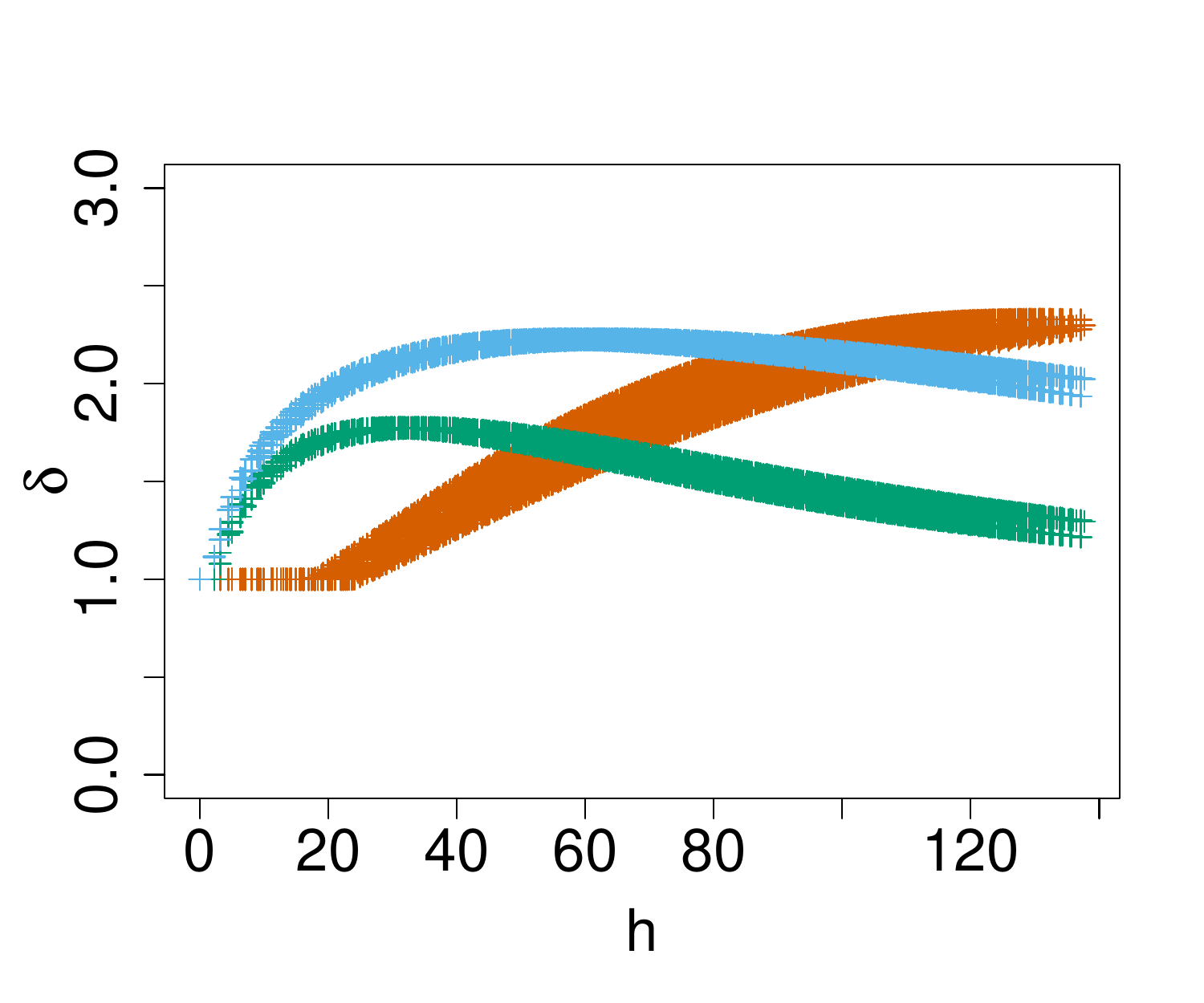} 
\end{minipage}
\begin{minipage}{0.32\linewidth}
\centering
\includegraphics[width=\linewidth]{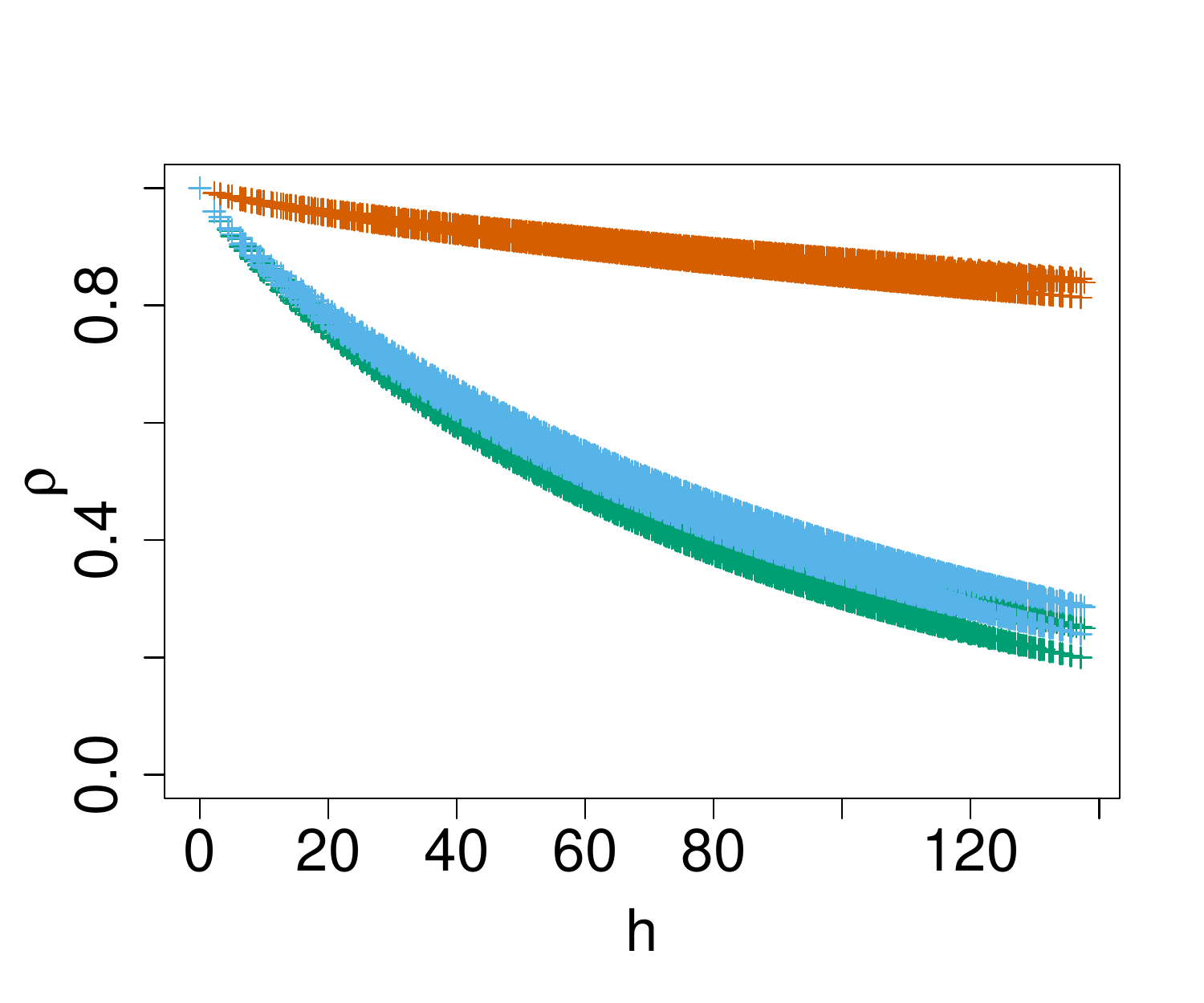} 
\end{minipage}
\caption{Estimates of extremal dependence functions evaluated at $h(s_i,s_O)$ for $i=1,\dots,d$, i.e., anisotropic distances, against great-circle distances $h:=h_*(s_i,s_O)$, which are given in km. The conditioning site $s_O$ is in the centre of the spatial domain $\mathcal{S}$ and the colours correspond to the estimates for the different spatial processes; these are green, orange and blue for convective (green), non-convective (orange) and all rainfall (blue), respectively. Dependence functions $\alpha$ and $\beta$ determine the strength of extremal dependence within the standardised process $X(s)$, whilst $(\mu,\sigma,\delta)$ and $\rho$ control the marginal and dependence properties, respectively, of the residual process $Z(s|s_O)$.}
\label{depestcompare}
\end{figure}

In Figure~\ref{depestcompare}, we fix a conditioning site $s_O$ in the centre of the domain and evaluate all dependence functions for each process at each pairwise anisotropic distance, i.e., $h(s_i,s_O)$ for $s_i\in \boldsymbol{S}$ and $i=1,\dots,d$; these are then plotted against the great-circle distances in the original coordinate system, denoted $h_*\{s_i,s_O\}$; as the anisotropy transformations differ for the three processes, we cannot compare estimates of their respective dependence functions in the original space, i.e., as functions of $h\{s_i,s_O\}$. Figure~\ref{depestcompare} indicates that there are similarities within the extremal dependence structures of $X_\mathcal{C}$ and $X_\mathcal{E}$, as most of the functions are approximately equal with slight differences for $\mu$ and $\delta$. In contrast, there are widely different estimated structures for the $\alpha$ and $\beta$ functions for $X_\mathcal{N}$, with $\alpha$ decaying much slower and independence not being achieved for any pairs of sites within $\mathcal{S}$. \par
The estimate of $\Delta=6.81$km given in Table~\ref{parEst1} suggests that the process $X_\mathcal{N}$ is asymptotically dependent up to this distance in the anisotropic setting, which corresponds to roughly 7.04km in the original coordinate system. Evidence of asymptotic dependence, even at short-range, has not been found previously using the spatial conditional extremes modelling approach \citep{wadsworth2018spatial,huser2020advances, simpson2020conditional, shooter2021spatial}. As extremal dependence is predominantly exhibited through the $\alpha$ function, this suggests that extreme realisations of $X_{\mathcal{N}}$ are much smoother than extreme events drawn from the other two processes, which agrees with our understanding of convective and non-convective precipitation. \cite{keef2013} identify additional constraints for the \cite{heff2004} framework, which lessen the risk of both parameter identifiability and inconsistency issues, and which in the spatial context correspond to $\beta(h)=0$ whenever $\alpha(h)=1$; our fitted model breaks this constraint with $0<\beta< 0.003$ for $0 <h\leq \Delta$, but this minor deviation is not problematic.\par
\begin{figure}[h!]
\centering
\begin{minipage}{0.35\linewidth}
\centering
\includegraphics[width=\linewidth]{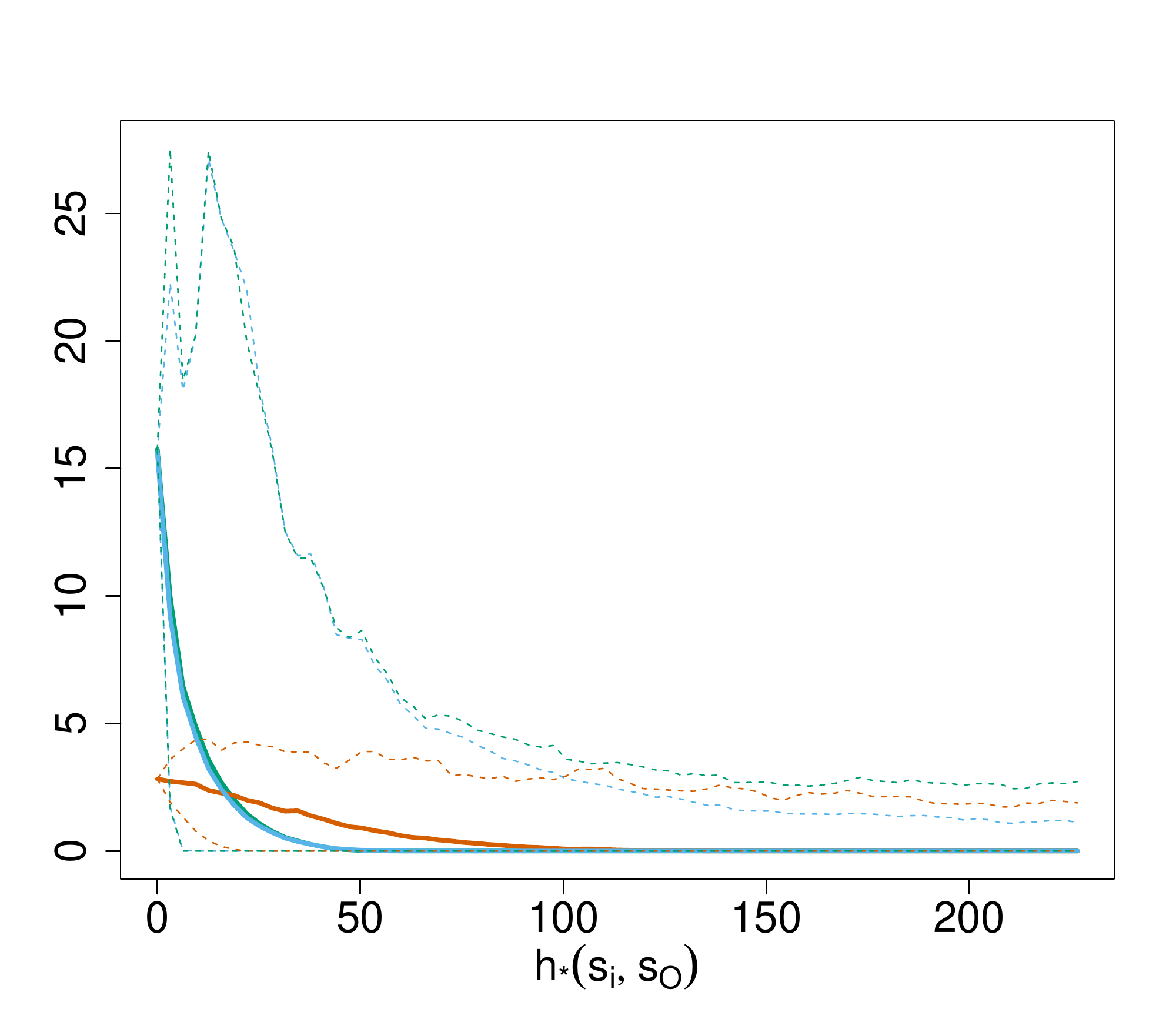} 
\end{minipage}
\begin{minipage}{0.35\linewidth}
\centering
\includegraphics[width=\linewidth]{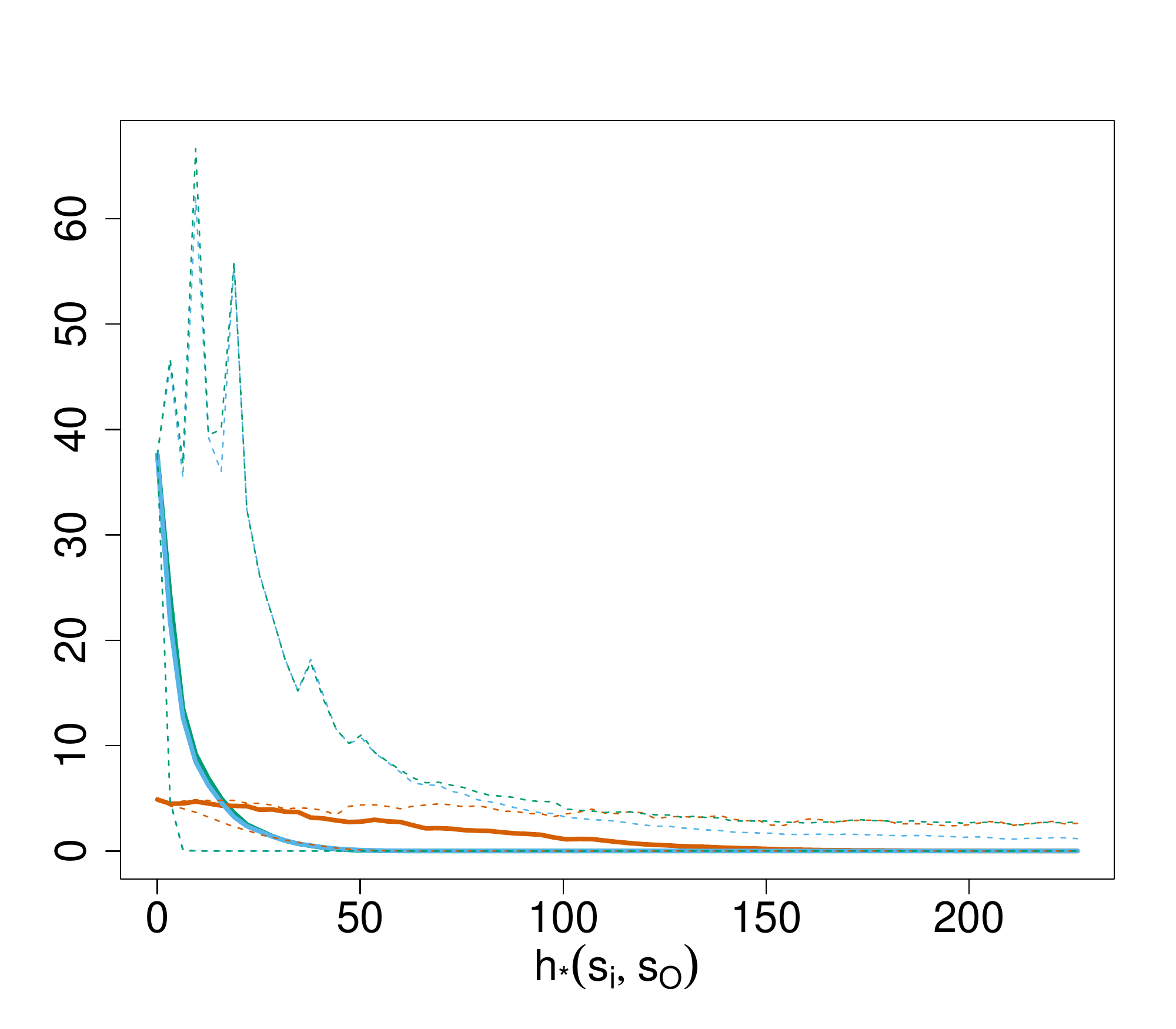} 
\end{minipage}
\begin{minipage}{0.28\linewidth}
\centering
\includegraphics[width=\linewidth]{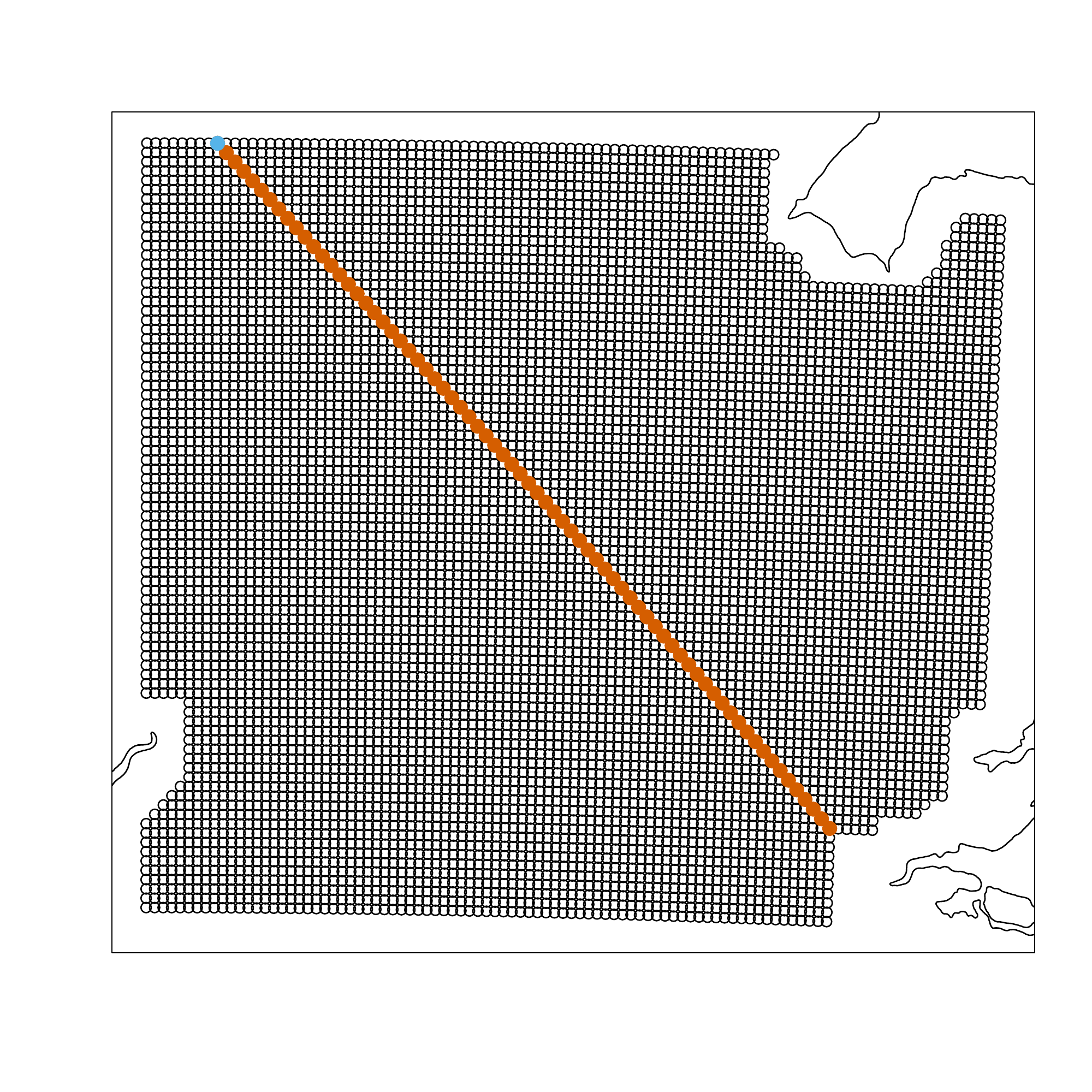} 
\end{minipage}
\caption{Summary statistics for $\{Y(s): s \in \mathcal{S}\}\;|\;\{Y(s_O)=y^{(l)}\}$ (mm/hr) against distance $h_*(s,s_O)$ (km), with $y^{(l)}$ the $l=1$ and $l=50$ years return level in the left and centre panels, respectively; the process is considered along a transect of points (right) only. Solid lines correspond to estimates for conditional medians, dashed lines denote $95\%$ confidence intervals. Lines are coloured green, brown and blue for convective, non-convective and all rainfall, respectively. Right: brown and blue points denote the transect and $s_O$, respectively.  }
\label{transmedians_fig}
\end{figure}

To compare the model fits of $Y_\mathcal{C}$, $Y_\mathcal{N}$ and $Y_\mathcal{E}$, we propose the following diagnostic that we present for a generic process. We investigate how the ``conditional process" \[\{Y(s):s \in \mathcal{S}\}\;|\;(Y(s_O)=y^{(l)})\] changes with distance $h_*(s,s_O)$ for $s,s_O \in \mathcal{S}$, where $y^{(l)}$ denotes the $l$-year return level for $Y(s_O)$. Figure~\ref{transmedians_fig} gives point-wise estimates for the median, and the $2.5\%$ and $97.5\%$ marginal quantiles, for $l=1$ and $50$; these are obtained from $50,000$ realisations from the fitted model. To achieve the largest possible distances, $s_O$ is chosen to be on the boundary of $\mathcal{S}$ and, to ease comparison, we consider a transect of points in $\mathcal{S}$ only. We take into account the respective length of observation periods when evaluating the respective $l-$year return levels, e.g., for $Y_\mathcal{C}$ we take $y^{(l)}=F^{-1}_{Y_\mathcal{C}(s_O)}\{1-1/(l\times |\mathcal{C}|/20)\}$ and similarly for $Y_\mathcal{N}$ and $Y_\mathcal{E}$. The diagnostics for $Y_\mathcal{C}$ and $Y_\mathcal{E}$ are almost identical, with the conditional medians of the respective processes both decaying quickly with distance. For $Y_\mathcal{N}$, the conditional median decays fairly slowly, with non-zero values at even the largest distances. \par
\begin{figure}[h!]
\centering
\begin{minipage}{0.32\linewidth}
\centering
\includegraphics[width=\linewidth]{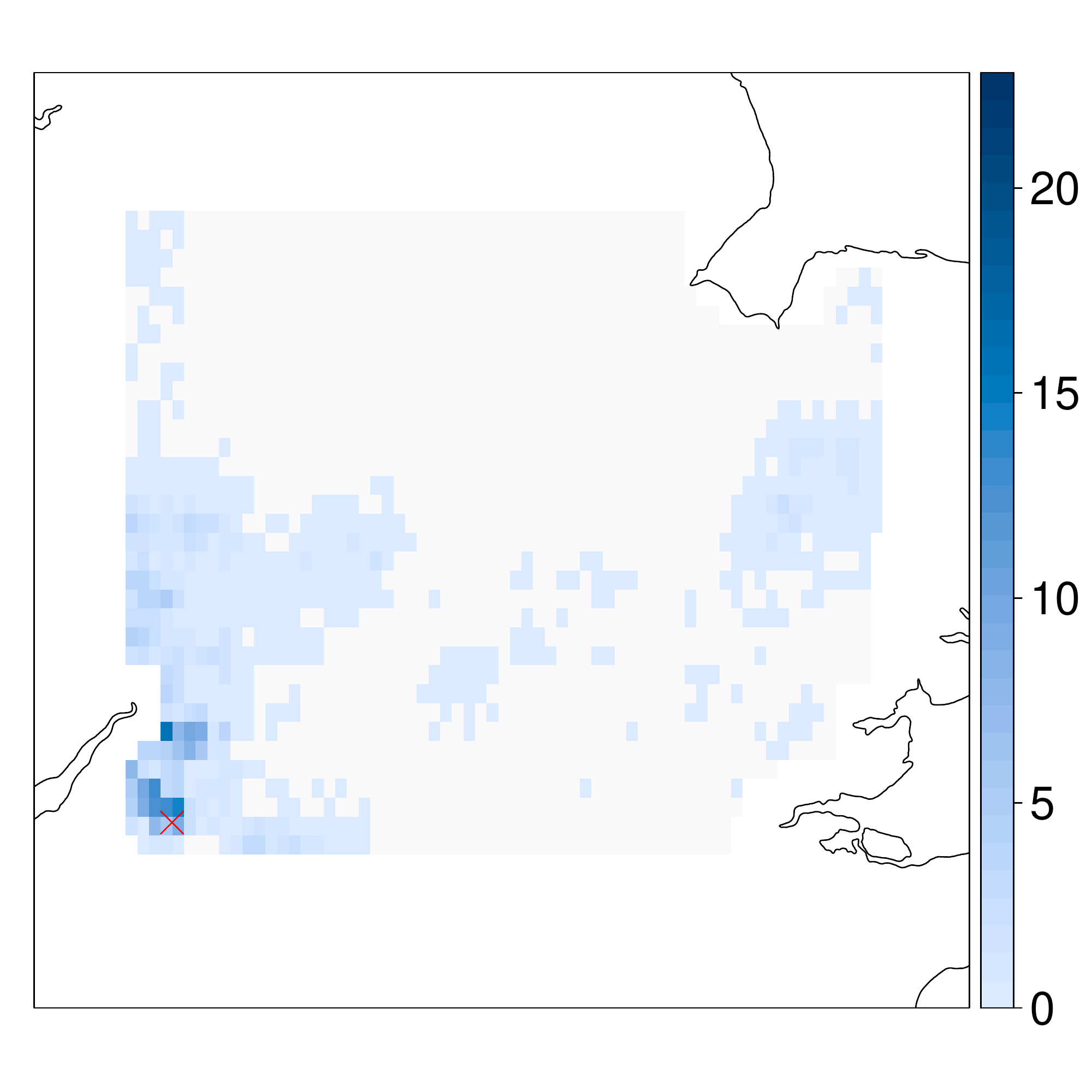} 
\end{minipage}
\begin{minipage}{0.32\linewidth}
\centering
\includegraphics[width=\linewidth]{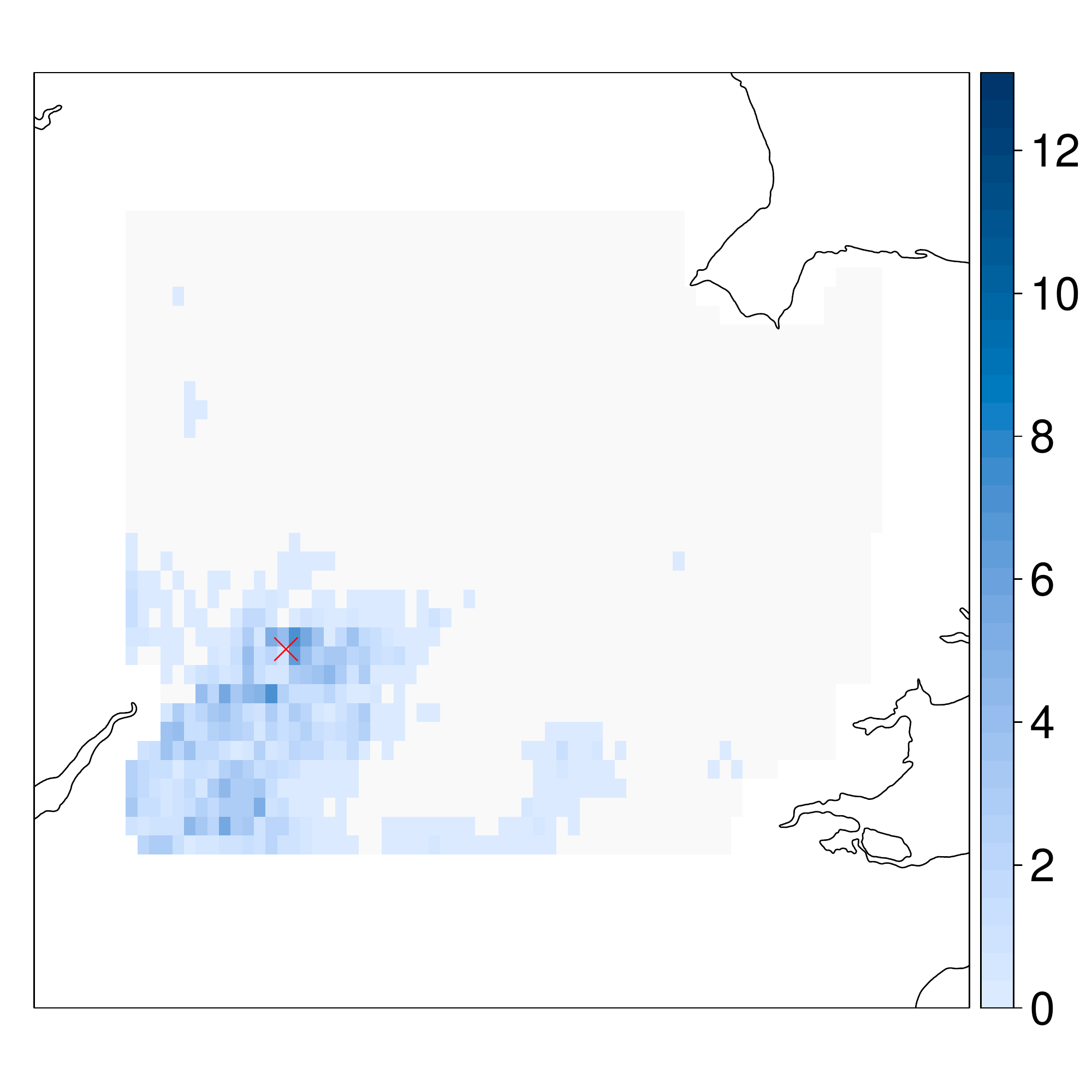} 
\end{minipage}
\begin{minipage}{0.32\linewidth}
\centering
\includegraphics[width=\linewidth]{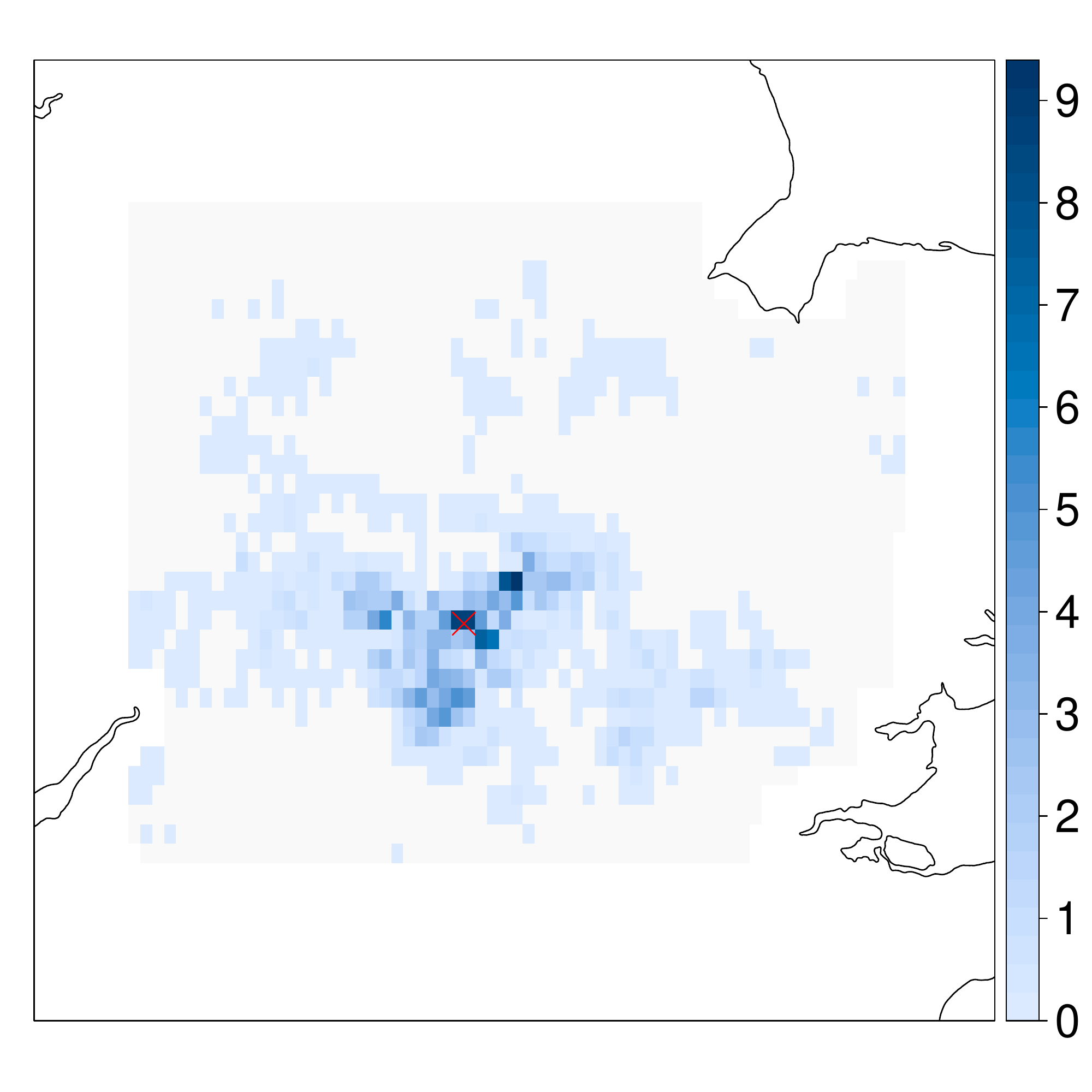} 
\end{minipage}
\begin{minipage}{0.32\linewidth}
\centering
\includegraphics[width=\linewidth]{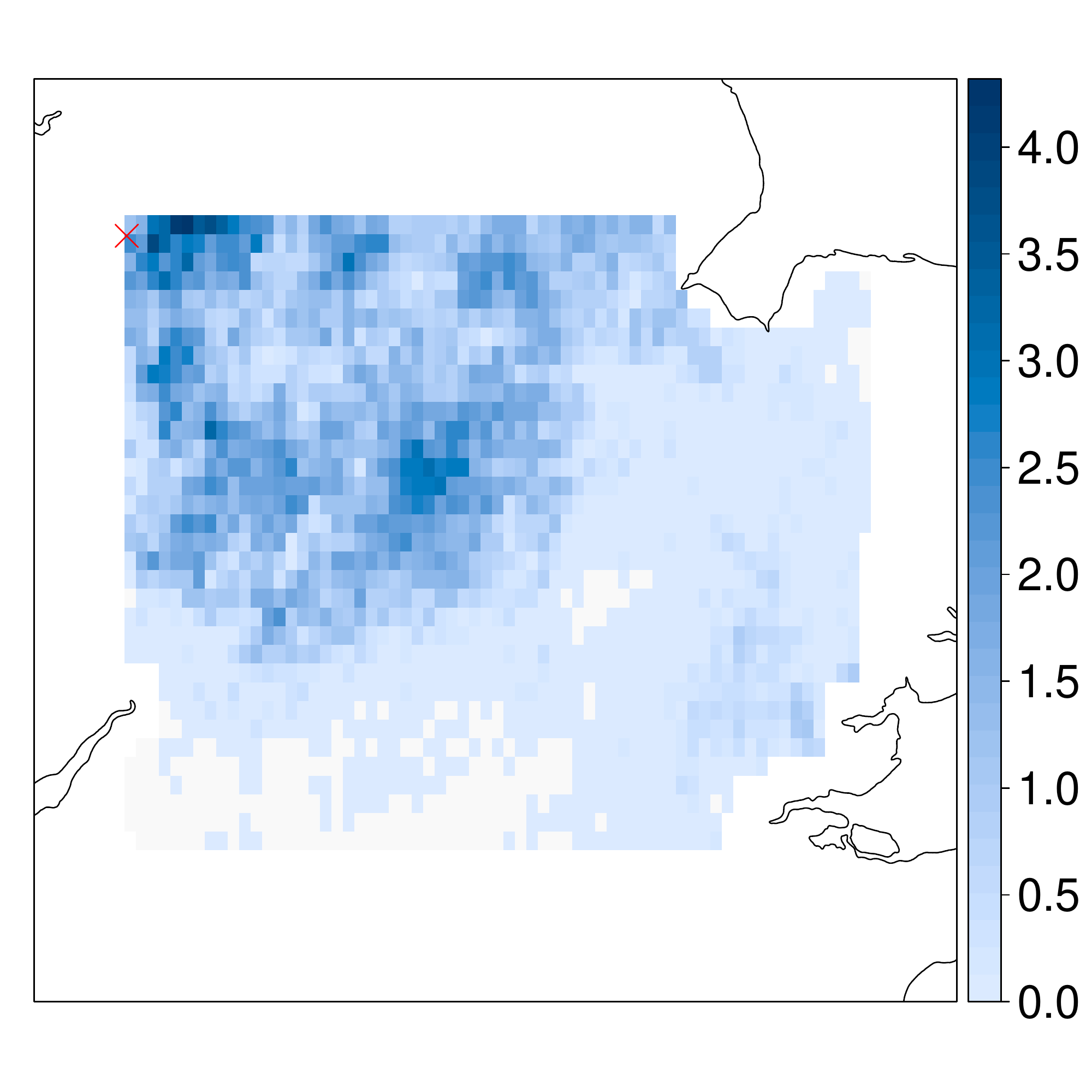} 
\end{minipage}
\begin{minipage}{0.32\linewidth}
\centering
\includegraphics[width=\linewidth]{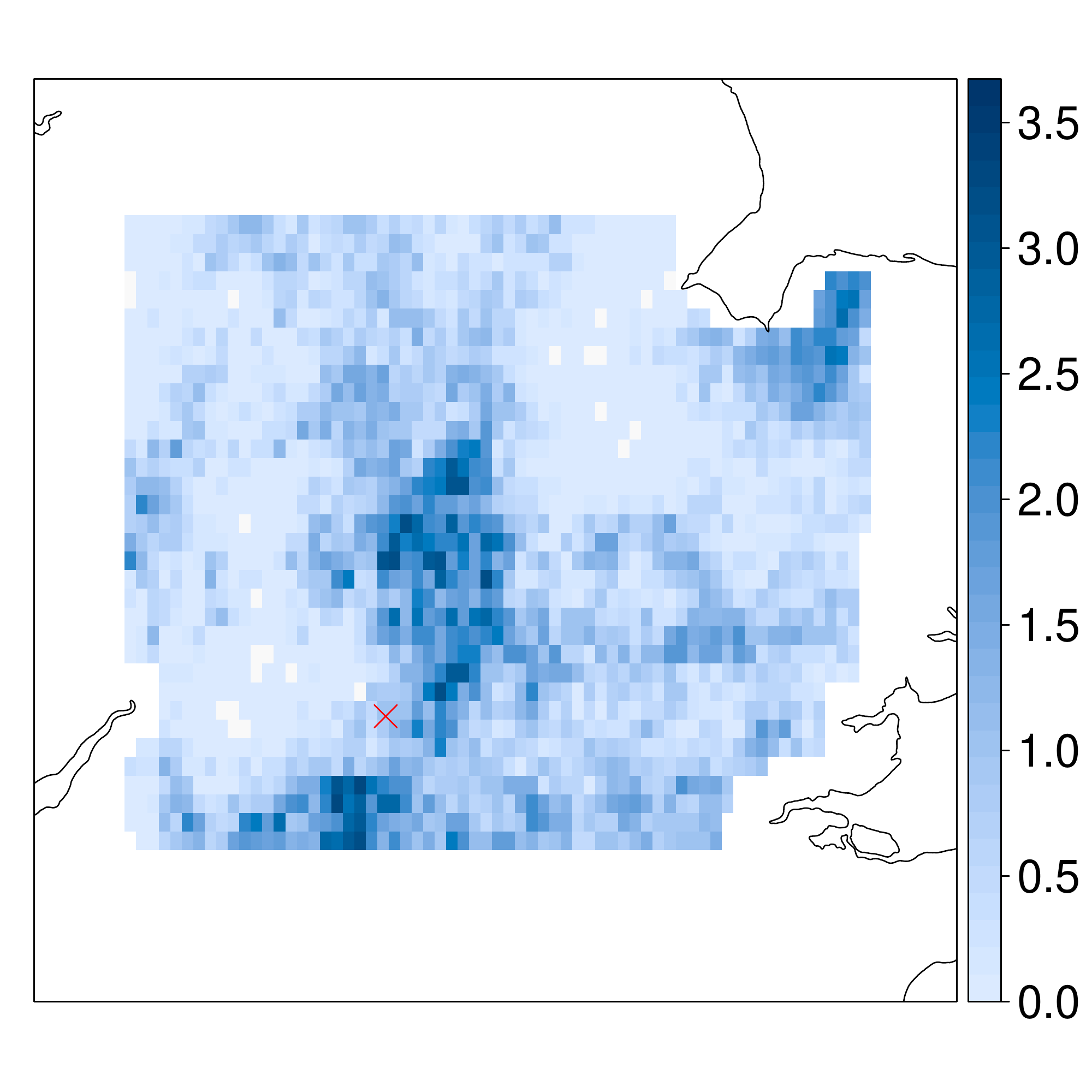} 
\end{minipage}
\begin{minipage}{0.32\linewidth}
\centering

\includegraphics[width=\linewidth]{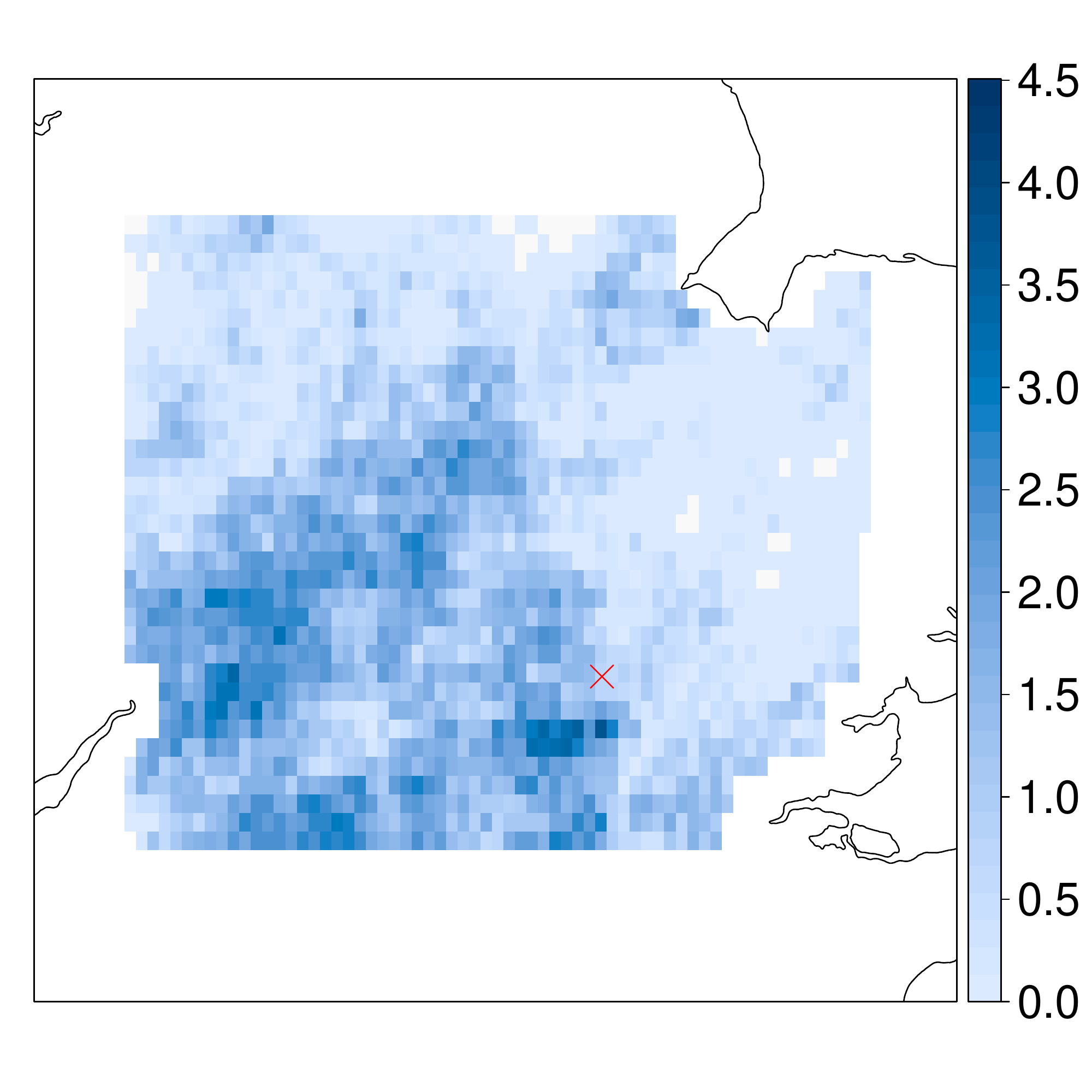} 
\end{minipage}

\caption{Extreme precipitation fields (mm/hr). Realisations from the
fitted models for exceedances over the 0.99 quantile at conditioning site $s_O$ for convective (top) and non-convective (bottom) events. The conditioning sites $s_O$ are given by the red crosses. Scales differ within each panel and row.}
\label{figrealise}
\end{figure}

Figure~\ref{figrealise} presents realisations of $\{Y(s):s \in \mathcal{S}\}\;|\;\{Y(s_O)>F^{-1}_{Y(s_O)}(0.99)\}$ for $Y_\mathcal{C}$ and $Y_\mathcal{N}$, where each $s_O$ was uniformly sampled over $\mathcal{S}$, with their locations identified; similar plots are given for $Y_\mathcal{E}$ in Figure~\ref{figrealise_sup}. There are similarities for $Y_\mathcal{C}$ and $Y_\mathcal{E}$, with characteristics we anticipated for convective rainfall, i.e., high intensity, spatially localised events with a large proportion of the domain $\mathcal{S}$ being dry. In contrast, $Y_\mathcal{N}$ produces events that are much smoother spatially, cover a much large area, and are lower in their intensity.
\subsection{Inference on spatial aggregates}
\label{appl-inferaggs}
 For each process, we draw $b=5.5\times 10^5$ realisations using Algorithm~\ref{chap6-sim-algo}, with the number of population replicates, $b^{'}$, being increased until the estimated quantiles of $\bar{R}_\mathcal{A}$ were stable; we used $b^{'}:= 8b$ for $Y_\mathcal{C}$ and $Y_\mathcal{E}$, and $b^{'} = 8b$ for $Y_\mathcal{N}$. Regions $\boldsymbol{S}_\tau, \boldsymbol{A}_c$ and $\boldsymbol{A}$, described in Section~\ref{choose_sec}, are illustrated in Figure~\ref{agg_diags_locs} with $\boldsymbol{S}:=\mathbf{s}$; the sets $\boldsymbol{A}$ and $\boldsymbol{S}_\tau$ are not changed over the processes, but we take $\boldsymbol{A}_c=\boldsymbol{S}_\tau$ for $Y_\mathcal{C}$ and $Y_\mathcal{E}$ and for $Y_\mathcal{N}$ we create $\boldsymbol{A}_c$ using the heuristic described in Section~\ref{choose_sec}, with $\boldsymbol{S}^+:=\boldsymbol{A}_c\setminus\boldsymbol{S}_\tau$ illustrated by the purple points in Figure~\ref{agg_diags_locs}. Using these regions, we create samples of $\{Y_\mathcal{M}(s):s \in \boldsymbol{S}_\tau\}$ for different regions $\boldsymbol{A}$ (see Figure~\ref{agg_diags_locs})
 by drawing from $\{Y_\mathcal{C}(s):s \in \boldsymbol{S}_\tau\}$ and $\{Y_\mathcal{N}(s):s \in \boldsymbol{S}_\tau\}$ with estimated probability $\hat{p}_\mathcal{C}$ and $1-\hat{p}_\mathcal{C}$, respectively.\par
\begin{figure}[h!]
\begin{minipage}{0.48\textwidth}
\centering
\includegraphics[page=1,width=\textwidth]{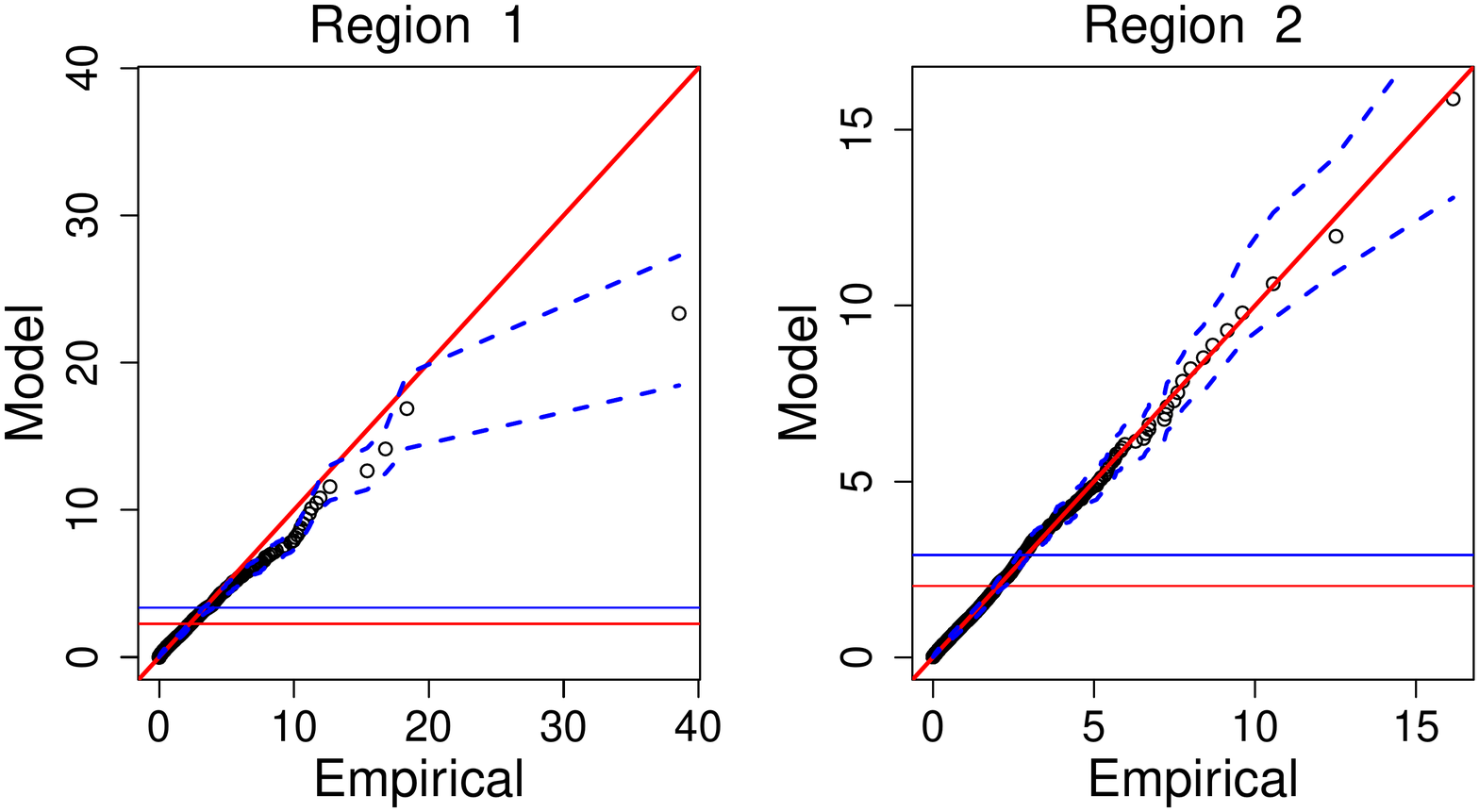} 
\end{minipage}
\hfill
\begin{minipage}{0.48\textwidth}
\centering
\includegraphics[page=2,width=\textwidth]{figures/Agg_Both_boot.pdf} 
\end{minipage}
\begin{minipage}{0.48\textwidth}
\centering
\includegraphics[page=1,width=\textwidth]{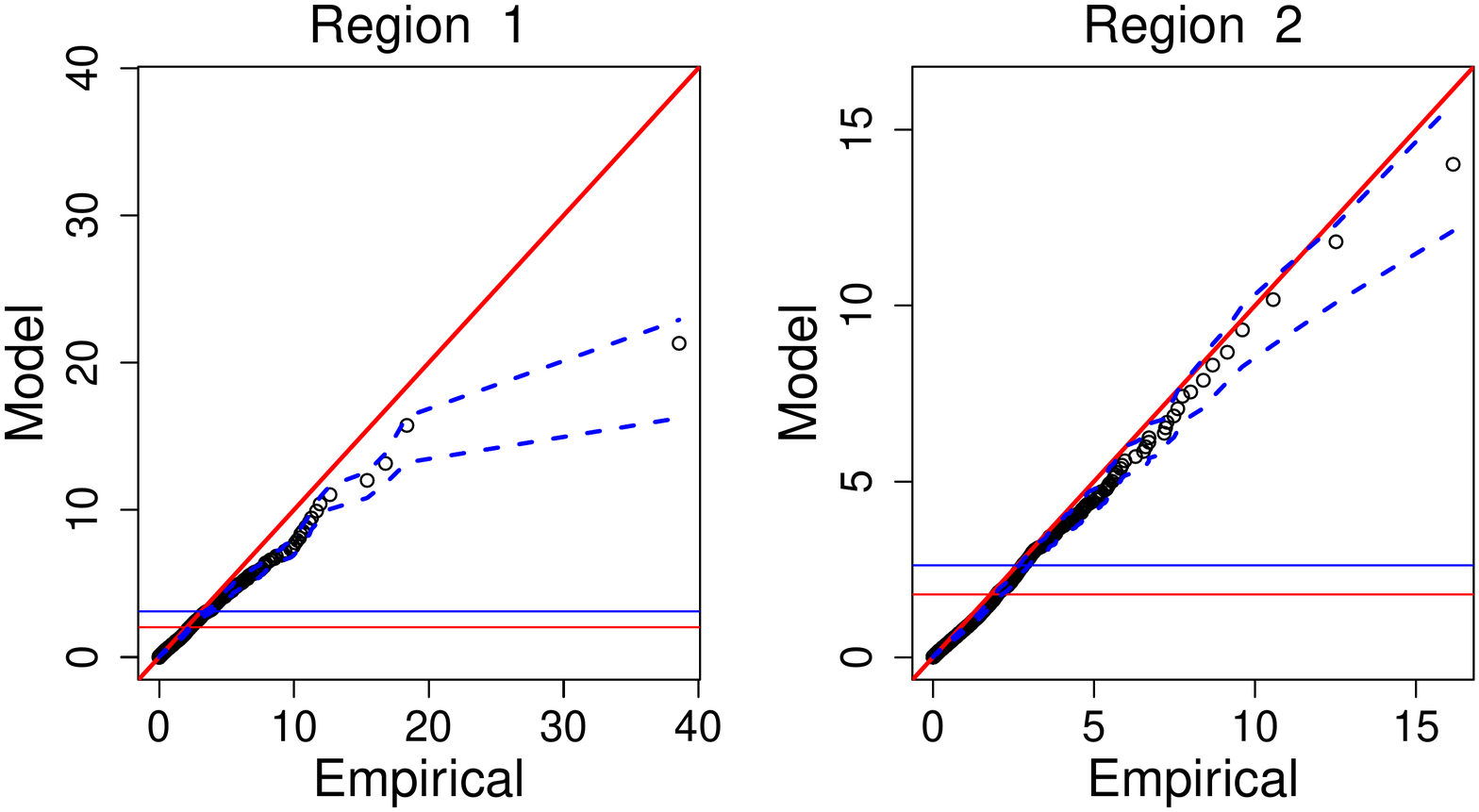} 
\end{minipage}
\hfill
\begin{minipage}{0.48\textwidth}
\centering
\includegraphics[page=2,width=\textwidth]{figures/Agg_All_boot.pdf} 
\end{minipage}

\caption{Q-Q plots for the model, and empirical, of average rainfall rate (mm/hr) over regions of increasing size. Top: new approach with $\bar{R}_{\mathcal{M},\mathcal{A}}$, bottom: existing approach with $\bar{R}_{\mathcal{E},\mathcal{A}}$.
The considered quantiles range from the 0.8-quantile to a value corresponding to the respective 20 year return level. Blue dashed lines denote point-wise $95\%$ quantile estimates. Regions 1--4 correspond to those illustrated in Figure~\ref{agg_diags_locs}. The blue and red horizontal lines denote the $99\%$ and $99.5\%$ quantiles of the respective simulated aggregates.}
\label{agg_diags2}
\end{figure}

 We take $\bar{R}_{\mathcal{M},\mathcal{A}}$ to denote samples from $\bar{R}_{\mathcal{A}}$ created using our new modelling approach and $\bar{R}_{\mathcal{E},\mathcal{A}}$ for samples using the single process approach of \cite{richards2021modelling}. To compare mixture and non-mixture approaches, we present Q-Q plots in Figures~\ref{agg_diags2} and \ref{agg_diags1} for spatial aggregates. In Figure~\ref{agg_diags1}, we assess the fit of individual mixture components $Y_\mathcal{C}$ and $Y_\mathcal{N}$ by using Q-Q plots of the variables $\bar{R}_{\mathcal{C},\mathcal{A}}:=|\mathcal{A}|^{-1}\int_\mathcal{A}Y_\mathcal{C}(s)\mathrm{d}s\approx |\boldsymbol{A}|^{-1}\sum_{s\in\boldsymbol{A}}Y_\mathcal{C}(s)$ and similarly for $\bar{R}_{\mathcal{N},\mathcal{A}}$. For Figure~\ref{agg_diags2}, we compare Q-Q plots for both $\bar{R}_{\mathcal{M},\mathcal{A}}$ and $\bar{R}_{\mathcal{E},\mathcal{A}}$. Point-wise confidence interval estimates for the quantiles on these figures are created using a bootstrap technique. Specifically, we obtain a sample of $50$ parameter estimates for each of model $X_\mathcal{C}$ and $X_\mathcal{N}$  by first applying a stationary bootstrap \citep{politis1994stationary} to observations of the process with expected block size of $48$ hours, and then fitting the extremal dependence model of Section~\ref{dep_sec}, i.e., with the same $h_{max}$ and $d_s$, but using different triples of sampling locations. For each of the $50$ parameter estimates for $X_\mathcal{C}$ and $X_\mathcal{N}$, we draw $5.5\times 10^5$ realisations of $\{Y_\mathcal{M}(s):s \in \boldsymbol{A}_c\}$, with $\boldsymbol{A}_c$ and $\boldsymbol{S}_\tau$ consistent across parameter estimates. Our implementation does not account for the classification uncertainty as we believe that incorporating its uncertainty does not justify the computational expense given the precision with which $p_\mathcal{C}$ is estimated.
\par
 Figures~\ref{agg_diags2} and \ref{agg_diags1} show good fits for all components of both modelling approaches, with slight underestimation for the smaller quantiles of $\bar{R}_{\mathcal{N},\mathcal{A}}$. It appears that $\bar{R}_{\mathcal{M},\mathcal{A}}$ provides slightly better fits than the corresponding $\bar{R}_{\mathcal{E},\mathcal{A}}$ for most of the tail, particularly for the largest regions (3 and 4); indicating an improvement over the approach of \cite{richards2021modelling}. To help understand these observed differences in fit, we estimate the proportion $p_\mathcal{M}(r^{(l)},\mathcal{A})$ of non-convective events that contribute to $\bar{R}_{\mathcal{M},\mathcal{A}}>r^{(l)}$, where here $r^{(l)}$ denotes the $99\%$ and $99.5\%$ quantile of the simulated $\bar{R}_{\mathcal{M},\mathcal{A}}$. The proportion $p_\mathcal{M}(r^{(l)},\mathcal{A})$ increases with the size of $\mathcal{A}$; for the $99\%$ quantile, $p_\mathcal{M}(r^{(l)},\mathcal{A})$ increases from $0.029$ to $0.052$ for the smallest and largest regions, respectively, and for the $99.5\%$ quantile, the values increase from $0.004$ to $0.040$. Neither the models for $\bar{R}_{\mathcal{M},\mathcal{A}}$ nor $\bar{R}_{\mathcal{E},\mathcal{A}}$ are able to capture the very largest empirical quantiles when $\mathcal{A}$ is the smallest region, region 1. This discrepancy was caused by two potentially spurious events that produced large values for the average over region~1, at consecutive hours (Whitall, M., 2021, personal communication).\par
As the differences in Figure~\ref{agg_diags2} are difficult to observe by eye, we better quantify our assertion that model $\mathcal{M}$ fits better than model $\mathcal{E}$, through a comparison using a metric adapted from a measure of \cite{varty2021inference}. Let $Q(p;W):[0,1]\mapsto \mathbb{R}_+$ denote the sample $p$-th quantile of random variable $W$ and $\tilde{R}_\mathcal{A}$ denote the observed spatial aggregate; we define measures of the expected deviance in the Q-Q plot for $W$ against $\tilde{R}_\mathcal{A}$ from the line $y=x$,
\begin{equation}
\label{qqdeviationdiag}
\Lambda_i(W)=\frac{1}{m}\sum^m_{j=1}|Q(p_j;W)-Q(p_j;\tilde{R}_\mathcal{A})|^i,\;\;\text{for}\;i=1,2,
\end{equation}
described through the mean absolute and mean squared distances $\Lambda_1$ and $\Lambda_2$, respectively, where $\{0<p_1< \dots < p_m<1\}$ is a grid of $m\;(m\geq 2)$ equally spaced values with $p_m=1-(p_2-p_1)$. As our interest lines in the upper-tails of $\bar{R}_\mathcal{A}$, we set $p_1$ close to one. We evaluate $\Lambda_1$ and $\Lambda_2$ for $W:=\bar{R}_{\mathcal{M},\mathcal{A}}$ and $W:=\bar{R}_{\mathcal{E},\mathcal{A}}$, and for each region $\mathcal{A}$ illustrated in Figure~\ref{agg_diags_locs}, with $p_1=0.99$ and $m=432$, i.e., the number of observations of $\tilde{R}_\mathcal{A}$ that exceed the $0.99$-quantile; this is done for each sample of bootstrapped parameter estimates, see above. These goodness-of-fit diagnostics are given in Table~\ref{diagnosticTab}, with smaller values indicating a best fit under that metric; our approach provides better fits for all four regions.
\begin{table}[h]
\caption{Median estimates (0.025 and 0.975 quantile estimates) of aggregate diagnostics $\Lambda_1$ and $\Lambda_2$, defined in eq.\ \eqref{qqdeviationdiag}, over all bootstrap samples (2 decimal places). }
 \centering
 \begin{tabular}{p{2cm}cccc} 
\toprule
&\multicolumn{4}{c}{$\mathcal{A}$}\\
\cmidrule(lr){2-5}
&1&2&3&4\\
\hline
\parbox{2cm}{$\Lambda_1(\bar{R}_{\mathcal{M},\mathcal{A}})$}&${0.46\;(0.23,0.73)}$&${0.12\;(0.07,0.31)}$&${0.15\;(0.09,0.37)}$&${0.12\;(0.07,0.32)}$\\
\parbox{2cm}{$\Lambda_1(\bar{R}_{\mathcal{E},\mathcal{A}})$}&${0.78\;(0.52,1.02)}$&${0.32\;(0.11,0.53)}$&${0.22\;(0.11,0.41)}$&${0.23\;(0.10,0.40)}$\\
\hline
\parbox{2cm}{$\Lambda_2(\bar{R}_{\mathcal{M},\mathcal{A}})$}&${1.12\;(0.45,1.79)}$&${0.04\;(0.01,0.15)}$&${0.05\;(0.03,0.21)}$&${0.03\;(0.01,0.18)}$\\
\parbox{1cm}{$\Lambda_2(\bar{R}_{\mathcal{E},\mathcal{A}})$}&${1.70\;(1.17,2.80)}$&${0.13\;(0.02,0.43)}$&${0.08\;(0.03,0.26)}$&${0.08\;(0.02,0.21)}$\\
\hline
  \end{tabular}
 \label{diagnosticTab}
 \end{table}
\section{Discussion}
\label{sec_discuss}
We have presented a simple but effective mixture model extension of the approach proposed by \cite{richards2021modelling} for modelling the extremes of spatial aggregates of precipitation over a wide range of spatial scales. We classify observed fields as being either convective or non-convective, fit separate spatial models to each dataset and then simulate from both models and combine samples to explore the upper-tail behaviour of $\bar{R}_{\mathcal{A}}$. Fits of our spatial models reveal major differences between the extremal dependence structures of convective and non-convective precipitation events, with evidence to suggest that  the latter process may exhibit short-range asymptotic dependence but long-range asymptotic independence. Our mixture modelling approach was compared against an approach where the mixture structure is ignored and we found that the former was able to better capture the extremal behaviour of aggregates over very large spatial regions. We now discuss some further extensions that can be made to improve the model.\par
In Section~\ref{appl-inferaggs}, we identified two consecutive convective events that provided the two largest values of the empirical average rainfall $\bar{R}_\mathcal{A}$ over the smallest aggregate region $\mathcal{A}$, denoted region~1; neither the new nor old model for $\bar{R}_{\mathcal{A}}$ were able to capture the characteristics of these two events. The consecutive nature of the values suggests that there may be a temporal aspect to the data that we cannot currently capture with the model. The data themselves are aggregates of continuous-in-time precipitation over a temporal interval of one hour and a spatial grid-box. Hence, two spatially identical storms can produce different hourly values simply by moving at different speeds through a grid-box. Potential model extensions would incorporate the speed of a storm as a covariate in the model or build a conditional spatial-temporal model \citep{simpson2020conditional} adapted for precipitation.\par
We classified entire observed fields as convective or not regardless of the amount of convective rainfall properties across $\mathcal{S}$, or the presence/proportion of non-convective rainfall within the same spatial domain. The classification of the whole field into one of two processes appears to work well as we observe distinctly different structures for the two identified components. A more realistic approach is to accommodate spatial mixing of convective and non-convective rainfall across $\mathcal{S}$ within a single field. This may require modelling of extremal dependence between rain types and a framework for simulating different types of event within the same region.\par
 The classification technique is deterministic and so improvements may be achieved by adopting a more probabilistic approach with labelling of mixture components conducted within inference. In this vein, \cite{tendijck2021modeling} have developed mixture models in a conditional extremes setting by allowing the $\alpha$ and $\beta$ functions to change with the mixture component, but only in a bivariate context. For the residual process $Z(s|s_O)$, a Dirichlet mixture of Gaussian processes \citep{Duan2007} could be used, rather than a single process. This would have the added benefit of the model not being limited to only two mixture components. Instead, a number of mixture components could be used, each with their own dependence structure; that is, we could model more than two ``types" of extreme rainfall. In the context of modelling spatial extremes, \cite{hazra2019estimating} advocate the use of Dirichlet mixtures of Student's-$t$ processes. \par This work is an important step towards protecting society from the impacts of extreme rainfall through the facilitation of more precise estimation of rainfall hazards from short-run climate model simulations. For this work to be appropriate in the future, we must accommodate future climate change into both the marginal and extremal dependence structures of the underlying process, which could be achieved through the use of linear or additive regression models. Further non-stationarity in the extremal dependence structure may be handled through the inclusion of other meteorological covariates, e.g., wind-speed, which is likely to have a large impact on the anisotropy in $Y_\mathcal{M}$. Finally, we recognise that it is not only the extremes of hourly-spatial aggregates that contribute to river flooding, but also the extremes of aggregates taken over temporal windows of varying length. In order to adapt our approach to allow for modelling of spatio-temporal aggregates, we would require a conditional spatial-temporal dependence model \citep{simpson2020conditional}.
\section*{Acknowledgments} Richards and Tawn gratefully acknowledge funding through the STOR-i Doctoral Training Centre and Engineering and Physical Sciences Research Council (grant EP/L015692/1). This publication is based upon work supported by the King Abdullah University of Science and Technology (KAUST) Office of Sponsored Research (OSR) under Award No. OSR-CRG2020-4394. Support from the KAUST Supercomputing Laboratory is gratefully acknowledged. Brown was supported by the Met Office Hadley Centre Climate Programme funded by BEIS and Defra. The data used in our analyses can be downloaded from the CEDA data catalogue, see \cite{datacite}. Code that supports our findings can be found in the \href{http:/github.com/Jbrich95/scePrecip}{\texttt{Jbrich95/scePrecip}} repository on GitHub.

%%%%%%%%%%%%%%%%%%%%%%%%%%%%%%%%%
%%%%%%%%%%%%%%%%%%%%%%%%%%%%%%%%%
%%%%%%%%%%%%%%%%%%%%%%%%%%%%%%%%% 

\baselineskip=14pt
\begingroup
\setstretch{0.75}

\bibliographystyle{apalike}
{\small
\bibliography{Ref}

\begin{thebibliography}{}

\bibitem[Berg et~al., 2013]{berg2013strong}
Berg, P., Moseley, C., and Haerter, J.~O. (2013).
\newblock Strong increase in convective precipitation in response to higher
  temperatures.
\newblock {\em Nature Geoscience}, 6(3):181--185.

\bibitem[Brown, 2018]{Brown2018}
Brown, S.~J. (2018).
\newblock {The drivers of variability in UK extreme rainfall}.
\newblock {\em International Journal of Climatology}, 38(S1):e119--e130.

\bibitem[Catto and Pfahl, 2013]{catto2013importance}
Catto, J.~L. and Pfahl, S. (2013).
\newblock The importance of fronts for extreme precipitation.
\newblock {\em Journal of Geophysical Research: Atmospheres}, 118(19):10--791.

\bibitem[Chan et~al., 2014]{chan2014value}
Chan, S.~C., Kendon, E.~J., Fowler, H.~J., Blenkinsop, S., Roberts, N.~M., and
  Ferro, C.~A. (2014).
\newblock {The value of high-resolution Met Office regional climate models in
  the simulation of multihourly precipitation extremes}.
\newblock {\em Journal of Climate}, 27(16):6155--6174.

\bibitem[Coles, 1993]{coles1993}
Coles, S.~G. (1993).
\newblock Regional modelling of extreme storms via max-stable processes.
\newblock {\em Journal of the Royal Statistical Society: Series B
  (Methodological)}, 55(4):797--816.

\bibitem[Coles, 2001]{Coles2001book}
Coles, S.~G. (2001).
\newblock {\em An Introduction to Statistical Modeling of Extreme Values}.
\newblock Springer-Verlag, London.

\bibitem[Coles and Tawn, 1996]{coles1996}
Coles, S.~G. and Tawn, J.~A. (1996).
\newblock Modelling extremes of the areal rainfall process.
\newblock {\em Journal of the Royal Statistical Society. Series B
  (Methodological)}, 58(2):329--347.

\bibitem[Cooley et~al., 2007]{cooley2007}
Cooley, D., Nychka, D., and Naveau, P. (2007).
\newblock Bayesian spatial modeling of extreme precipitation return levels.
\newblock {\em Journal of the American Statistical Association},
  102(479):824--840.

\bibitem[Cooley and Sain, 2010]{cooley2010spatial}
Cooley, D. and Sain, S.~R. (2010).
\newblock Spatial hierarchical modeling of precipitation extremes from a
  regional climate model.
\newblock {\em Journal of Agricultural, Biological, and Environmental
  Statistics}, 15(3):381--402.

\bibitem[de~Fondeville and Davison, 2022]{deFondeville2020functional}
de~Fondeville, R. and Davison, A.~C. (2022).
\newblock Functional peaks-over-threshold analysis.
\newblock {\em Journal of the Royal Statistical Society: Series B
  (Methodology)}, 84(4):1392--1422.

\bibitem[Duan et~al., 2007]{Duan2007}
Duan, J.~A., Guindani, M., and Gelfand, A.~E. (2007).
\newblock Generalized spatial {D}irichlet process models.
\newblock {\em Biometrika}, 94(4):809--825.

\bibitem[Fasiolo et~al., 2021]{qgam}
Fasiolo, M., Wood, S.~N., Zaffran, M., Nedellec, R., and Goude, Y. (2021).
\newblock Fast calibrated additive quantile regression.
\newblock {\em Journal of the American Statistical Association},
  116(535):1402--1412.

\bibitem[Gregersen et~al., 2013]{gregersen2013assessing}
Gregersen, I.~B., S{\o}rup, H. J.~D., Madsen, H., Rosbjerg, D., Mikkelsen,
  P.~S., and Arnbjerg-Nielsen, K. (2013).
\newblock Assessing future climatic changes of rainfall extremes at small
  spatio-temporal scales.
\newblock {\em Climatic Change}, 118(3):783--797.

\bibitem[Hazra and Huser, 2021]{hazra2019estimating}
Hazra, A. and Huser, R. (2021).
\newblock {Estimating high-resolution Red Sea surface temperature hotspots,
  using a low-rank semiparametric spatial model}.
\newblock {\em The Annals of Applied Statistics}, 15(2):572 -- 596.

\bibitem[Heffernan and Tawn, 2001]{heffernan2001extreme}
Heffernan, J.~E. and Tawn, J.~A. (2001).
\newblock {Extreme value analysis of a large designed experiment: A case study
  in bulk carrier safety}.
\newblock {\em Extremes}, 4(4):359--378.

\bibitem[Heffernan and Tawn, 2004]{heff2004}
Heffernan, J.~E. and Tawn, J.~A. (2004).
\newblock A conditional approach for multivariate extreme values (with
  discussion).
\newblock {\em Journal of the Royal Statistical Society: Series B
  (Methodology)}, 66(3):497--546.

\bibitem[Hosseinzadehtalaei et~al., 2020]{hosseinzadehtalaei2020climate}
Hosseinzadehtalaei, P., Tabari, H., and Willems, P. (2020).
\newblock {Climate change impact on short-duration extreme precipitation and
  intensity--duration--frequency curves over Europe}.
\newblock {\em Journal of Hydrology}, 590:125249.

\bibitem[Houze~Jr, 1997]{houze1997stratiform}
Houze~Jr, R.~A. (1997).
\newblock Stratiform precipitation in regions of convection: A meteorological
  paradox?
\newblock {\em Bulletin of the American Meteorological Society},
  78(10):2179--2196.

\bibitem[Huser and Wadsworth, 2020]{huser2020advances}
Huser, R. and Wadsworth, J.~L. (2020).
\newblock Advances in statistical modeling of spatial extremes.
\newblock {\em Wiley Interdisciplinary Reviews: Computational Statistics},
  14:e1537.

\bibitem[{Institute of Hydrology}, 1999]{feh1999}
{Institute of Hydrology} (1999).
\newblock {\em Flood Estimation Handbook}.
\newblock Wallingford, UK.

\bibitem[Joe, 1997]{joe1997multivariate}
Joe, H. (1997).
\newblock {\em Multivariate Models and Multivariate Dependence Concepts}.
\newblock CRC Press, New York.

\bibitem[Keef et~al., 2013]{keef2013}
Keef, C., Papastathopoulos, I., and Tawn, J.~A. (2013).
\newblock {Estimation of the conditional distribution of a multivariate
  variable given that one of its components is large: Additional constraints
  for the Heffernan and Tawn model}.
\newblock {\em Journal of Multivariate Analysis}, 115:396--404.

\bibitem[Kendon et~al., 2019]{ukcp18CPMscience}
Kendon, E., Fosser, G., Murphy, J., Chan, S., Clark, R., Harris, G., Lock, A.,
  Lowe, J., Martin, G., Pirret, J., Roberts, N., Sanderson, M., and Tucker, S.
  (2019).
\newblock {UKCP Convection-permitting model projections: Science report}.
\newblock Technical report, The Met Office.

\bibitem[Kendon et~al., 2021]{kendon2021}
Kendon, E., Short, C., Pope, J., Chan, S., Wilkinson, J., Tucker, S., Bett, P.,
  and Harris, G. (2021).
\newblock {Update to UKCP Local (2.2km) projections}.
\newblock Technical report, The Met Office.

\bibitem[Kendon et~al., 2012]{kendon2012realism}
Kendon, E.~J., Roberts, N.~M., Senior, C.~A., and Roberts, M.~J. (2012).
\newblock Realism of rainfall in a very high-resolution regional climate model.
\newblock {\em Journal of Climate}, 25(17):5791--5806.

\bibitem[Martins and Stedinger, 2000]{martins2000generalized}
Martins, E.~S. and Stedinger, J.~R. (2000).
\newblock Generalized maximum-likelihood generalized extreme-value quantile
  estimators for hydrologic data.
\newblock {\em Water Resources Research}, 36(3):737--744.

\bibitem[{Met Office Hadley Centre}, 2019]{datacite}
{Met Office Hadley Centre} (2019).
\newblock {UKCP Convection-Permitting Model Projections for the UK at 2.2km
  resolution. NERC EDS Centre for Environmental Data Analysis.}
\newblock
  \url{https://catalogue.ceda.ac.uk/uuid/ad2ac0ddd3f34210b0d6e19bfc335539}.
\newblock Last accessed 14/09/2022.

\bibitem[Padoan et~al., 2010]{padoan2010likelihood}
Padoan, S.~A., Ribatet, M., and Sisson, S.~A. (2010).
\newblock Likelihood-based inference for max-stable processes.
\newblock {\em Journal of the American Statistical Association},
  105(489):263--277.

\bibitem[Palacios-Rodr{\'\i}guez et~al., 2020]{palacios2020generalized}
Palacios-Rodr{\'\i}guez, F., Toulemonde, G., Carreau, J., and Opitz, T. (2020).
\newblock {Generalized Pareto processes for simulating space-time extreme
  events: an application to precipitation reanalyses}.
\newblock {\em Stochastic Environmental Research and Risk Assessment},
  34(12):2033--2052.

\bibitem[Politis and Romano, 1994]{politis1994stationary}
Politis, D.~N. and Romano, J.~P. (1994).
\newblock The stationary bootstrap.
\newblock {\em Journal of the American Statistical Association},
  89(428):1303--1313.

\bibitem[Reich and Shaby, 2012]{reich2012hierarchical}
Reich, B.~J. and Shaby, R.~A. (2012).
\newblock A hierarchical max-stable spatial model for extreme precipitation.
\newblock {\em The Annals of Applied Statistics}, 6(4):1430.

\bibitem[Richards et~al., 2022]{richards2021modelling}
Richards, J., Tawn, J.~A., and Brown, S. (2022).
\newblock Modelling extremes of spatial aggregates of precipitation using
  conditional methods.
\newblock {\em The Annals of Applied Statistics}, 16(4):2693--2713.

\bibitem[Sang and Gelfand, 2010]{sang2010continuous}
Sang, H. and Gelfand, A.~E. (2010).
\newblock Continuous spatial process models for spatial extreme values.
\newblock {\em Journal of Agricultural, Biological and Environmental
  Statistics}, 15(1):49--65.

\bibitem[Saunders et~al., 2017]{SAUNDERS201717}
Saunders, K., Stephenson, A.~G., Taylor, P.~G., and Karoly, D. (2017).
\newblock The spatial distribution of rainfall extremes and the influence of
  {El Niño Southern Oscillation}.
\newblock {\em Weather and Climate Extremes}, 18:17--28.

\bibitem[Schroeer et~al., 2018]{Schroeer2018}
Schroeer, K., Kirchengast, G., and O, S. (2018).
\newblock Strong dependence of extreme convective precipitation intensities on
  gauge network density.
\newblock {\em Geophysical Research Letters}, 45(16):8253--8263.

\bibitem[Shooter et~al., 2021a]{shooter2021spatial}
Shooter, R., Ross, E., Ribal, A., Young, I., and Jonathan, P. (2021a).
\newblock {Spatial dependence of extreme seas in the North East Atlantic from
  satellite altimeter measurements}.
\newblock {\em Environmetrics}, 32(4):e2674.

\bibitem[Shooter et~al., 2019]{shooter2019}
Shooter, R., Ross, E., Tawn, J.~A., and Jonathan, P. (2019).
\newblock On spatial conditional extremes for ocean storm severity.
\newblock {\em Environmetrics}, 30(6):e2562.

\bibitem[Shooter et~al., 2021b]{shooterinpress}
Shooter, R., Tawn, J.~A., Ross, E., and Jonathan, P. (2021b).
\newblock Basin-wide spatial conditional extremes for severe ocean storms.
\newblock {\em Extremes}, 24(2):241--265.

\bibitem[Simpson et~al., 2022]{simpson2020highdimensional}
Simpson, E.~S., Opitz, T., and Wadsworth, J.~L. (2022).
\newblock {High-dimensional modeling of spatial and spatio-temporal conditional
  extremes using INLA and Gaussian Markov random fields}.
\newblock {\em arXiv e-prints}, arXiv:2011.04486.

\bibitem[Simpson and Wadsworth, 2021]{simpson2020conditional}
Simpson, E.~S. and Wadsworth, J.~L. (2021).
\newblock {Conditional modelling of spatio-temporal extremes for Red Sea
  surface temperatures}.
\newblock {\em Spatial Statistics}, 41:100482.

\bibitem[Simpson et~al., 2020]{simpson2020determining}
Simpson, E.~S., Wadsworth, J.~L., and Tawn, J.~A. (2020).
\newblock Determining the dependence structure of multivariate extremes.
\newblock {\em Biometrika}, 107(3):513--532.

\bibitem[Tawn et~al., 2018]{Tawn2018}
Tawn, J.~A., Shooter, R., Towe, R., and Lamb, R. (2018).
\newblock Modelling spatial extreme events with environmental applications.
\newblock {\em Spatial Statistics}, 28:39--58.

\bibitem[Tendijck et~al., 2021]{tendijck2021modeling}
Tendijck, S., Eastoe, E., Tawn, J., Randell, D., and Jonathan, P. (2021).
\newblock Modeling the extremes of bivariate mixture distributions with
  application to oceanographic data.
\newblock {\em Journal of the American Statistical Association},
  doi:10.1080/01621459.2021.1996379.

\bibitem[Thibaud et~al., 2013]{thibaud2013threshold}
Thibaud, E., Mutzner, R., and Davison, A.~C. (2013).
\newblock Threshold modeling of extreme spatial rainfall.
\newblock {\em Water Resources Research}, 49(8):4633--4644.

\bibitem[Varty et~al., 2021]{varty2021inference}
Varty, Z., Tawn, J.~A., Atkinson, P.~M., and Bierman, S. (2021).
\newblock Inference for extreme earthquake magnitudes accounting for a
  time-varying measurement process.
\newblock {\em arXiv preprint arXiv:2102.00884}.

\bibitem[Wadsworth and Tawn, 2022]{wadsworth2018spatial}
Wadsworth, J. and Tawn, J. (2022).
\newblock Higher-dimensional spatial extremes via single-site conditioning.
\newblock {\em Spatial Statistics}, 51:100677.

\bibitem[Weller et~al., 2013]{weller2013two}
Weller, G.~B., Cooley, D., Sain, S.~R., Bukovsky, M.~S., and Mearns, L.~O.
  (2013).
\newblock Two case studies on {NARCCAP} precipitation extremes.
\newblock {\em Journal of Geophysical Research: Atmospheres}, 118(18):10--475.

\bibitem[Westra and Sisson, 2011]{westra2011detection}
Westra, S. and Sisson, S.~A. (2011).
\newblock Detection of non-stationarity in precipitation extremes using a
  max-stable process model.
\newblock {\em Journal of Hydrology}, 406(1-2):119--128.

\bibitem[Wood, 2006]{wood2006generalized}
Wood, S. (2006).
\newblock {\em Generalized Additive Models: An Introduction with R}.
\newblock Chapman \& Hall/CRC, New York. 2nd edition.

\bibitem[Youngman, 2019]{youngman2019}
Youngman, B.~D. (2019).
\newblock Generalized additive models for exceedances of high thresholds with
  an application to return level estimation for {U.S.} wind gusts.
\newblock {\em Journal of the American Statistical Association},
  114(528):1865--1879.

\bibitem[Zheng et~al., 2015]{Zheng2015}
Zheng, F., Thibaud, E., Leonard, M., and Westra, S. (2015).
\newblock Assessing the performance of the independence method in modeling
  spatial extreme rainfall.
\newblock {\em Water Resources Research}, 51(9):7744--7758.

\end{thebibliography}
}
\endgroup
\baselineskip 10pt
\renewcommand{\theequation}{S.\arabic{equation}}
\renewcommand{\thefigure}{S\arabic{figure}}
\renewcommand{\thetable}{S\arabic{table}}
\renewcommand{\thesection}{S\arabic{section}}
\setcounter{figure}{0}
\setcounter{table}{0}
\setcounter{equation}{0}
\pagebreak
\begin{appendix}
 \begin{center} {\Large{\bf Supplement to ``Joint Estimation of Extreme Spatially Aggregated Precipitation at Different Scales through Mixture Modelling"}}
\end{center}
 \section{Model specification for $X_\mathcal{C}$ and $X_\mathcal{E}$}
\label{model_spec_sup}
For $X_\mathcal{C}$ and $X_\mathcal{E}$, we take the following parametric forms for $\beta$, $\mu$, $\sigma$ and $\delta$; for $h\geq 0$,
\begin{align}
\label{depend_funcs}
\beta(h)&=
\kappa_{\beta_3}\exp\{-(h/\kappa_{\beta_1})^{\kappa_{\beta_2}}\},&\kappa_{\beta_1} >0, \kappa_{\beta_2} >0, \kappa_{\beta_3}\in[0,1],\\
\mu(h)&=\kappa_{\mu_1}h^{\kappa_{\mu_2}}\exp(-h/\kappa_{\mu_3}),&\kappa_{\mu_2} >0, \kappa_{\mu_3} >0,\nonumber\\
\sigma(h)&=\sqrt{2}\left[1-\exp\{-(h/\kappa_{\sigma_1})^{\kappa_{\sigma_2}}\}\right],&\kappa_{\sigma_1}>0, \kappa_{\sigma_2} >0,\nonumber\\
\delta(h)&=\max\{1,1+(\kappa_{\delta_1}h^{\kappa_{\delta_2}}-\kappa_{\delta_4})\exp(-h/\kappa_{\delta_3})\},\qquad&\kappa_{\delta_1} \geq 0, \kappa_{\delta_2} >0, \kappa_{\delta_3} >0, \kappa_{\delta_4}<1.\nonumber
\end{align}
For $X_\mathcal{N}$, we use $\mu$ and $\delta$ as above, but with the newly proposed $\beta_{\mathcal{N}}$ and $\sigma_{\mathcal{N}}$ functions  
\begin{align*}
\beta_{\mathcal{N}}(h)&=\frac{\kappa_{\beta_1}h^{\kappa_{\beta_2}}\exp(-h/\kappa_{\beta_3})}{\max_{h_*>0}\{h_*^{\kappa_{\beta_2}}\exp(-h_*/\kappa_{\beta_3})\}},\qquad&\kappa_{\beta_1}\in[0,1],\kappa_{\beta_2} >0, \kappa_{\beta_3} >0,\\
\sigma_{\mathcal{N}}(h)&=\kappa_{\sigma_3}\left[1-\exp\{-(h/\kappa_{\sigma_1})^{\kappa_{\sigma_2}}\}\right],\qquad&\kappa_{\sigma_1}>0, \kappa_{\sigma_2} >0,\kappa_{\sigma_3}>0.
\end{align*}\par
Spatial anisotropy in the extremal dependence structure of the standardised processes is incorporated using the standard transformation of coordinates
\begin{equation}
\label{anisoeq}
s^*=\begin{pmatrix}
1 & 0\\ 
0 & 1/L \end{pmatrix}\begin{pmatrix}
\cos\theta & -\sin\theta \\
\sin\theta & \cos\theta \end{pmatrix}s,
\end{equation}
where $\theta \in [-\pi/2, 0]$ and $L > 0$ control the rotation and coordinate stretching effect, respectively; with $L=1$ recovering the isotropic model. We define our distance metric $\|s_A-s_B\| = \|s^*_A-s^*_B\|_*$, where $\|\cdot\|_*$ denotes great-circle, or spherical, distance.
\section{Inference details}
\label{append-infer}
\subsection{Overview}
Here we provide details for fitting the extremal dependence model defined in Section~\ref{dep_sec} of the main text. Details are provided for a generic standardised process $\{X(s):s\in\mathbf{s}\}$, where $\mathbf{s}$ denotes the sampling locations, but can be applied, separately, to each of $X_\mathcal{C}$, $X_\mathcal{N}$ and $X_\mathcal{E}$; note that there are different parametric forms for $\beta$ and $\sigma$ for $X_\mathcal{N}$, see Section~\ref{model_spec_sup} above and Section~\ref{sec_differ} of the main text, which must considered when fitting this model.
\subsection{Full likelihood}
We employ a censored triplewise composite likelihood approach for model fitting, which pools information across different conditioning sites. Use of censoring is required to handle point masses in the marginals of $X(s)$ caused by zeroes in the marginals of $Y(s)$. The censoring threshold for $X(s)$, denoted $c(s)$, is found by transforming $p(s)$ in eq. \eqref{chap6-MargTransform} of the main text to the Laplace scale, i.e., $c(s)=F^{-1}_L\{p(s)\}$ where $F_L$ denotes the standard Laplace distribution function. \par
We define the likelihood contribution from a single conditioning site $s_i \in \mathbf{s}=\{s_1,\dots,s_d\}$, at time $t$, by defining the set of exceedance times ${\mathbf{T}^{(s_i)}~=\{t\in \{1,\dots, n\}: x_t(s_i)\geq u\}}$. Then for each site $s_j\in\mathbf{s}\setminus s_i$, the values of the residual process conditional on $s_i$ at time $t \in \mathbf{T}^{(s_i)}$ are
\[
z_t^{(s_i)}(s_j)=\begin{cases}
[x_t(s_j)-\alpha\{h(s_j,s_i)\}x_t(s_i)]\{x_t(s_i)\}^{-\beta\{h(s_j,s_i)\}}, \;\;&\text{if}\;\;x_t
(s_j) > c(s_j),\\
c^{(s_i)}_{t}(s_j), \;\;&\text{otherwise,}
\end{cases}
\]
where $
c^{(s_i)}_{t}(s_j)=[c(s_j)-\alpha\{h(s_j,s_i)\}x_t(s_i)]\{x_t(s_i)\}^{-\beta\{h(s_j,s_i)\}}$. Define $\boldsymbol{\psi}$ as the vector containing all parameters that control dependence functions \eqref{depend_funcs} and anisotropy transformation \eqref{anisoeq}; then the censored triplewise likelihood contribution for conditioning site $s_i$ at time $t \in \mathbf{T}^{(s_i)}$ is
$ \prod_{\forall j<k ; j \& k \neq i}L^{(s_j,s_k)|s_i}_{CL}\{\boldsymbol{\psi};{z}^{(s_i)}_t(s_j),z^{(s_i)}_t(s_k)\}$, where 
 \begin{equation}
 \label{CLsi}
 L^{(s_j,s_k)|s_i}_{CL}\{\boldsymbol{\psi};{z}^{(s_i)}_t(s_j),z^{(s_i)}_t(s_k)\}=\frac{g_{(s_j,s_k)|s_i}(z^{(s_i)}_t(s_j),z^{(s_i)}_t(s_k),c^{(s_i)}_{t}(s_j),c^{(s_i)}_{t}(s_k))}{\prod_{l\in\{j,k\}}\{x_t(s_i)\}^{\beta\{h(s_l,s_i)\}\mathrm{1}\{z_t(s_l)>c^{(s_i)}_{t}(s_l)\}}},
 \end{equation}
 where $\mathrm{1}\{\cdot\}$ is the indicator function and $g_{(s_j,s_k)|s_i}$ is a bivariate density to be defined in Section~\ref{bivdens}; eq. \eqref{CLsi} corresponds to the likelihood contribution from a triple of sites $(s_i,s_j,s_k)$, where $s_i$ is the conditioning site and $s_j\neq s_k\neq s_i$. The full composite likelihood is 
 \begin{equation}
 \label{full_lik}
 L_{CL}(\boldsymbol{\psi})=\prod^d_{i=1}\prod_{\forall j<k ; j \& k \neq i}\prod_{t \in \mathcal{T}^{(s_i)}}L^{(s_j,s_k)|s_i}_{CL}\{\boldsymbol{\psi};{z}^{(s_i)}_t(s_j),z^{(s_i)}_t(s_k)\}.
 \end{equation}
 \subsection{Stratified sampling regime}
\label{sec-strat}
We approximate the full triple-wise likelihood in eq. \eqref{full_lik} by \[
\prod_{(s_i,s_j,s_k)\in\mathbf{D}}\prod_{t \in \mathcal{T}^{(s_i)}}L^{(s_j,s_k)|s_i}_{CL}\{\boldsymbol{\psi};{z}^{(s_i)}_t(s_j),z^{(s_i)}_t(s_k)\},
\] 
where $\mathbf{D}$ is a subset of all possible triples of sites $(s_i,s_j,s_k)$. To create $\mathbf{D}$, we first uniformly sample a conditioning site $s_i \in \mathbf{s}$ and then draw $(s_j,s_k)$ uniformly at random from the set ${\{s_j,s_k \in \mathbf{s}\setminus s_i, j < k : \max\{h(s_j,s_i),h(s_k,s_i)\} < h_{max}\}}$ without replacement; this process is repeated until $d_s=|\mathbf{D}|$ triples are drawn. We choose the number of triples $d_s$ to be as large as is computationally feasible.
 \subsection{Bivariate residual density}
 \label{bivdens}
 To define the bivariate density $g_{(s_j,s_k)|s_i}$ in eq. \eqref{CLsi}, 
we first describe the dependence structure of the residual process $Z(s|s_O)$; consider the process \[\{W^*(s):s \in \mathcal{S}\}=\{W(s):s \in \mathcal{S}\;|\;(W(s_O)=0)\},\] where $\{W(s)\}$ is a standard Gaussian process with stationary Mat\'ern correlation function
\[
\rho(h)=2^{1-\kappa_{\rho_2}}\{\Gamma(\kappa_{\rho_2})\}^{-1}\left(2h\sqrt{\kappa_{\rho_2}}/\kappa_{\rho_1}\right)^{\kappa_{\rho_2}}K_{\kappa_{\rho_2}}\left(2h\sqrt{\kappa_{\rho_2}}/\kappa_{\rho_1}\right),\qquad\kappa_{\rho_1} > 0, \kappa_{\rho_2} >0,
\]
for $h\geq 0$ and where $K_{\kappa_{\rho_2}}(\cdot)$ is the modified Bessel function of the second kind of order $\kappa_{\rho_2}$. Recall that the marginal distribution of $Z(s|s_O)$ is delta-Laplace (see eq. \eqref{DL_dens} of the main text); we denote here its marginal density by $f_{1,s|s_O}$ and $F_{1,s|s_O}$ and $F^{-1}_{1,s|s_O}$ as the corresponding marginal distribution function and its inverse, respectively.
We build the residual process by setting ${Z(s|s_O)=F^{-1}_{1,s|s_O}[\Phi\{W^*(s)\}]}$ for all $s \in \mathcal{S}$, where $\Phi(\cdot)$ is the standard Gaussian distribution  function. The bivariate joint distribution of $Z(s|s_O)$ observed at a pair of sites $(s_j,s_k)$ with $s_j\neq s_k\neq s_O$ is defined through the copula model 
\[
F_{(s_j,s_k)|s_O}(z_j,z_k)=\Phi_{2}\left[\Phi^{-1}\{F_{1,s_j|s_O}(z_j)\},\Phi^{-1}\{F_{1,s_k|s_O}(z_k)\};\Sigma^{(s_j,s_k|s_O)}\right],
\]
where $\Phi_{2}(\cdot;\Sigma)$ is the distribution function of a standard bivariate Gaussian random variable with zero mean vector, unit variance and correlation matrix $\Sigma$. The correlation matrix $\Sigma^{(s_j,s_k|s_O)}$ has off-diagonal element
\[
\frac{\rho\{h(s_j,s_k)\}-\rho\{h(s_j,s_O)\}\rho\{h(s_k,s_O)\}}{(1-[\rho\{h(s_j,s_O)\}]^2)^{1/2}(1-[\rho\{h(s_k,s_O)\}]^2)^{1/2}},
\]
which accounts for the conditioning event $W(s_O)=0$.
\par
We then define the density $g_{(s_j,s_k)|s_i}$ in eq. \eqref{CLsi} by
\[
g_{(s_j,s_k)|s_i}(z_j,z_k,c_j,c_k)=\begin{cases}
f_{(s_j,s_k)|s_i}(z_j,z_k) \;\;&\text{if}\;\;z_j > c_j, z_k > c_k,\\
f_{1,s_j|s_i}(z_j)F_{(s_j|s_k)|s_i}(c_j,z_k)\;\;&\text{if}\;\;z_j > c_j, z_k \leq c_k,\\
f_{1,s_k|s_i}(z_k)F_{(s_k|s_j)|s_i}(c_j,z_j)\;\;&\text{if}\;\;z_j \leq c_j, z_k > c_k,\\
F_{(s_j,s_k)|s_i}(c_j,c_k)\;\;&\text{if}\;\;z_j \leq c_j, z_k \leq c_k,
\end{cases}
\]
where $f_{(s_j,s_k)|s_i}$ is the corresponding density of $F_{(s_j,s_k)|s_i}$ and the conditional distribution is
\[
F_{(s_j|s_k)|s_i}(c_j,z_k)=\Phi^*\left[\Phi^{-1}\{F_{1,s_j|s_i}(c_j)\};\mu_{j|k|i},\Sigma_{j|k|i}\right],
\]
where $\Phi^*(\cdot; \mu,c)$ is the Gaussian distribution function with mean $\mu$ and variance $c>0$; for $\Sigma^*:= \Sigma^{(s_j,s_k|s_O)}$, we have $\mu_{j|k|i}=\Sigma^*_{12}(\Sigma^*)^{-1}_{11}\Phi^{-1}\{F_{1,s_{k}|s_i}(z_k)\}$ and $\Sigma_{j|k|i}=\Sigma^*_{12}-(\Sigma^*_{12})^2(\Sigma^*)^{-1}_{11}$.
\pagebreak
\section{Model parameter estimates}
\label{append-deptable}
Table~\ref{parEst1} gives the estimates and standard errors for the extremal dependence model parameters for each of the three processes. Parameters without standard errors were treated as fixed in model inference.
\begin{table}[H]
\caption{Parameter estimates (standard errors) to 2 decimal places for $X_\mathcal{N}, X_\mathcal{C}$ and $X_\mathcal{E}$.}
 \centering
 \begin{tabular}{ccccccc} 
\toprule
&\multicolumn{3}{c}{$\alpha(h)$}& \multicolumn{3}{c}{$\beta(h)$}\\
\cmidrule(lr){2-4}\cmidrule(lr){5-7}
&$\kappa_{\alpha_1}$ & $\kappa_{\alpha_2}$ & $\Delta$ & $\kappa_{\beta_1}$ & $\kappa_{\beta_2}$& $\kappa_{\beta_3}$   \\
$\mathcal{N}$&170.07 (14.48) & 0.87 (0.06) & 6.81 (3.71)& 0.12 (0.04)& 1.36 (0.37) & 256.00 (1.34) \\
$\mathcal{C}$&1.60 (1.35) & 0.64 (0.11) & 0.00& 32.49 (4.64)& 0.73 (0.09) & 1.00\\
$\mathcal{E}$&0.65 (0.77) & 0.50 (0.06) & 0.00& 30.28 (3.37) & 0.63 (0.06)  & 1.00\\
\hline
& \multicolumn{3}{c}{$\mu(h)$}& \multicolumn{3}{c}{$\sigma(h)$}\\
 \cmidrule(lr){2-4}\cmidrule(lr){5-7}
& $\kappa_{\mu_1}$ & $\kappa_{\mu_2}$ &$\kappa_{\mu_3}$ & $\kappa_{\sigma_1}$ & $\kappa_{\sigma_2}$ & $\kappa_{\sigma_3}$ \\
$\mathcal{N}$&$-0.06$ (0.02) &0.40 (0.54) &15.50 (3.26)& 16.79 (2.59) &0.73 (0.09) &2.11 (0.14) \\
$\mathcal{C}$& 0.70 (0.18) &0.28 (0.11) &47.56 (19.83)&20.87 (2.97) &0.76 (0.03)&$\sqrt{2}$\\
$\mathcal{E}$& 0.96 (0.15) &1.19 (0.05) &81.83 (10.2)&18.62 (2.24) &0.85 (0.02)&$\sqrt{2}$\\
\hline  
\end{tabular}

  \begin{tabular}{ccccc} 
& \multicolumn{4}{c}{$\delta(h)$}\\
\cmidrule(lr){2-5}
&$\kappa_{\delta_1}$ & $\kappa_{\delta_2}$ &$\kappa_{\delta_3}$&$\kappa_{\delta_4}$\\
 \hline
 $\mathcal{N}$&1.84 $\times 10^{-3}$ ($1.93 \times 10^{-3})$ &1.65  (0.14) &90.57 (7.84) & 0.28 (0.21) \\
 $\mathcal{C}$&0.77 (0.06) &0.32  (0.05) &57.90 (8.94) &1.00\\
 $\mathcal{E}$&0.88 (0.09) &0.29 (0.04) &136.26 (0.49) &1.00\\
\hline
&\multicolumn{2}{c}{$\rho(h)$}&\multicolumn{2}{c}{$h(s,s_O)$}\\
\cmidrule(lr){2-3}\cmidrule(lr){4-5}

&$\kappa_{\rho_1}$ & $\kappa_{\rho_2}$&$\theta$&$L$\\
 \hline
$\mathcal{N}$&1496.36 (352.88)&0.36 (0.02) &$-0.32$ (0.08)&1.00 (0.03)\\
$\mathcal{C}$&104.81 (17.54)&0.41 (0.01) &$-0.15$ (0.02)&0.97 (0.01)\\
$\mathcal{E}$&121.52 (12.40)&0.40 (0.01) &$-0.14$ (0.02)&0.98 (0.01)\\
 \hline
 \end{tabular}
 \label{parEst1}
 \end{table}

\section{Supplementary figures}
 \label{append_supfigs}
 \begin{figure}[H]
\centering

\includegraphics[width=0.5\linewidth]{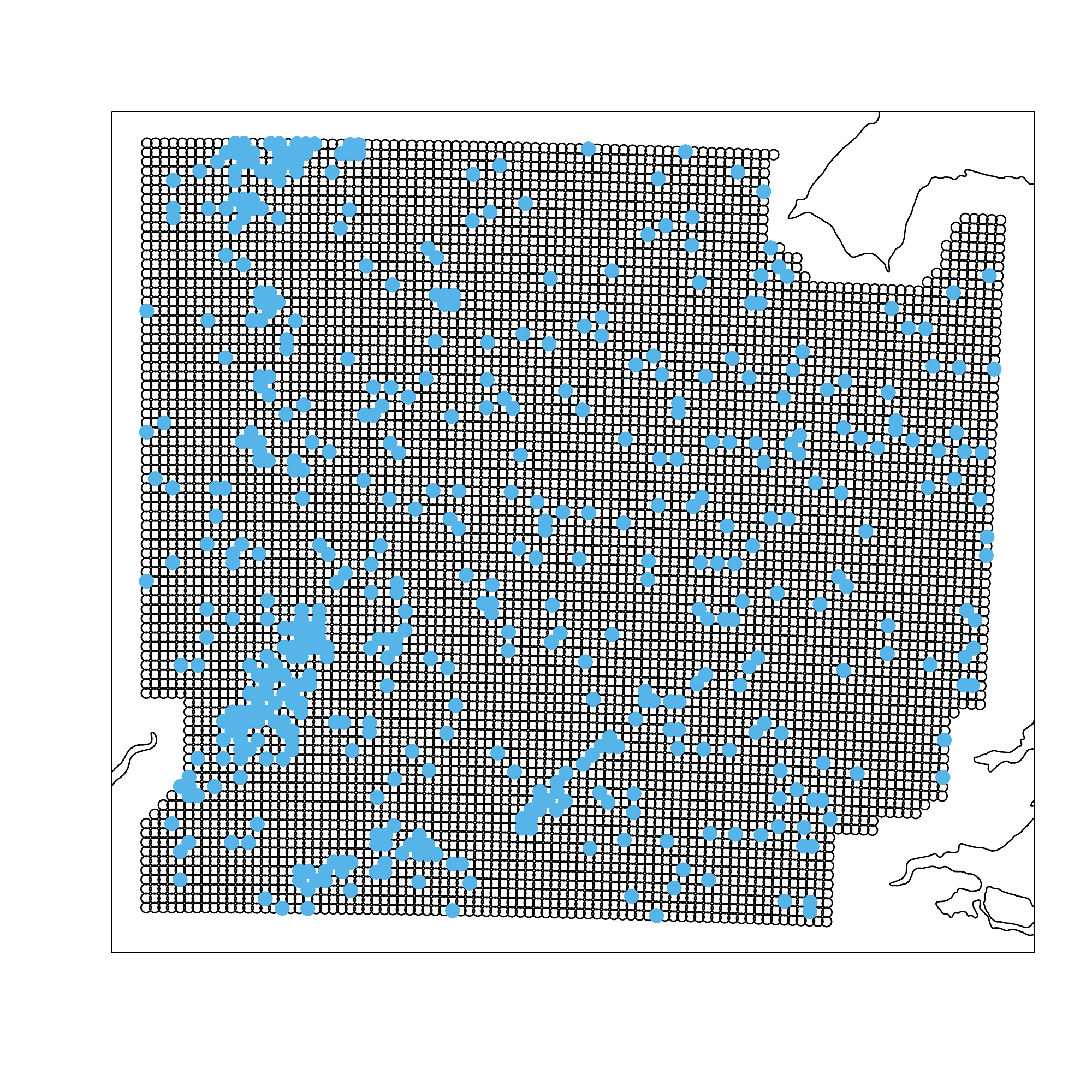} 
\caption{The subset of sites chosen to estimate the exceedance threshold $q_\lambda(s)$, given in blue.}
\label{GAM_locs}
\end{figure}
 \begin{figure}[H]
\centering
\begin{minipage}{0.49\linewidth}
\centering
\includegraphics[width=\linewidth]{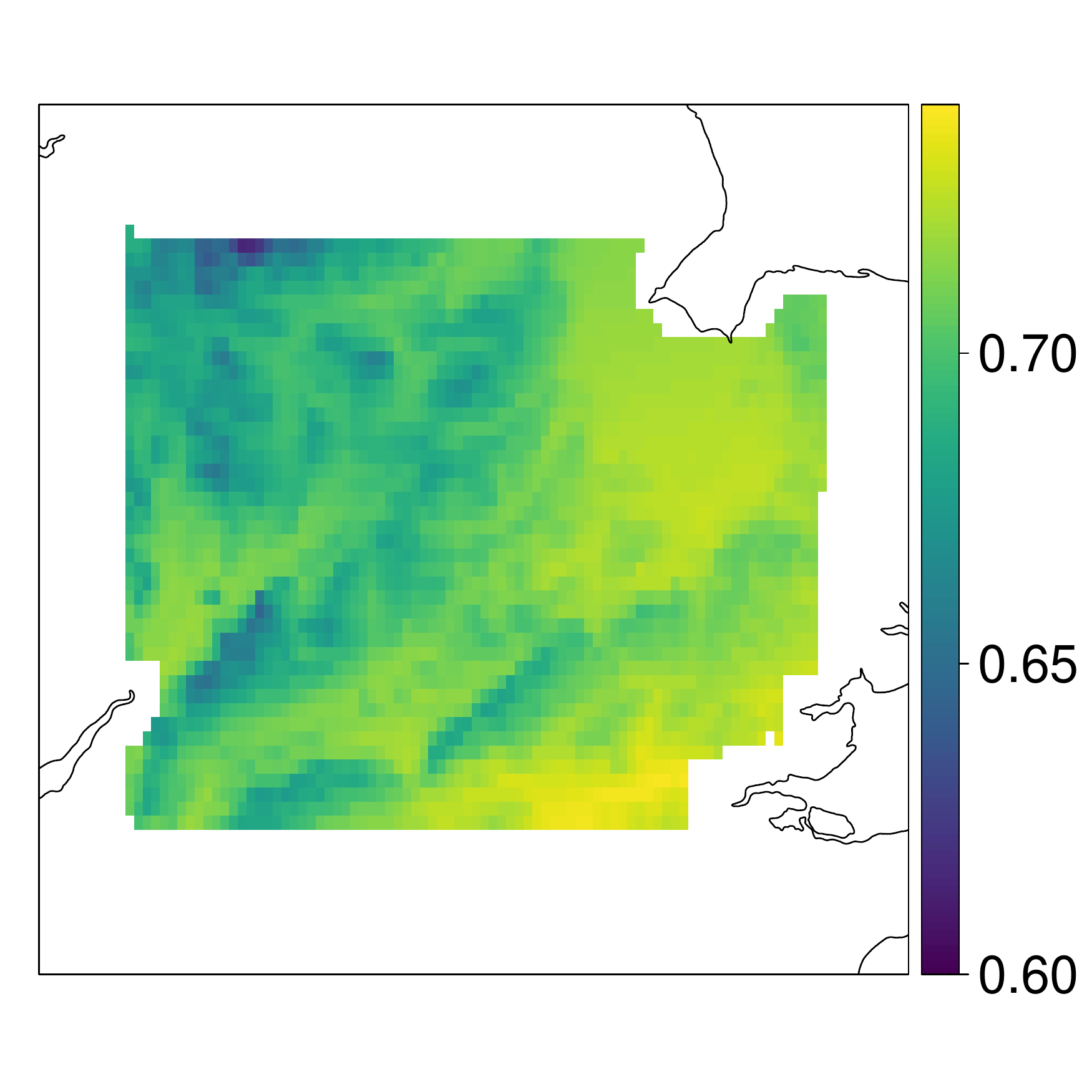} 
\end{minipage}
\begin{minipage}{0.49\linewidth}
\centering
\includegraphics[width=\linewidth]{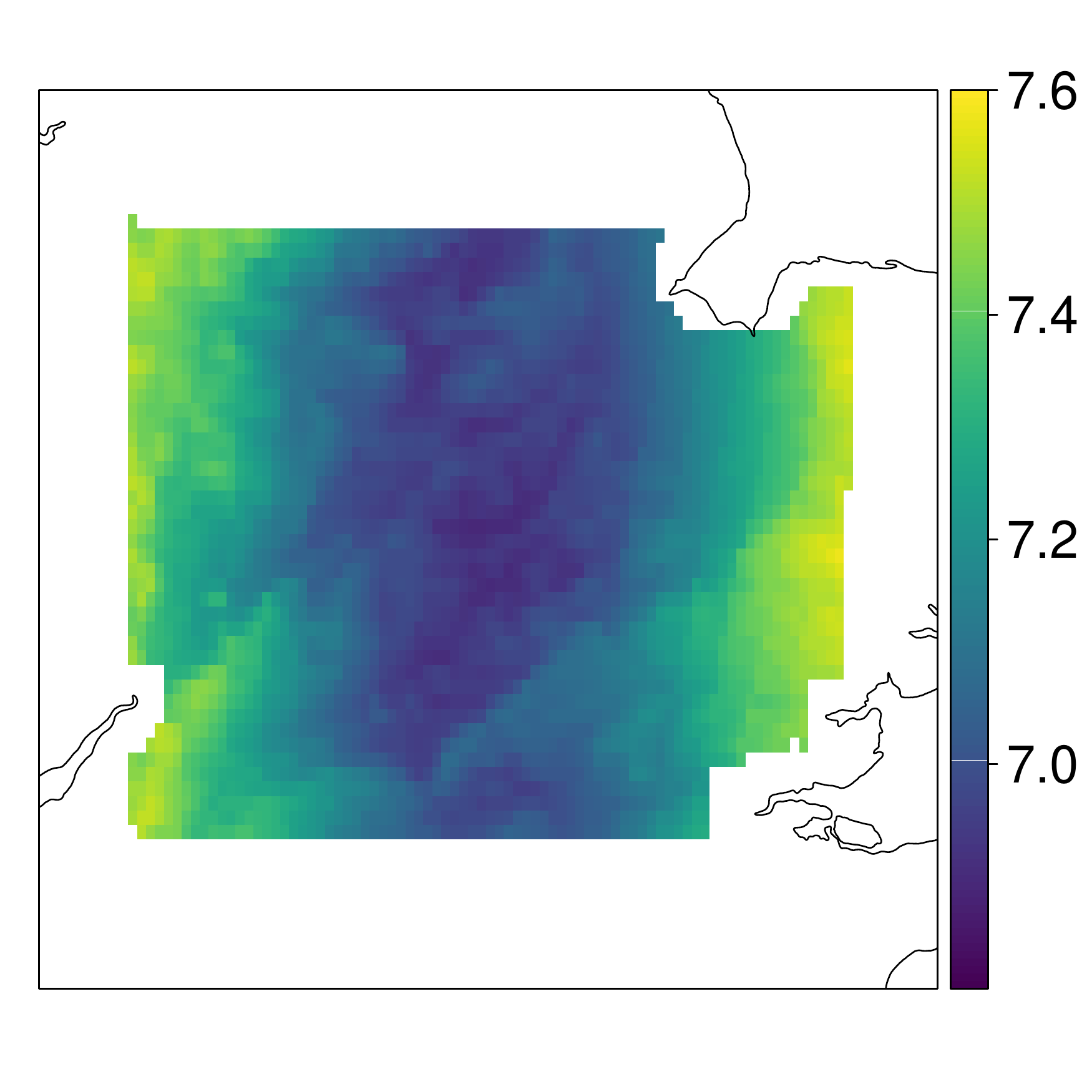} 
\end{minipage}
\begin{minipage}{0.49\linewidth}
\centering
\includegraphics[width=\linewidth]{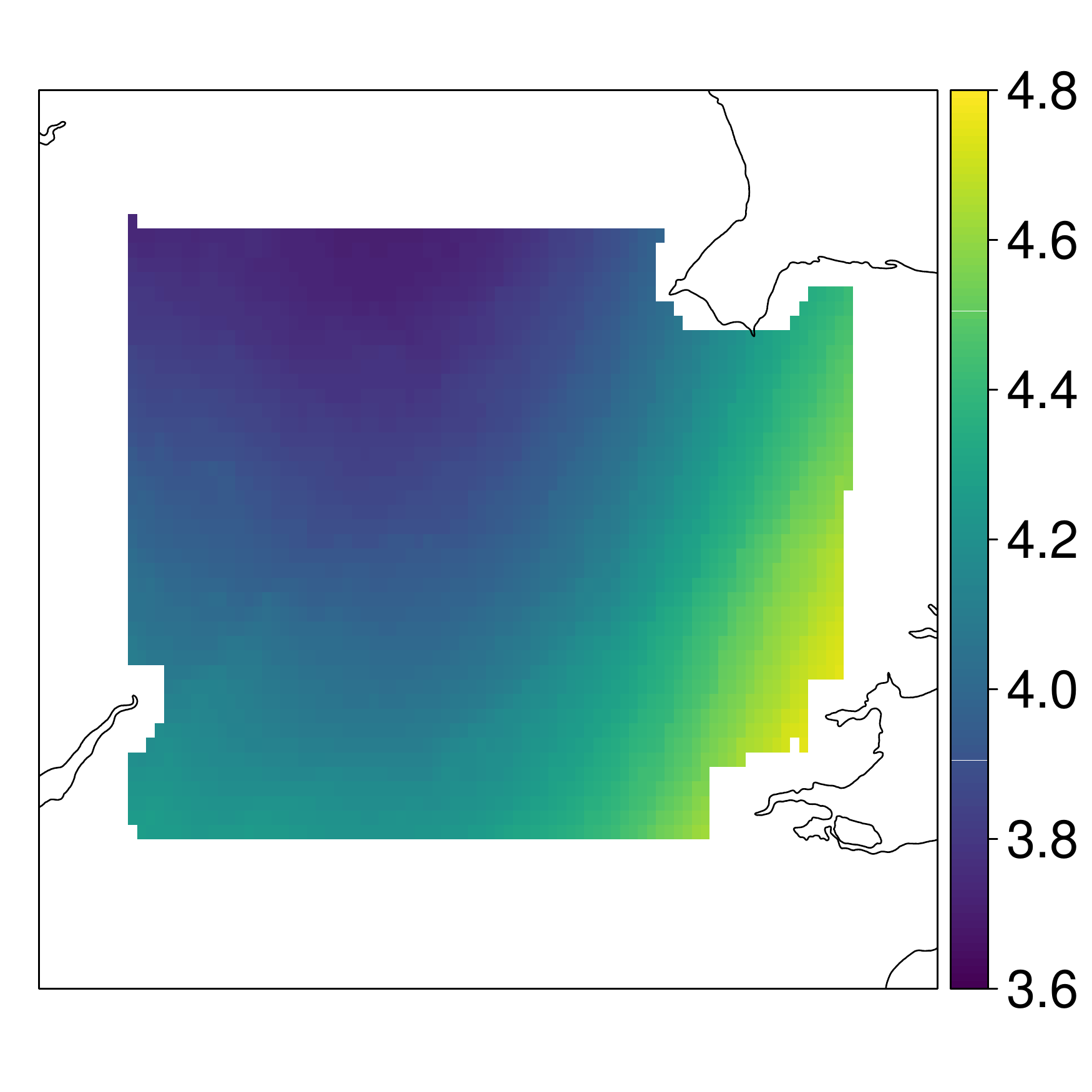} 
\end{minipage}
\begin{minipage}{0.49\linewidth}
\centering
\includegraphics[width=\linewidth]{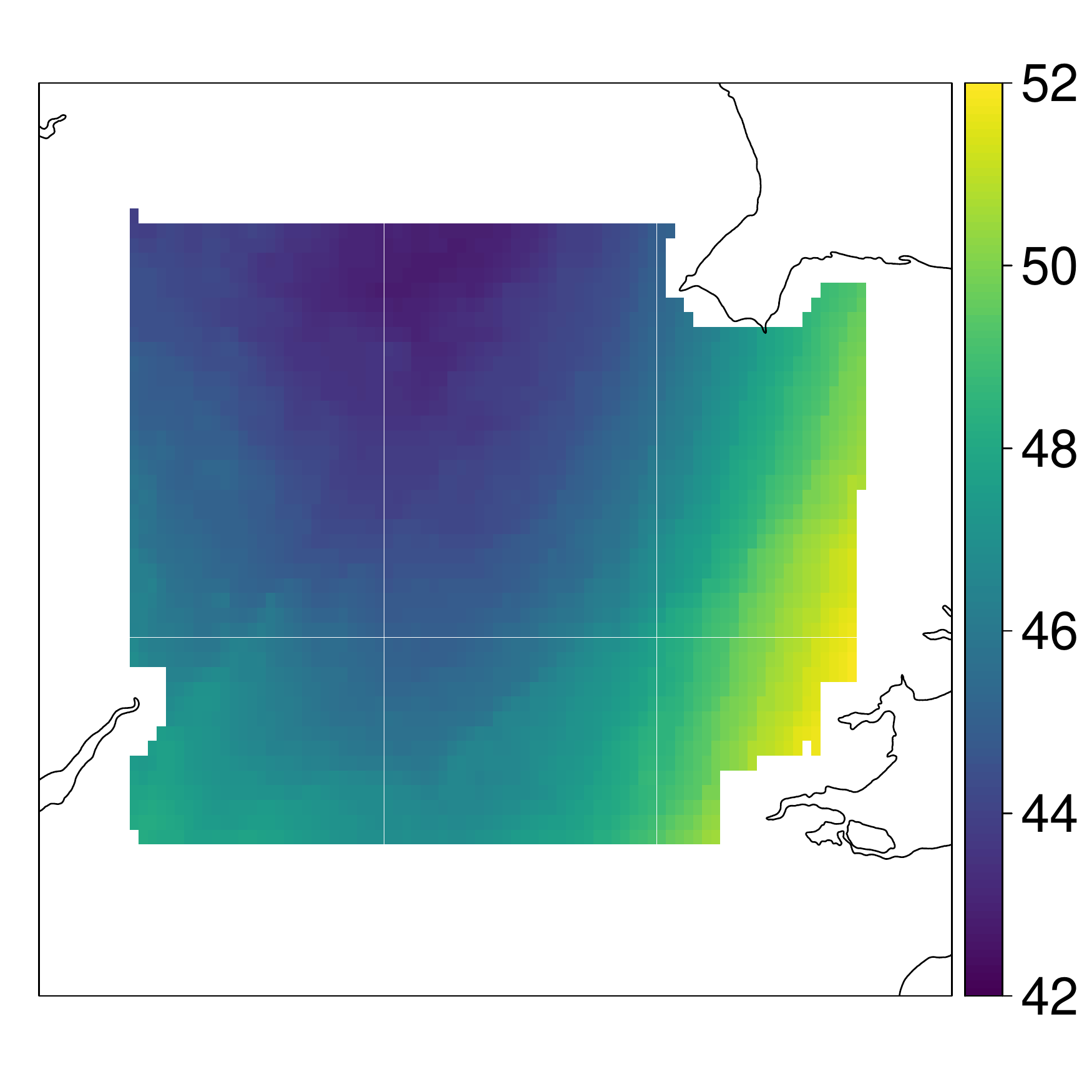} 
\end{minipage}
\caption{Spatially smoothed marginal distribution parameter estimates for convective rainfall, i.e., convective rainfall. Top left: dry probability $\hat{p}(s)$, top right: $0.995$-quantile $\hat{q_\lambda}(s)$, bottom left: generalised Pareto scale parameter $\hat{\upsilon}(s)$, bottom right: $20$-year return level estimate. Different image colour scales are used across each panel.}
\label{convGAMfig}
\end{figure}

\begin{figure}[H]
\centering
\begin{minipage}{0.49\linewidth}
\centering
\includegraphics[width=\linewidth]{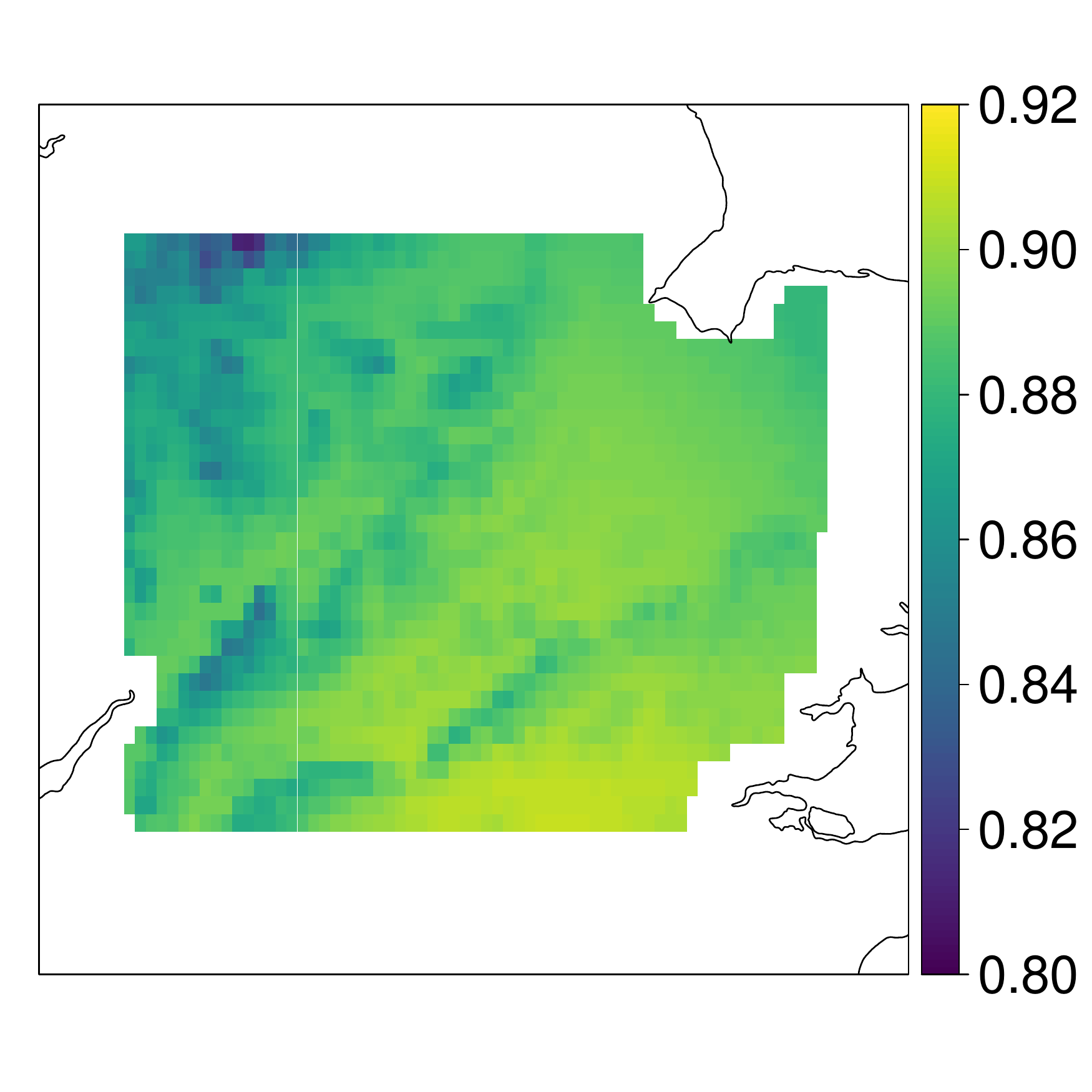} 
\end{minipage}
\begin{minipage}{0.49\linewidth}
\centering
\includegraphics[width=\linewidth]{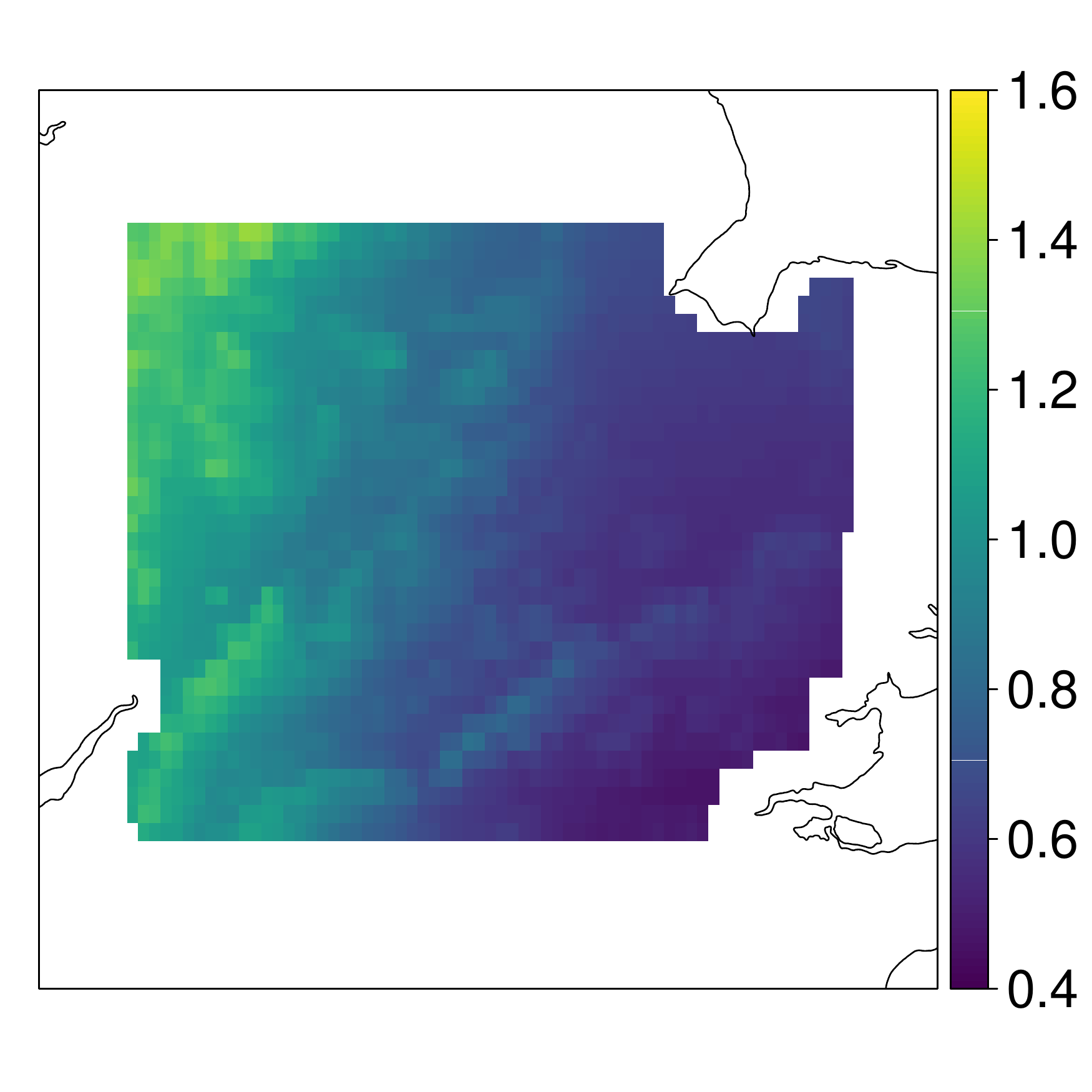} 
\end{minipage}
\begin{minipage}{0.49\linewidth}
\centering
\includegraphics[width=\linewidth]{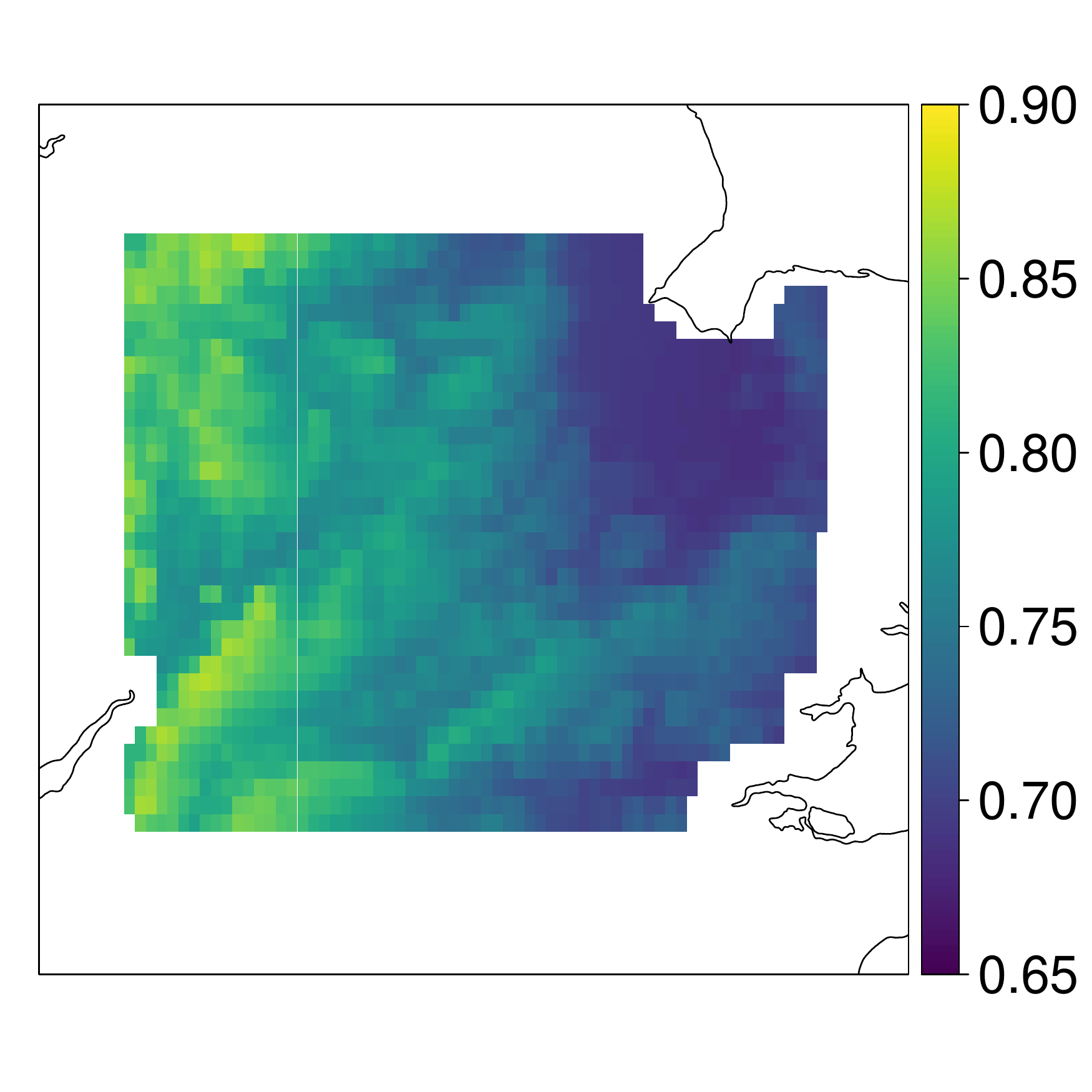} 
\end{minipage}
\begin{minipage}{0.49\linewidth}
\centering
\includegraphics[width=\linewidth]{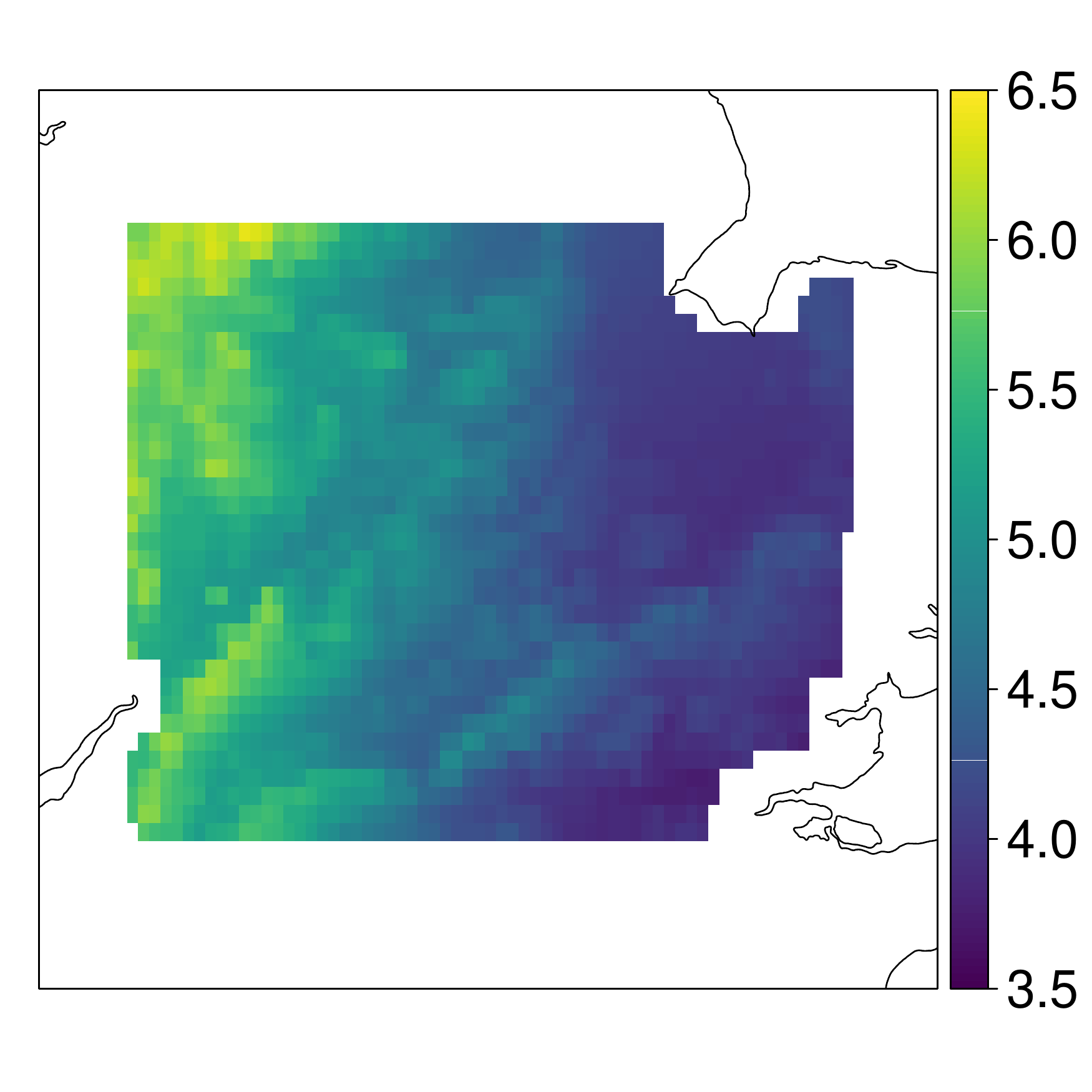} 
\end{minipage}
\caption{Spatially smoothed marginal distribution parameter estimates for non-convective rainfall, i.e., non-convective rainfall. Top left: dry probability $\hat{p}(s)$, top right: $0.995$-quantile $\hat{q_\lambda}(s)$, bottom left: generalised Pareto scale parameter $\hat{\upsilon}(s)$, bottom right: $20$-year return level estimate. Different image colour scales are used across each panel.}
\label{frontGAMfig}
\end{figure}
\begin{figure}[H]
\centering
\begin{minipage}{0.49\linewidth}
\centering
\includegraphics[width=\linewidth]{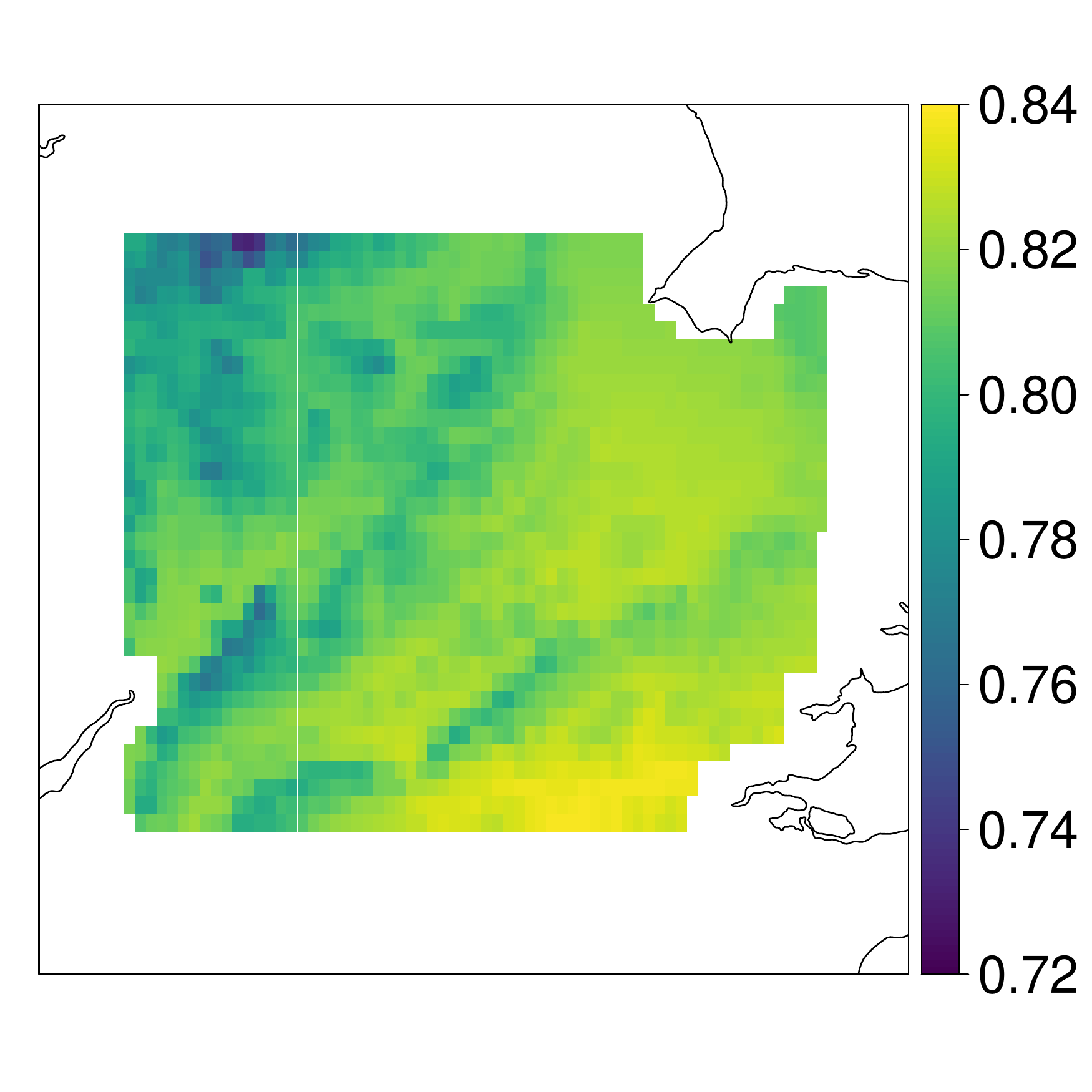} 
\end{minipage}
\begin{minipage}{0.49\linewidth}
\centering
\includegraphics[width=\linewidth]{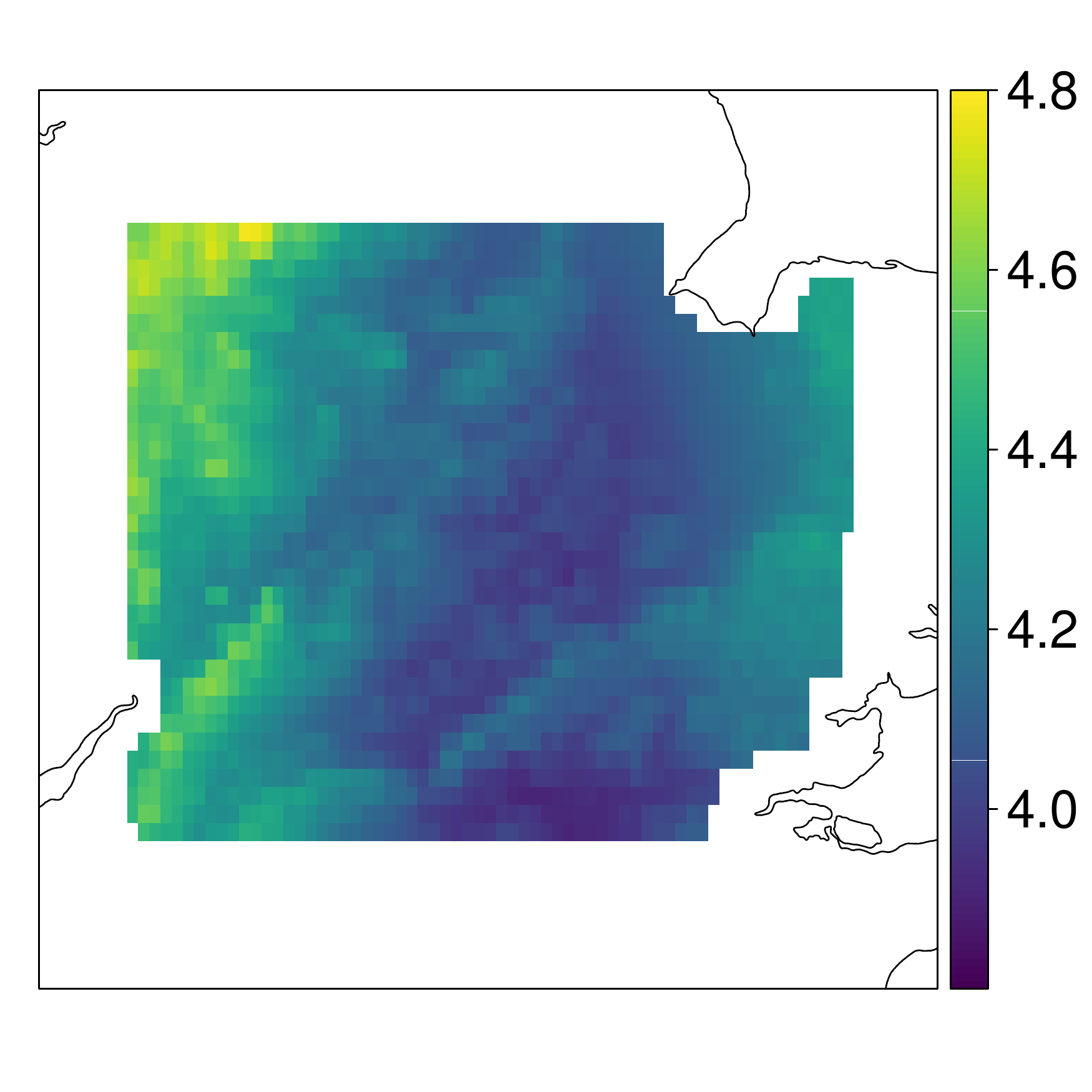} 
\end{minipage}
\begin{minipage}{0.49\linewidth}
\centering
\includegraphics[width=\linewidth]{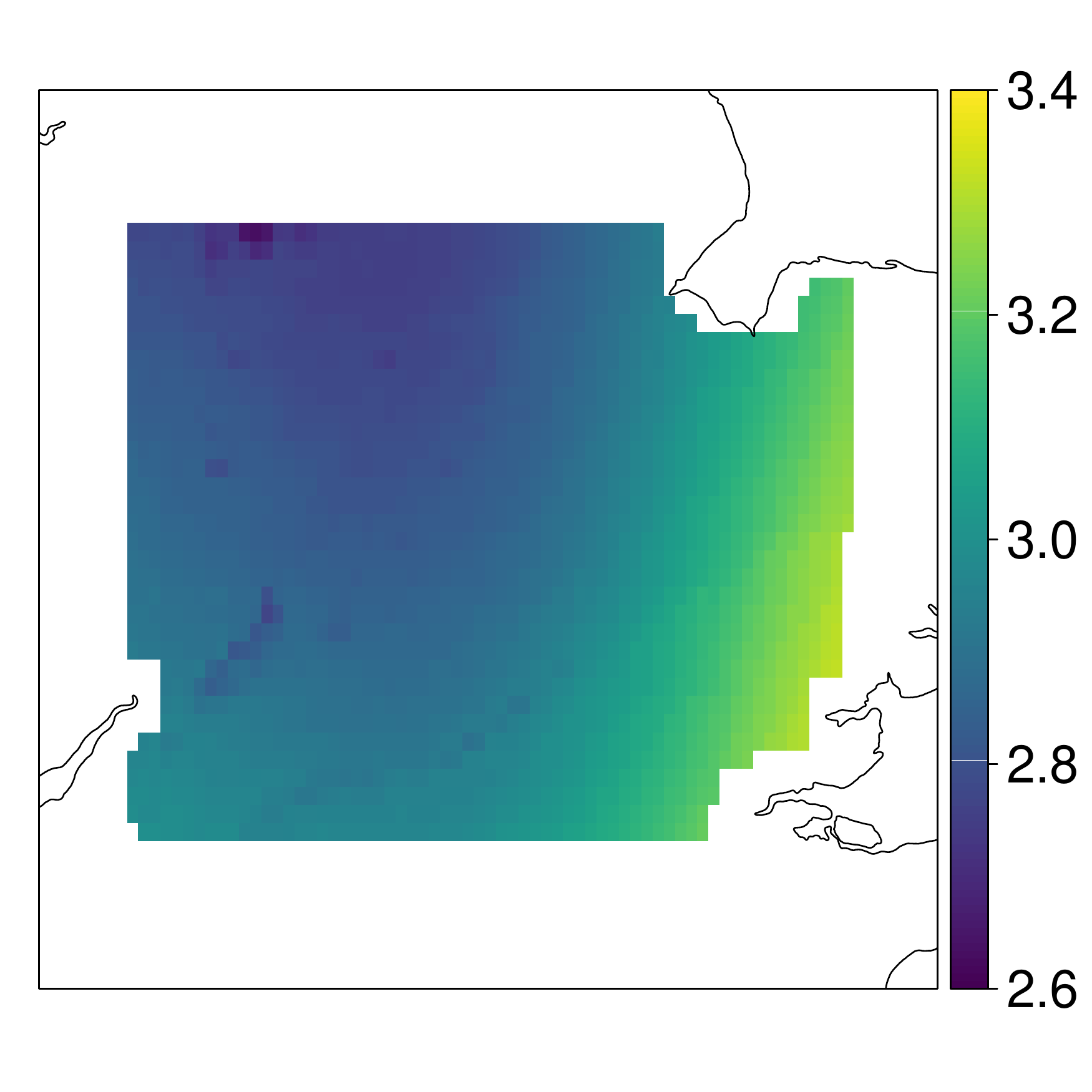} 
\end{minipage}
\begin{minipage}{0.49\linewidth}
\centering
\includegraphics[width=\linewidth]{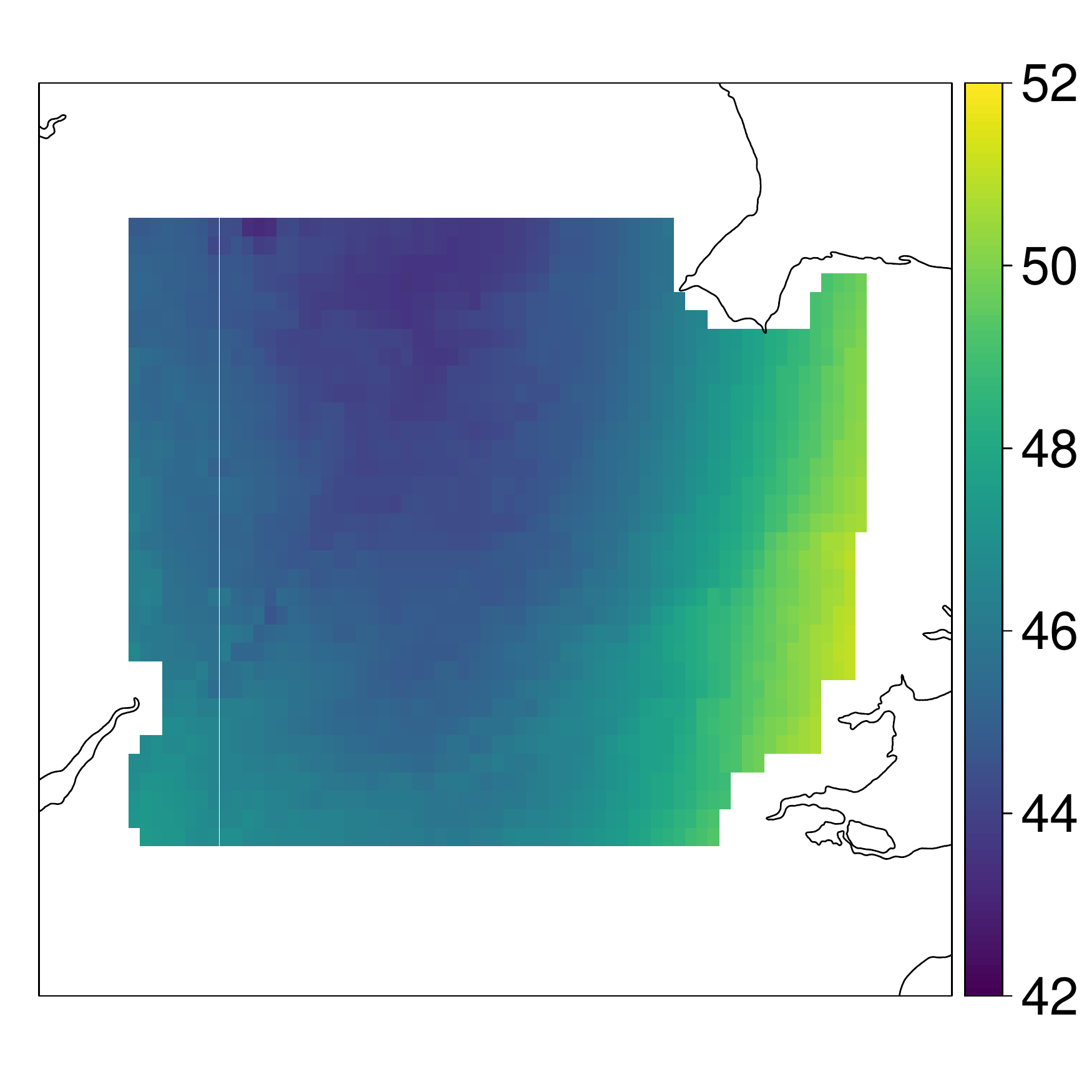} 
\end{minipage}
\caption{Spatially smoothed marginal distribution parameter estimates for all rainfall, i.e., $\{Y_\mathcal{E}(s)\}$. Top left: dry probability $\hat{p}(s)$, top right: $0.995$-quantile $\hat{q_\lambda}(s)$, bottom left: generalised Pareto scale parameter $\hat{\upsilon}(s)$, bottom right: $20$-year return level estimate. Different image colour scales are used across each panel.}
\label{BothGAMfig}
\end{figure}
 \begin{figure}[h]
\centering
\begin{minipage}{0.3\linewidth}
\centering
\includegraphics[width=\linewidth]{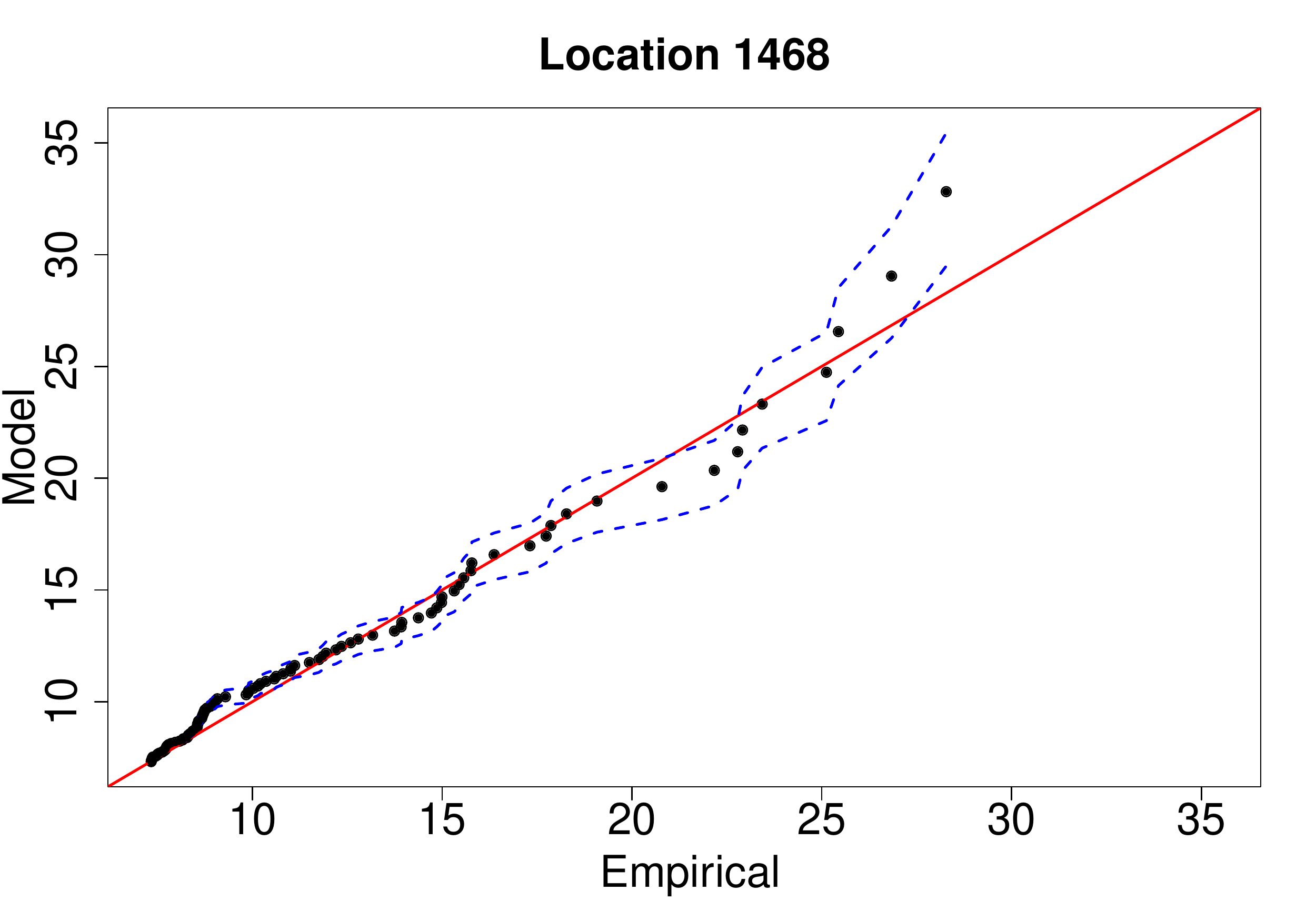} 
\end{minipage}
\begin{minipage}{0.3\linewidth}
\centering
\includegraphics[width=\linewidth]{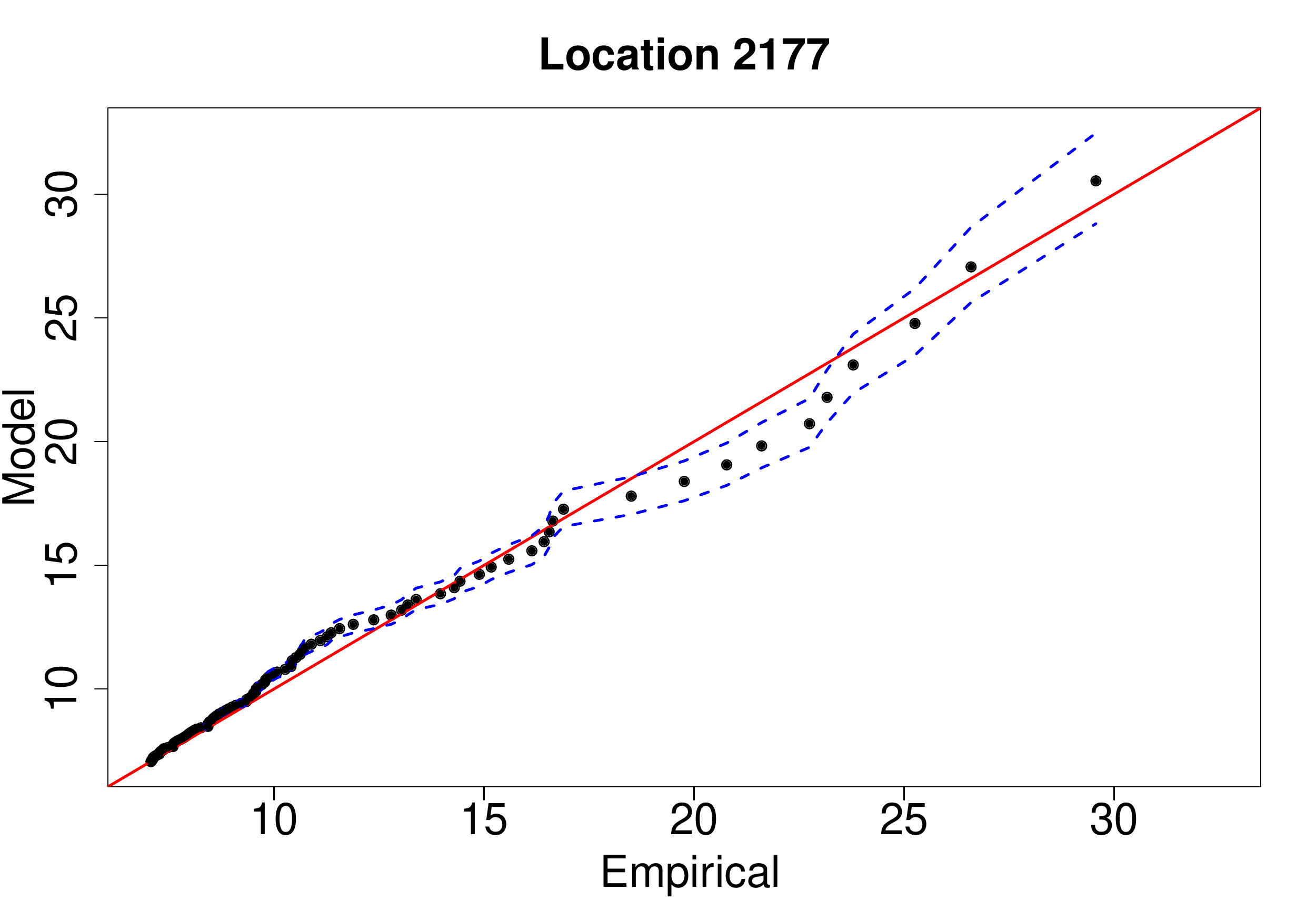} 
\end{minipage}
\begin{minipage}{0.3\linewidth}
\centering
\includegraphics[width=\linewidth]{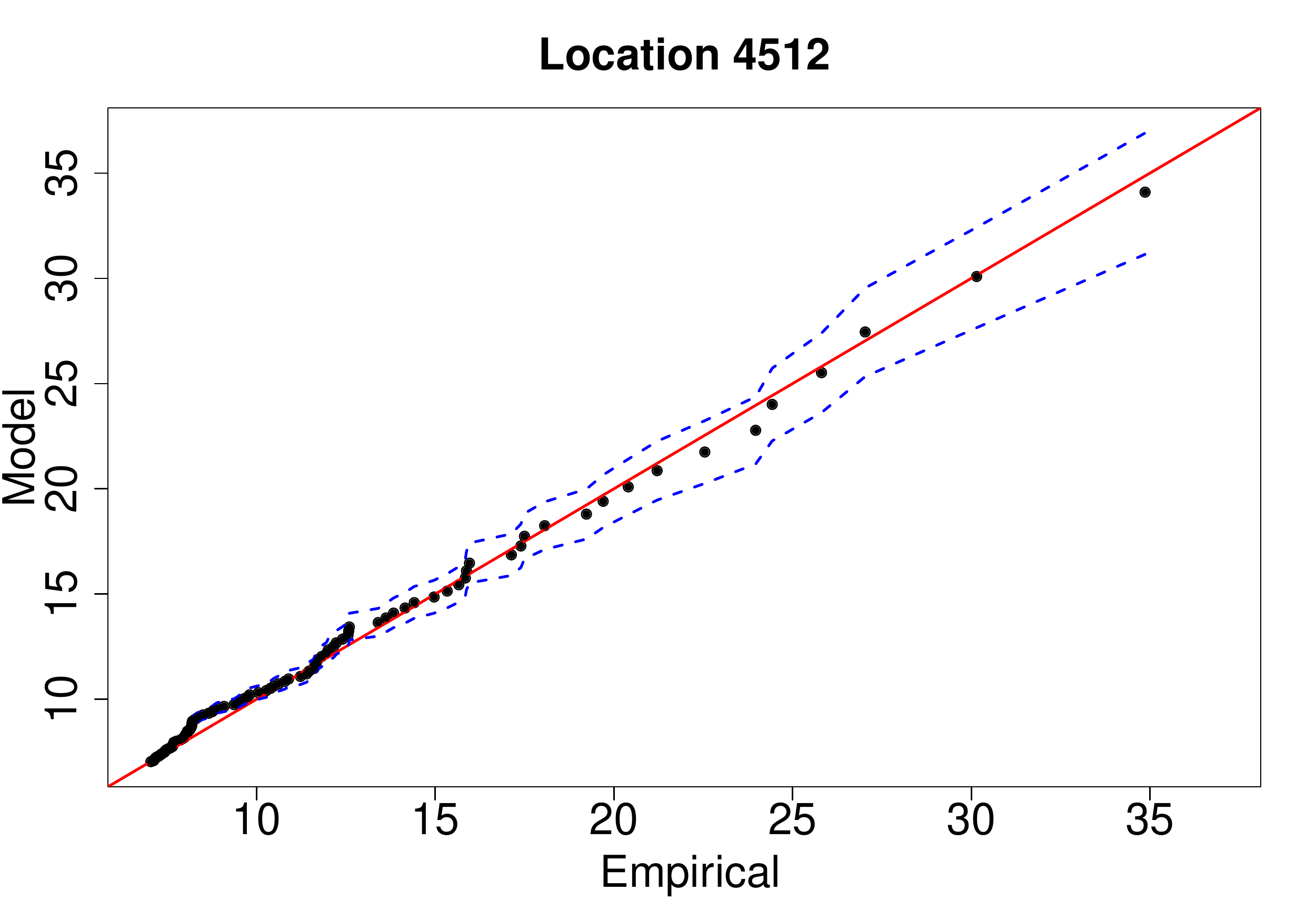} 
\end{minipage}
\begin{minipage}{0.3\linewidth}
\centering
\includegraphics[width=\linewidth]{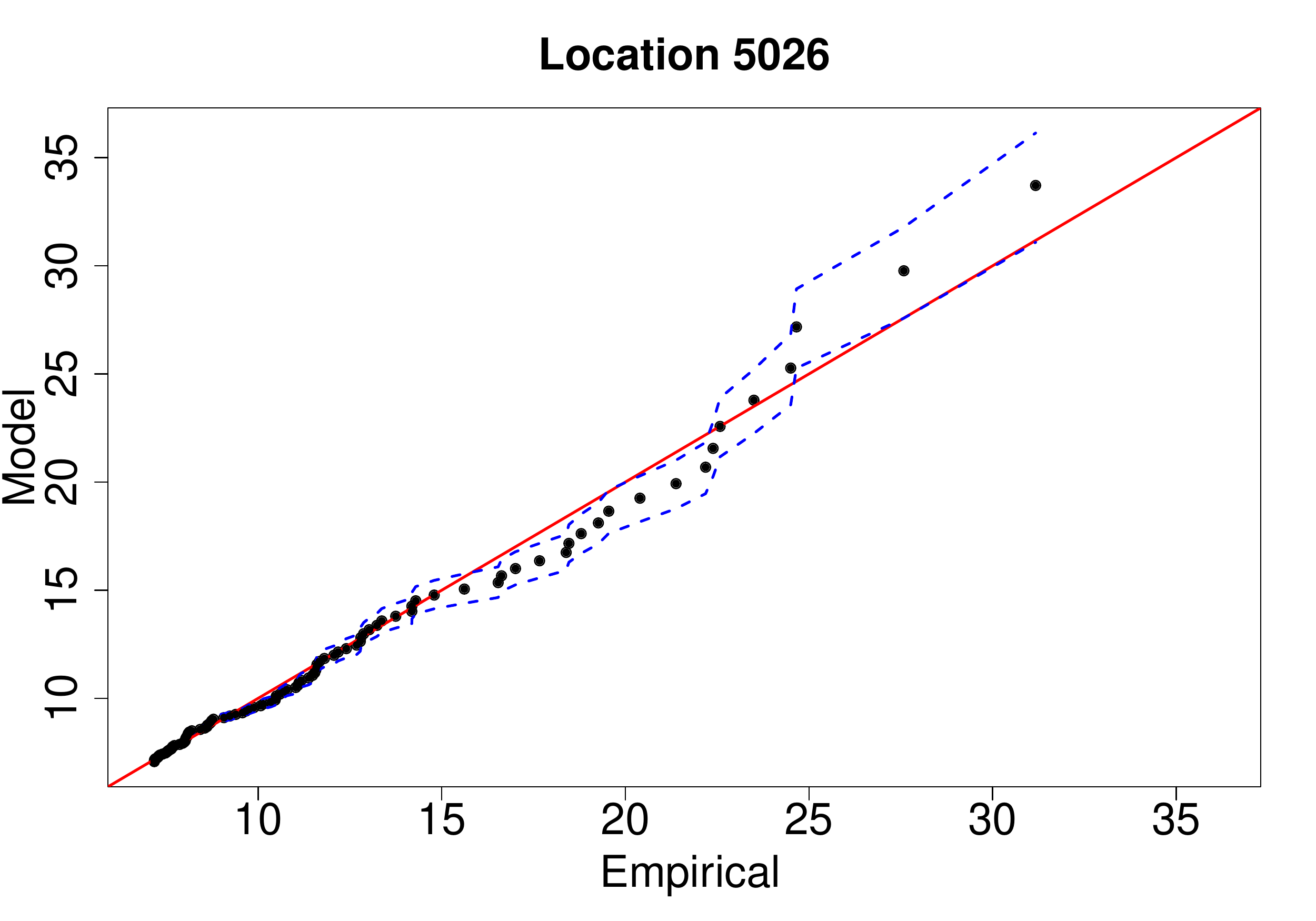} 
\end{minipage}
\begin{minipage}{0.3\linewidth}
\centering
\includegraphics[width=\linewidth]{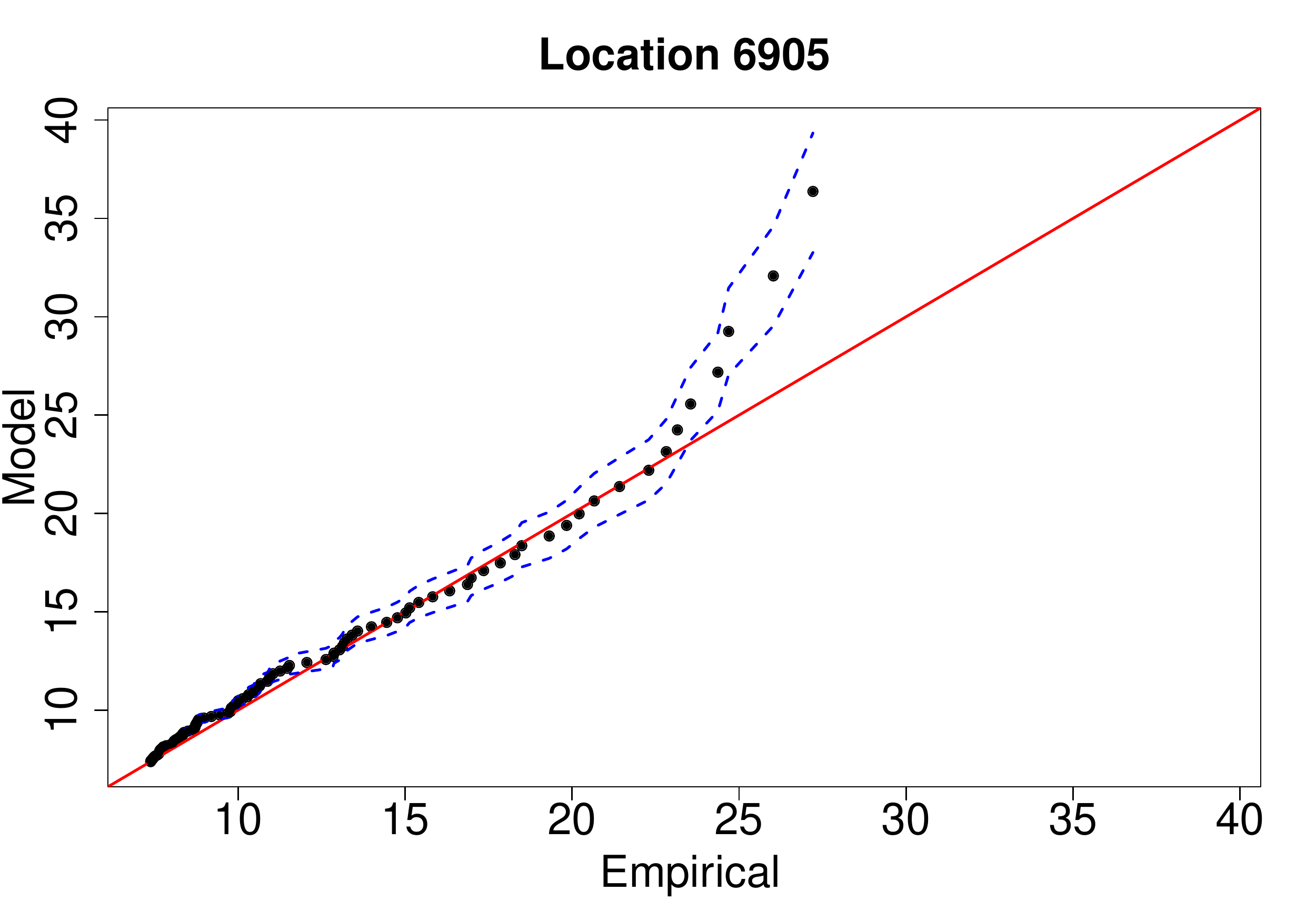} 
\end{minipage}
\begin{minipage}{0.3\linewidth}
\centering
\includegraphics[width=\linewidth]{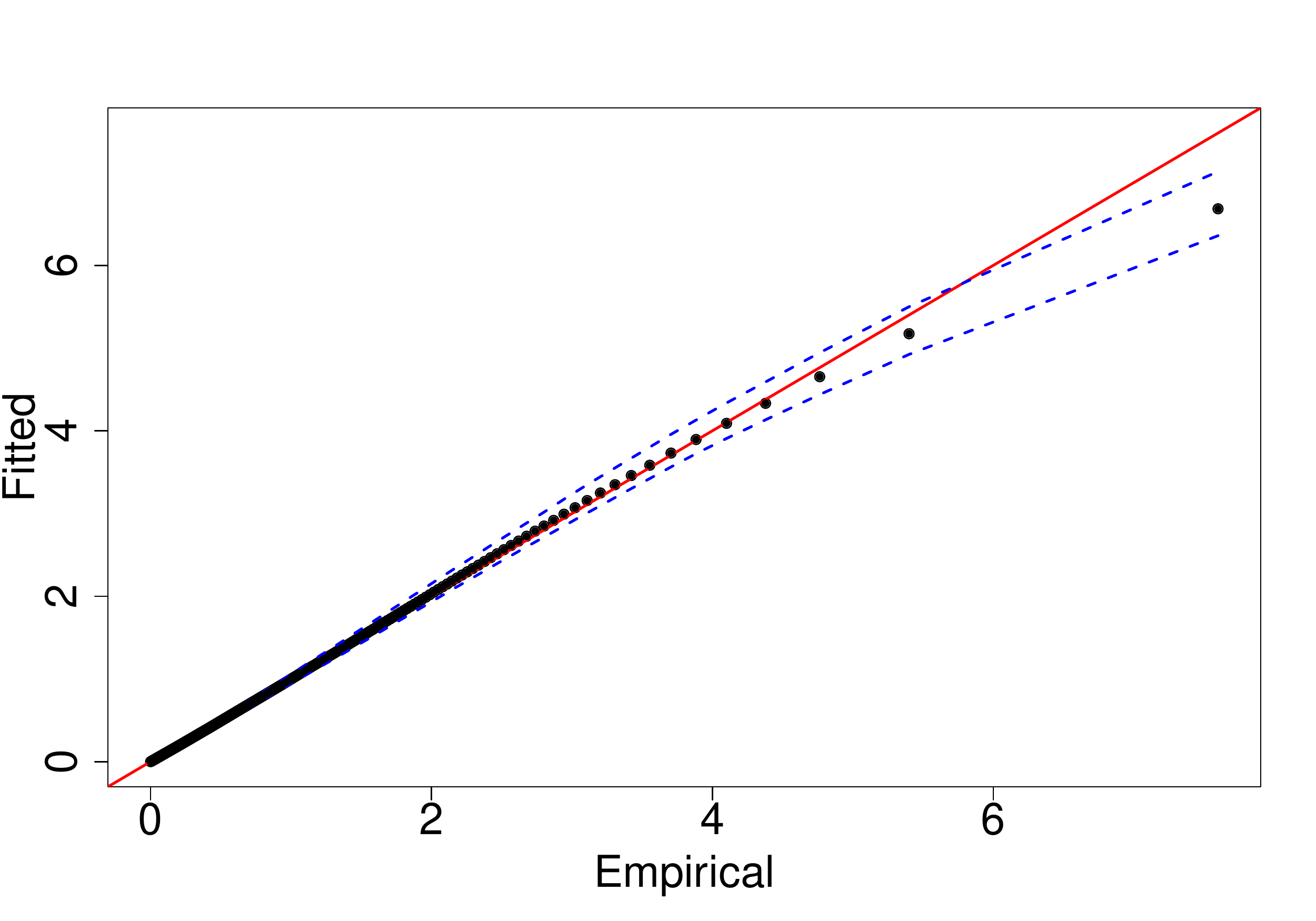} 
\end{minipage}
\caption{Q-Q plots for the generalised Pareto distribution additive model fits for convective rainfall at five randomly sampled sites in $\mathcal{S}$. Bottom right: Q-Q plot for pooled marginal transformation over all sites to standard exponential margins. The $95\%$ pointwise confidence intervals are given by the blue dashed lines.}
\label{conv_gamdiag}
\end{figure}
\begin{figure}[h]
\centering
\begin{minipage}{0.3\linewidth}
\centering
\includegraphics[width=\linewidth]{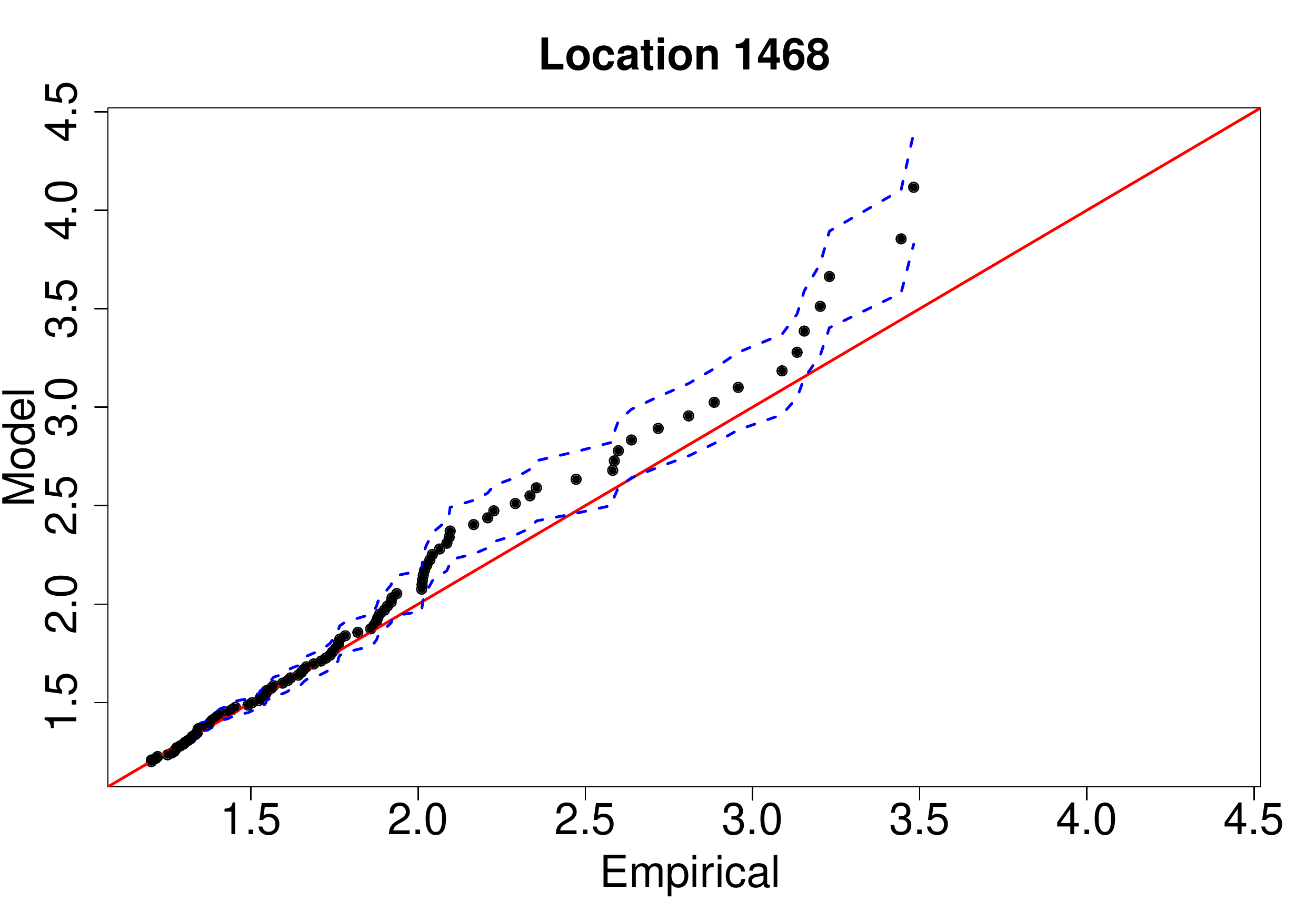} 
\end{minipage}
\begin{minipage}{0.3\linewidth}
\centering
\includegraphics[width=\linewidth]{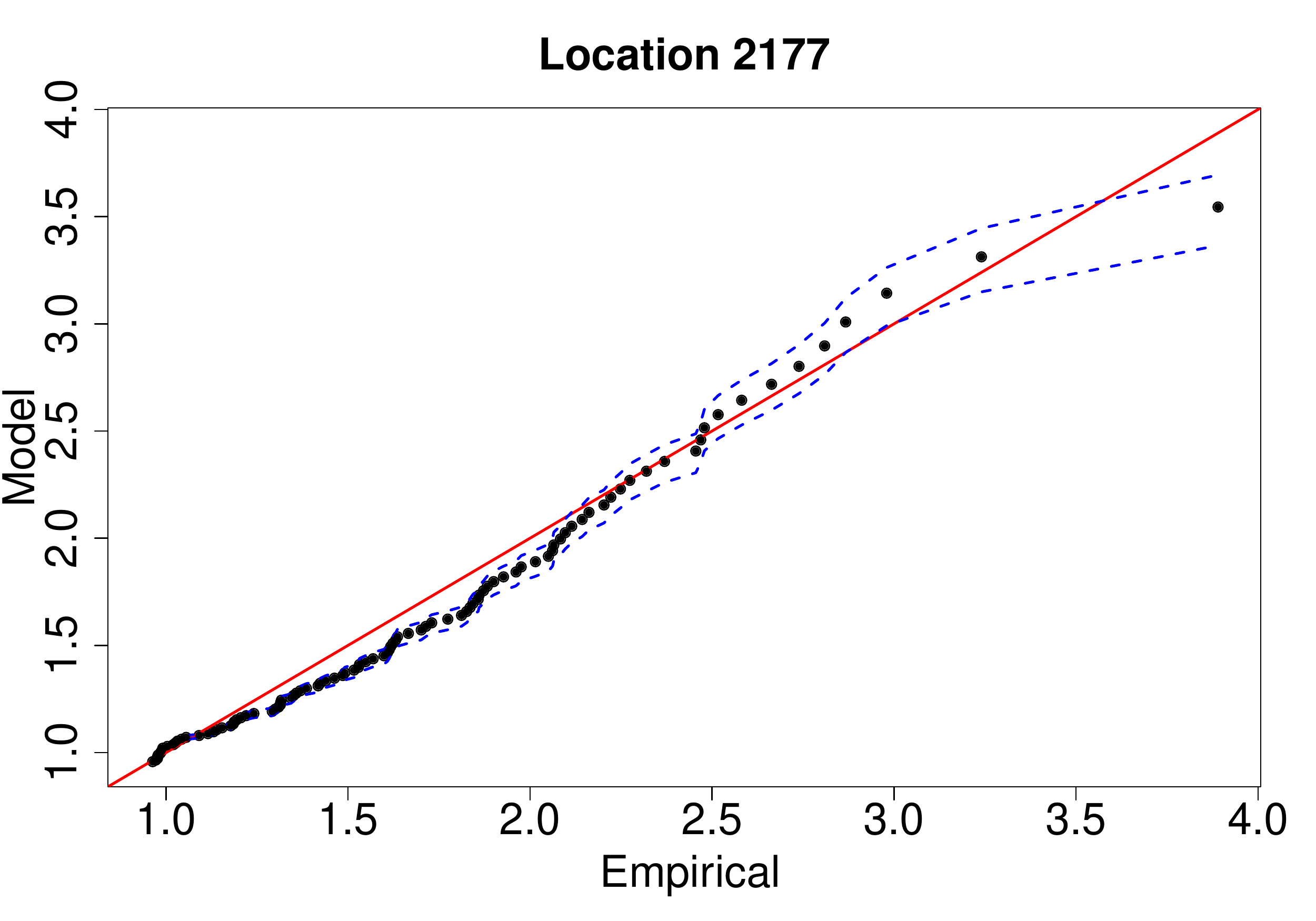} 
\end{minipage}
\begin{minipage}{0.3\linewidth}
\centering
\includegraphics[width=\linewidth]{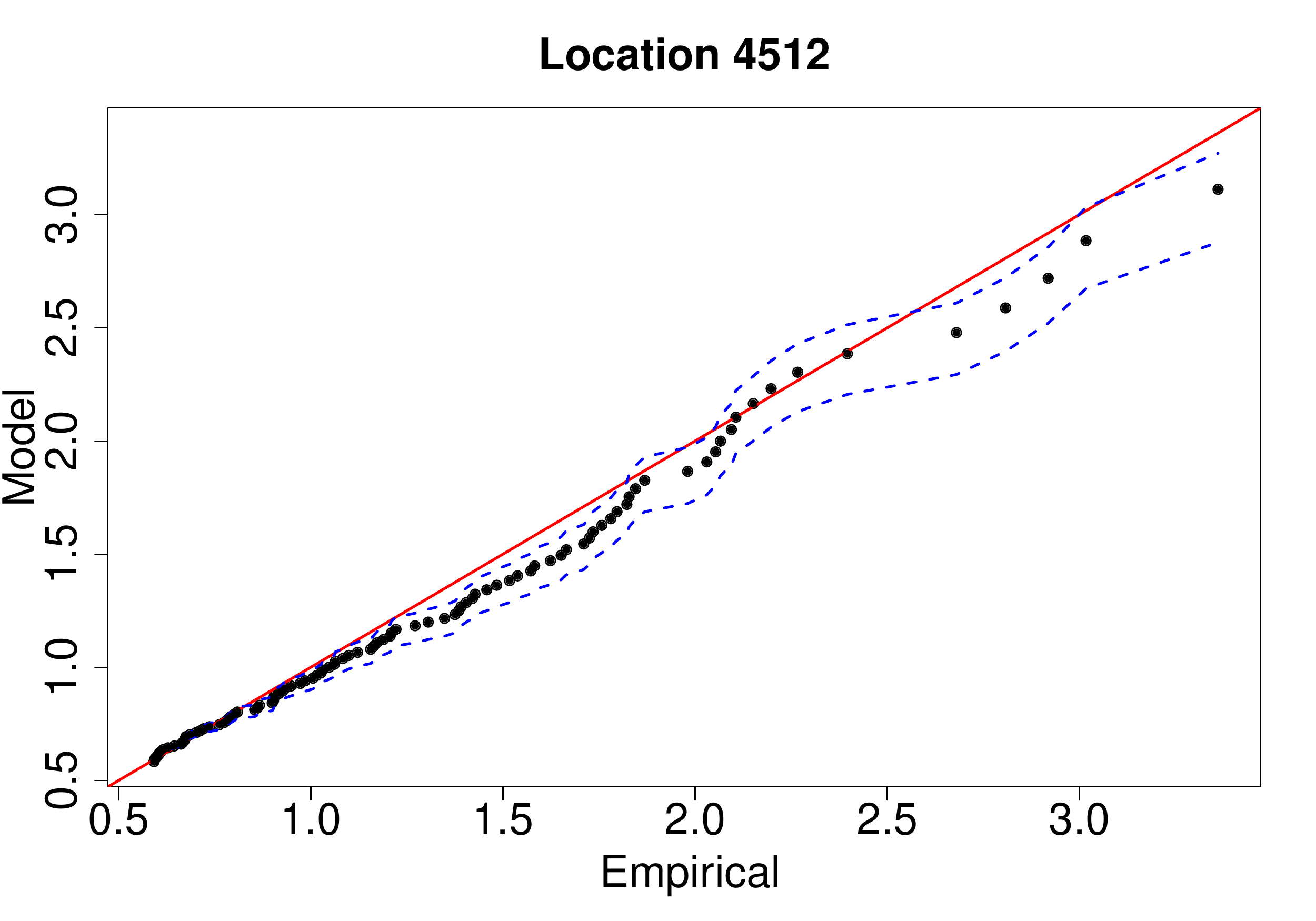} 
\end{minipage}
\begin{minipage}{0.3\linewidth}
\centering
\includegraphics[width=\linewidth]{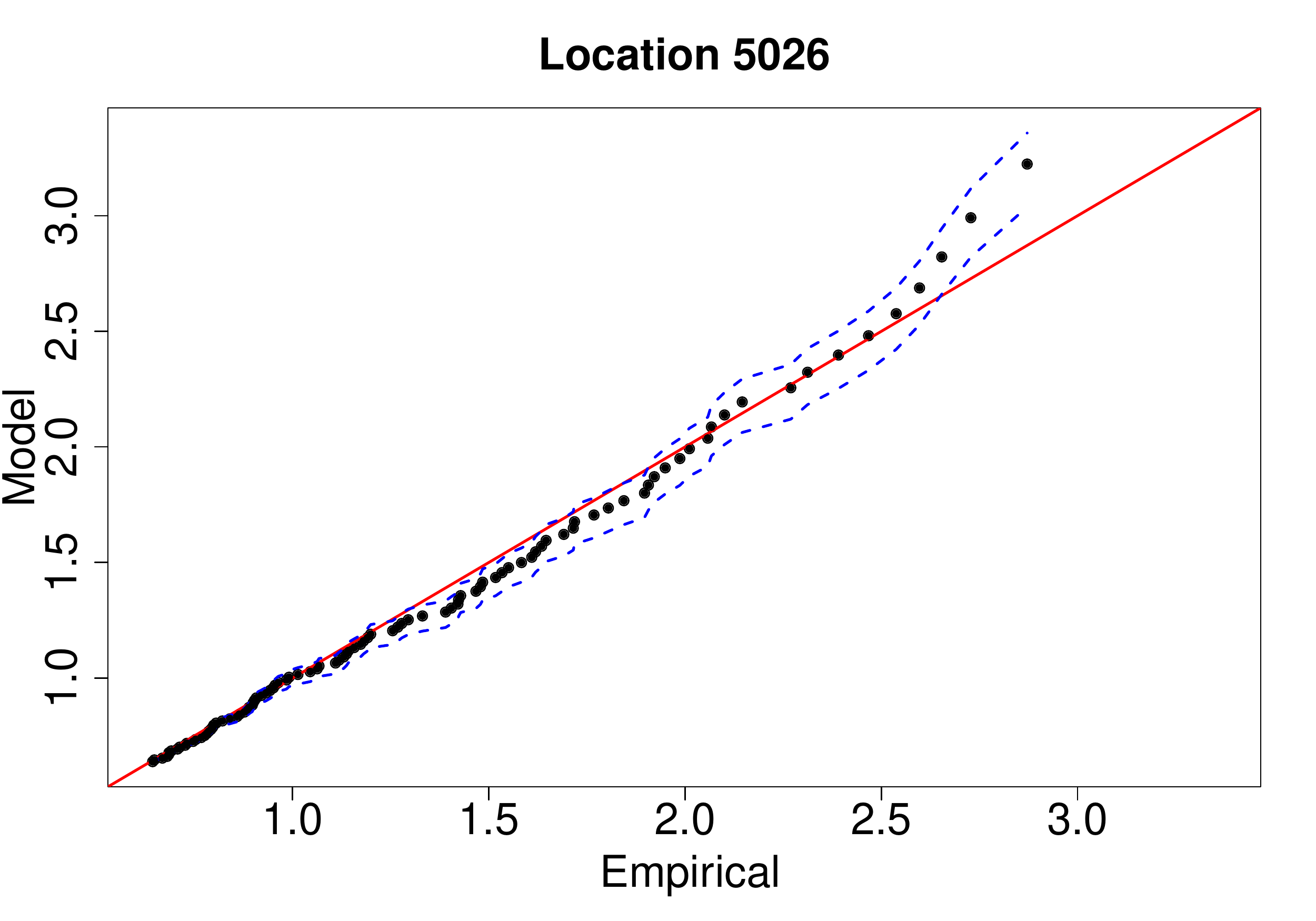} 
\end{minipage}
\begin{minipage}{0.3\linewidth}
\centering
\includegraphics[width=\linewidth]{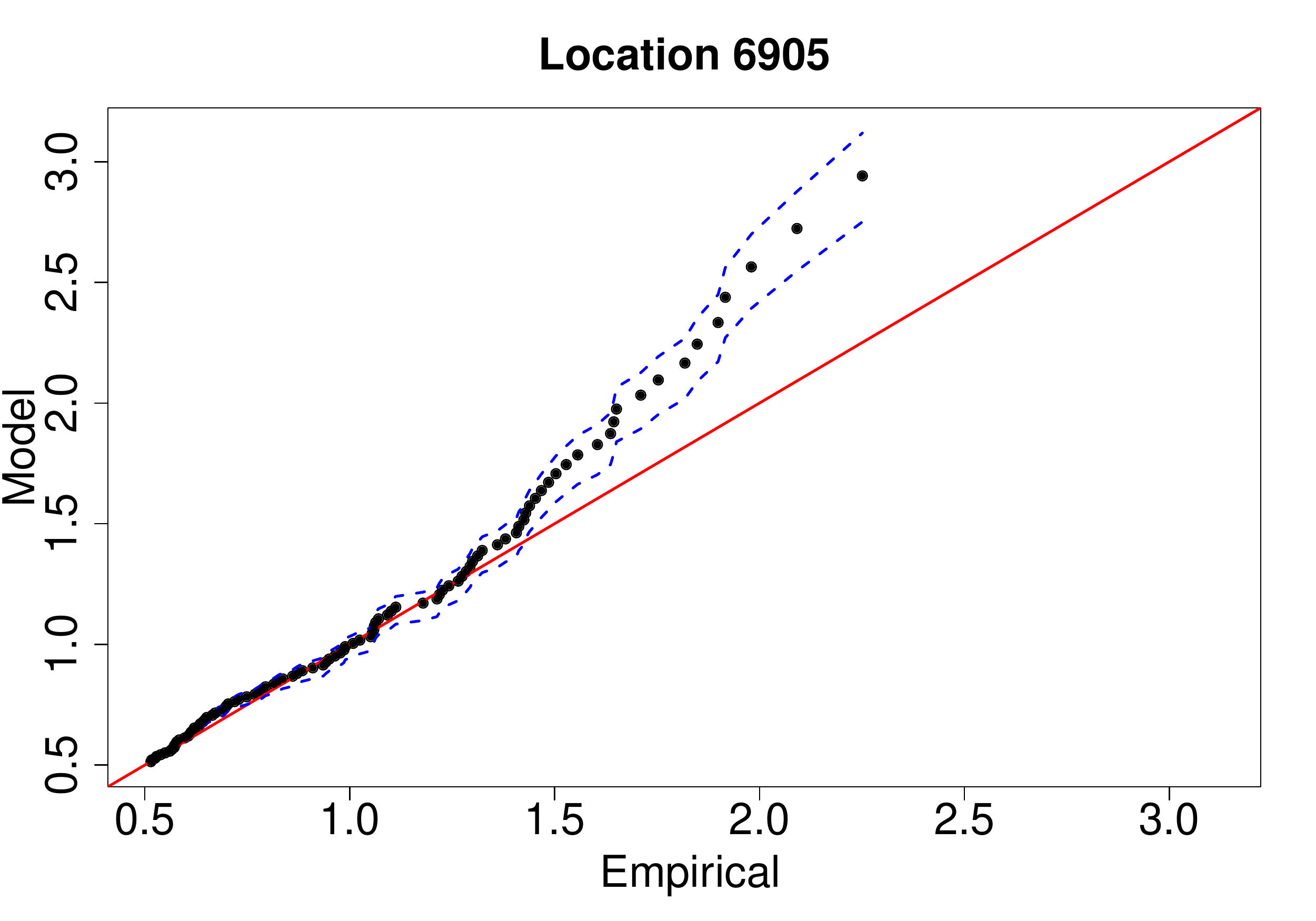} 
\end{minipage}
\begin{minipage}{0.3\linewidth}
\centering
\includegraphics[width=\linewidth]{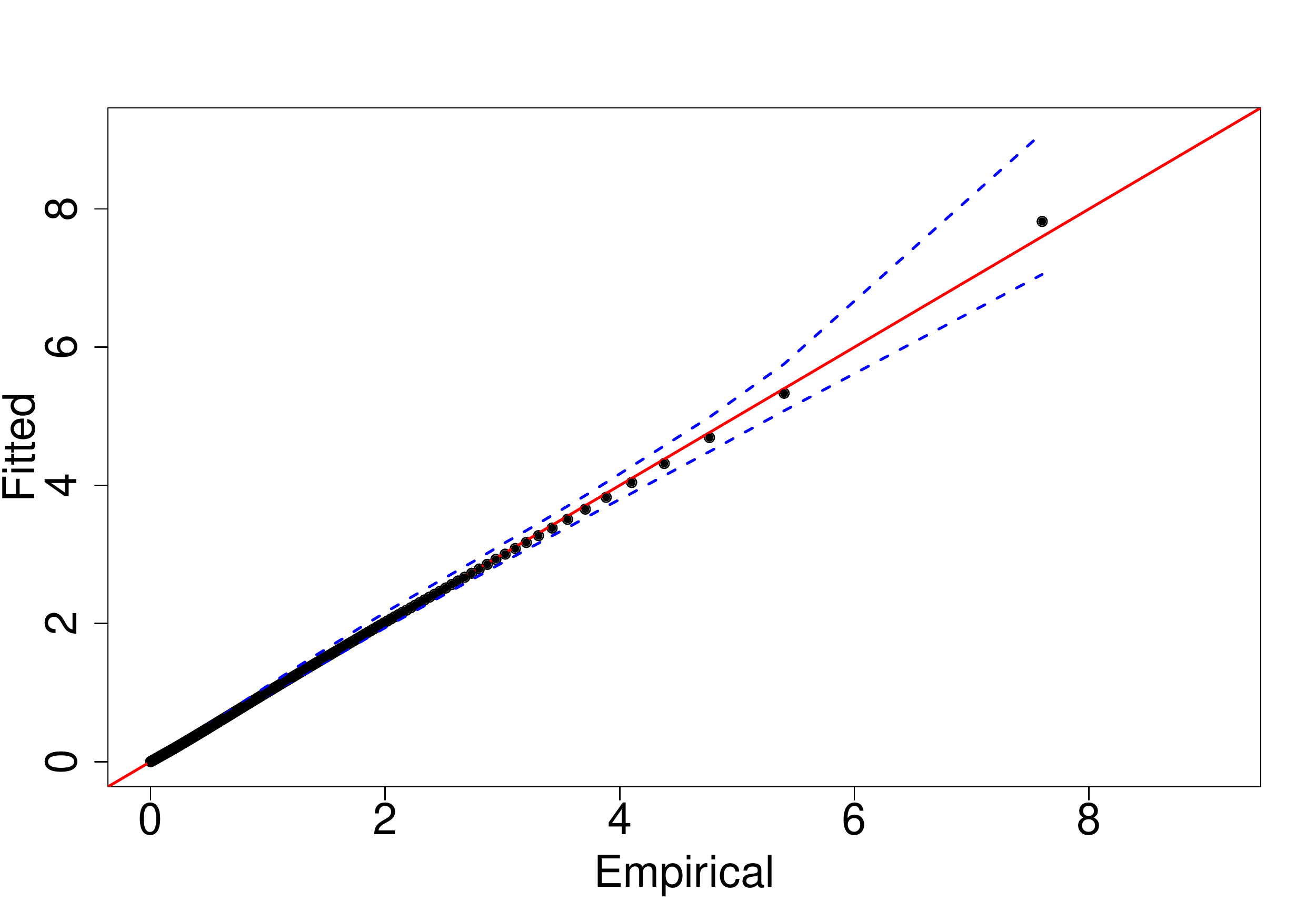} 
\end{minipage}
\caption{Q-Q plots for the generalised Pareto distribution additive model fits for non-convective rainfall at five randomly sampled sites in $\mathcal{S}$. Bottom right: Q-Q plot for pooled marginal transformation over all sites to standard exponential margins. The $95\%$ pointwise confidence intervals are given by the blue dashed lines.}
\label{front_gamdiag}
\end{figure}
\begin{figure}[h]
\centering
\begin{minipage}{0.3\linewidth}
\centering
\includegraphics[width=\linewidth]{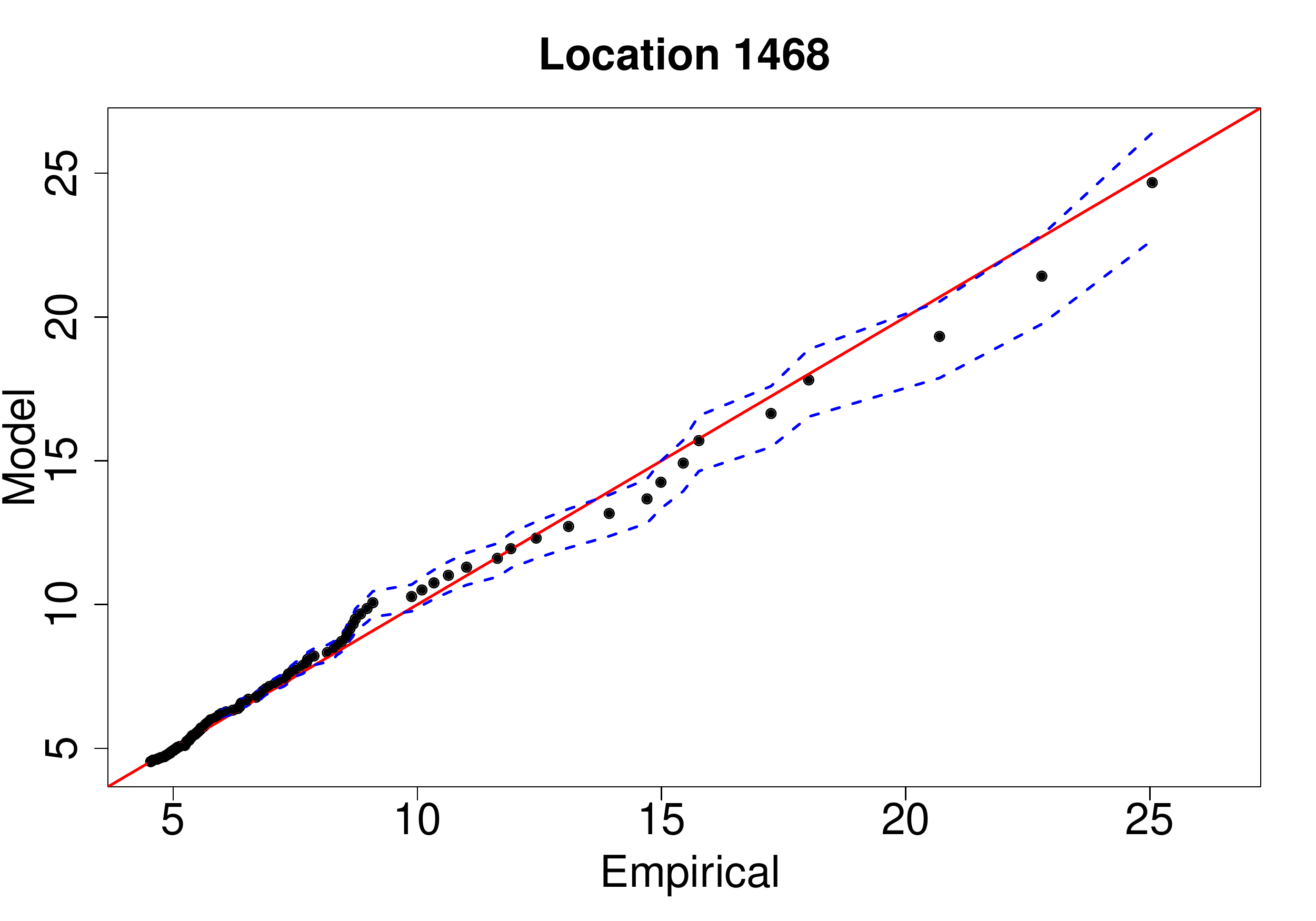} 
\end{minipage}
\begin{minipage}{0.3\linewidth}
\centering
\includegraphics[width=\linewidth]{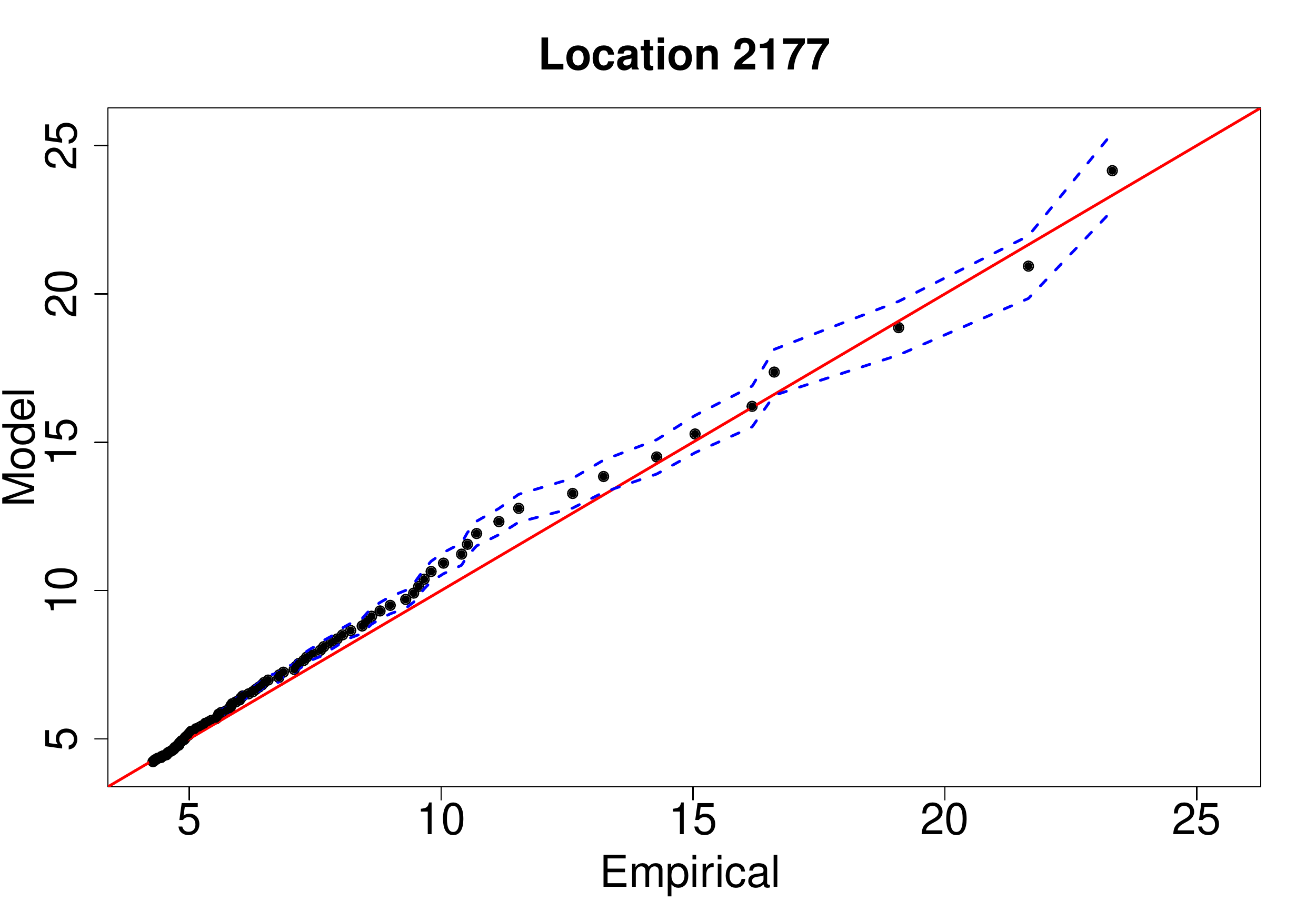} 
\end{minipage}
\begin{minipage}{0.3\linewidth}
\centering
\includegraphics[width=\linewidth]{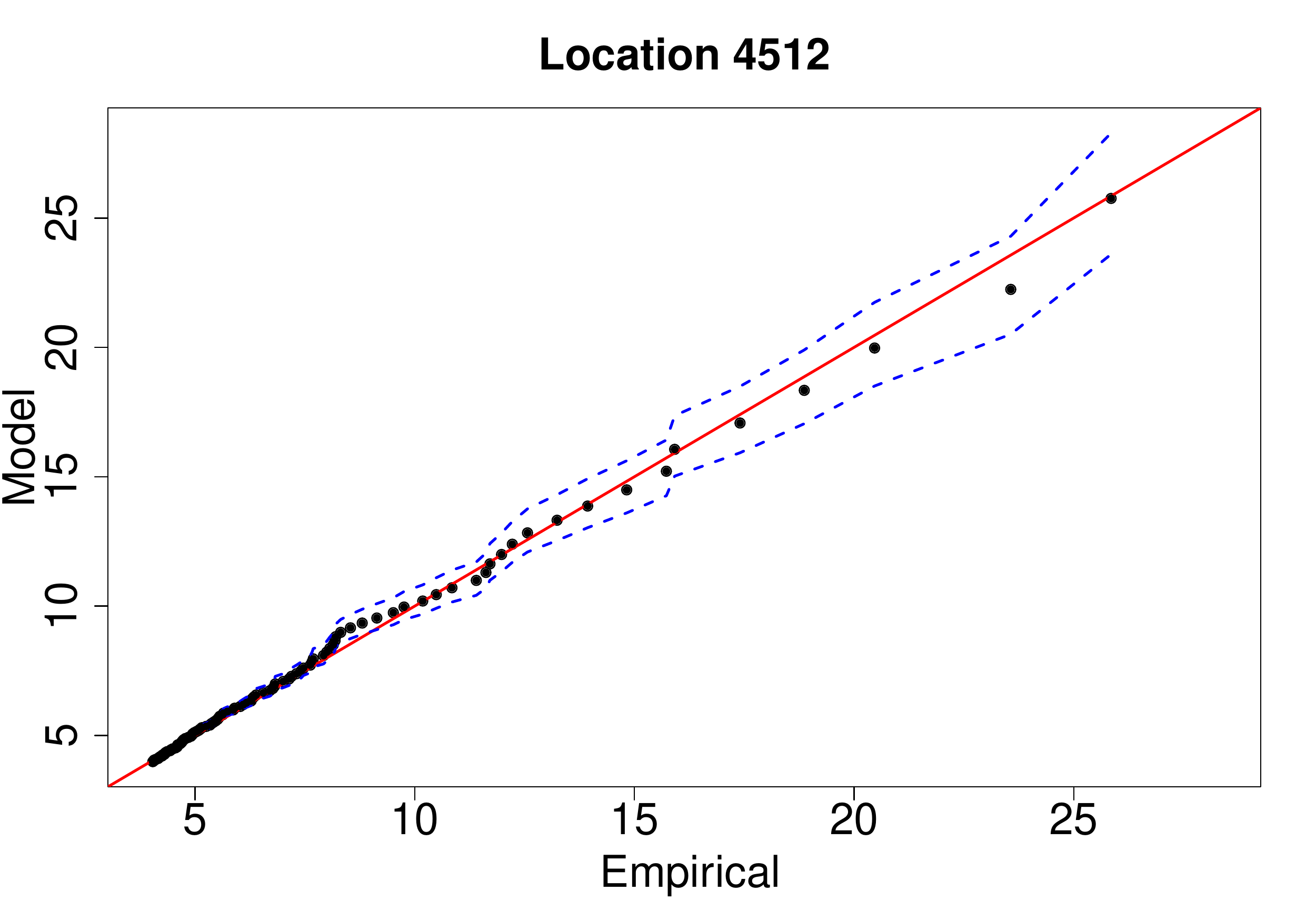} 
\end{minipage}
\begin{minipage}{0.3\linewidth}
\centering
\includegraphics[width=\linewidth]{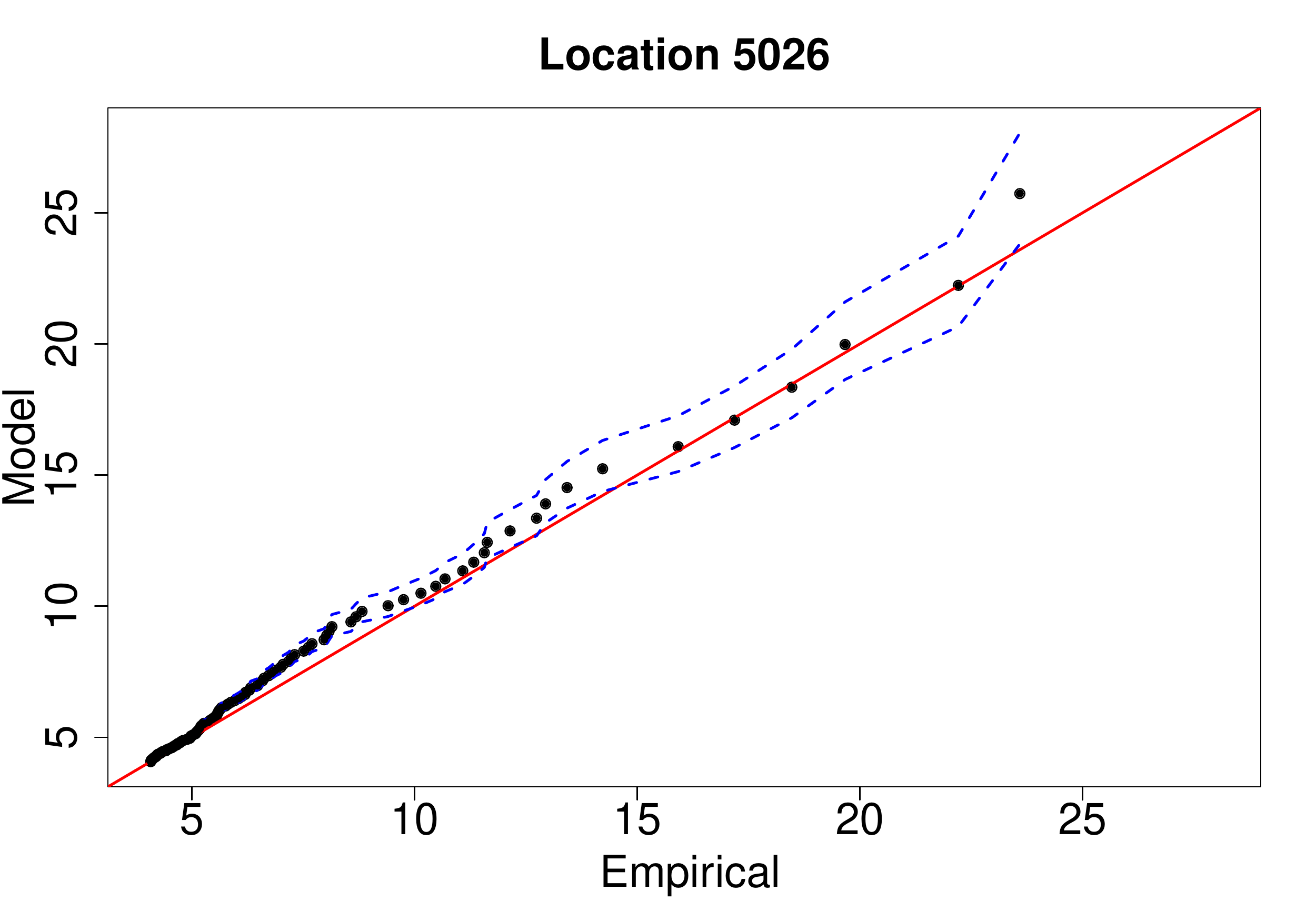} 
\end{minipage}
\begin{minipage}{0.3\linewidth}
\centering
\includegraphics[width=\linewidth]{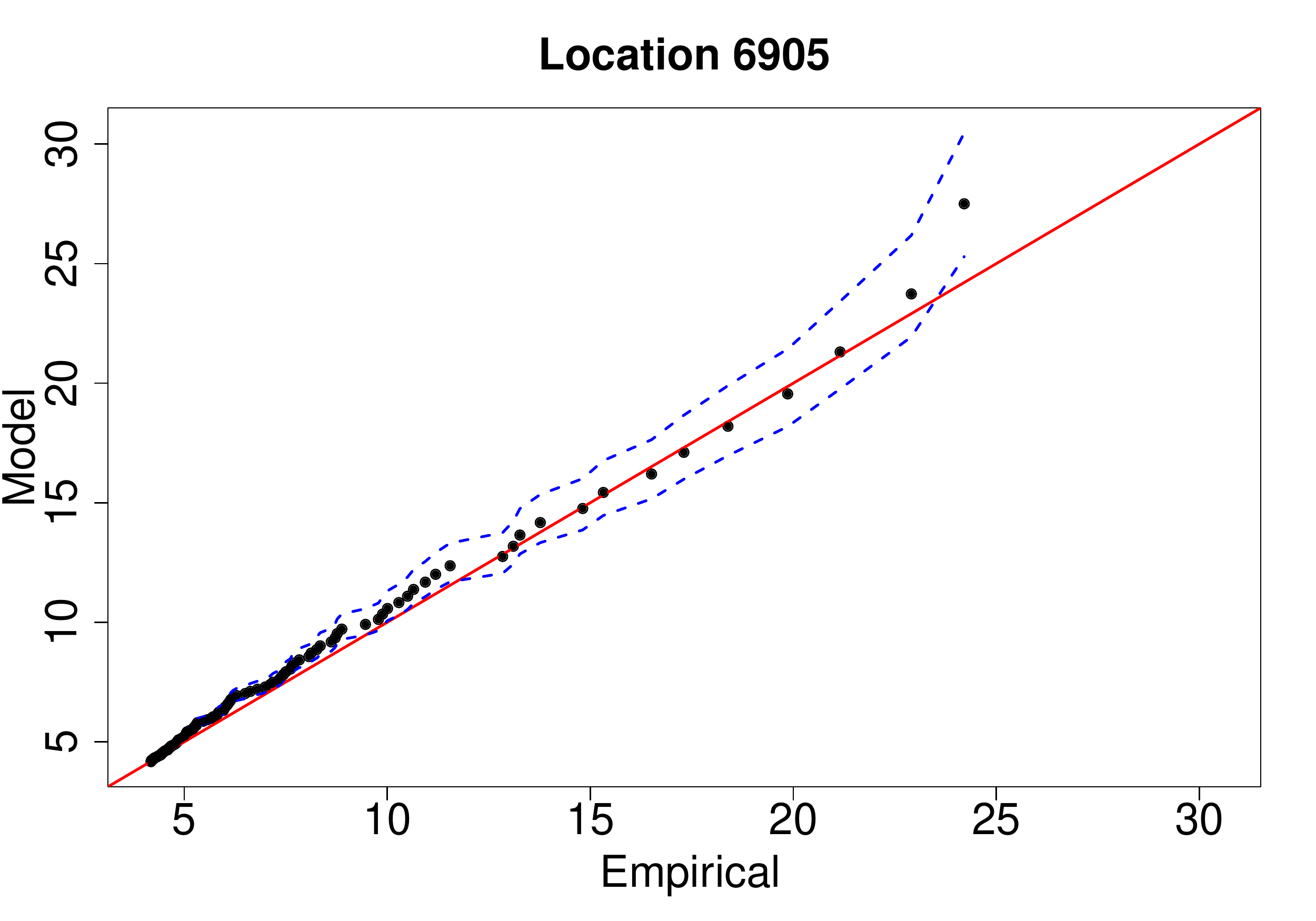} 
\end{minipage}
\begin{minipage}{0.3\linewidth}
\centering
\includegraphics[width=\linewidth]{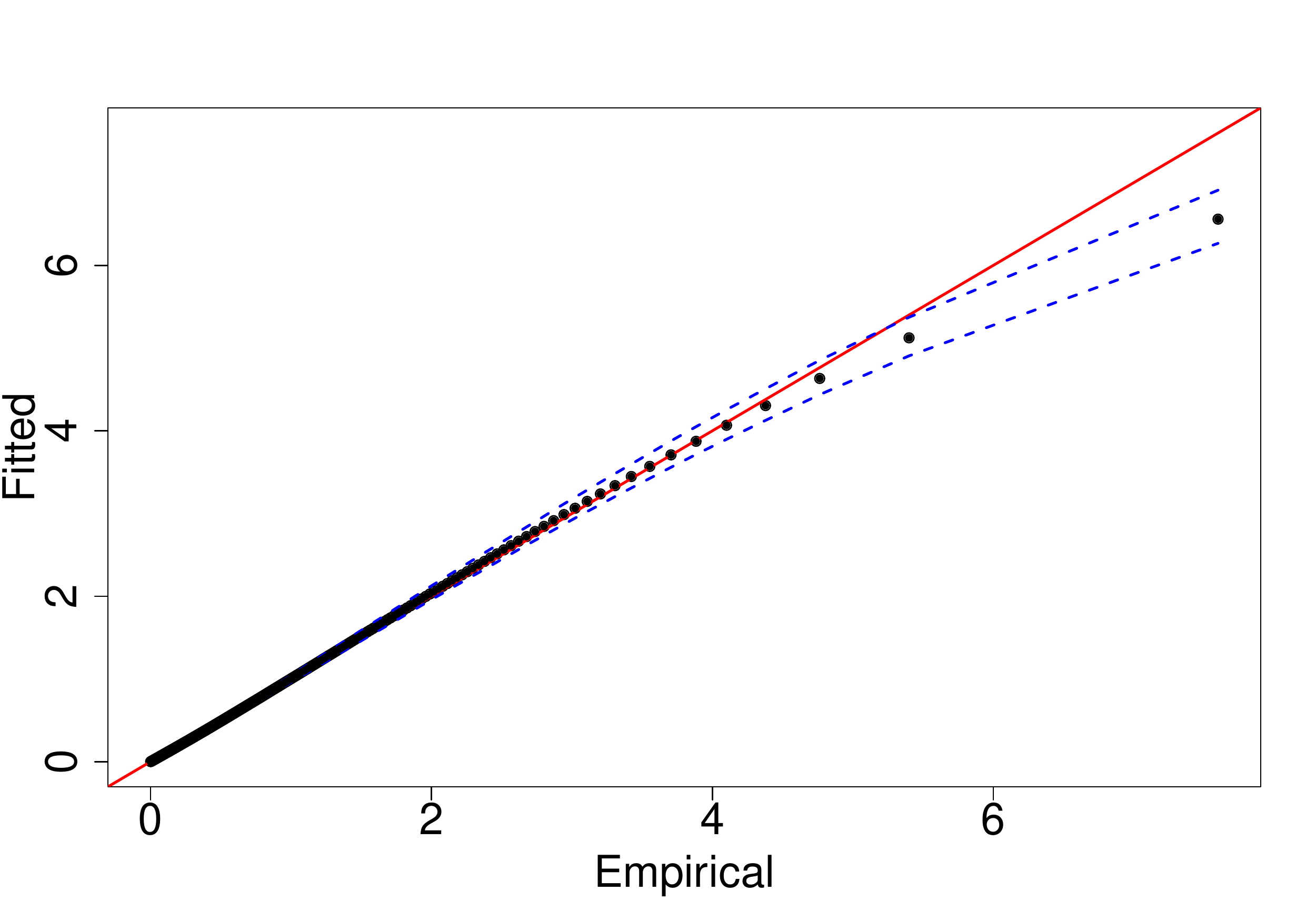} 
\end{minipage}
\caption{Q-Q plots for the generalised Pareto distribution additive model fits for all rainfall, i.e., $\{Y_\mathcal{E}(s)\}$, at five randomly sampled sites in $\mathcal{S}$. Bottom right: Q-Q plot for pooled marginal transformation over all sites to standard exponential margins. The $95\%$ pointwise confidence intervals are given by the blue dashed lines.}
\label{both_gamdiag}
\end{figure}
\begin{figure}[h!]
\begin{minipage}{0.32\linewidth}
\centering
\includegraphics[width=\linewidth]{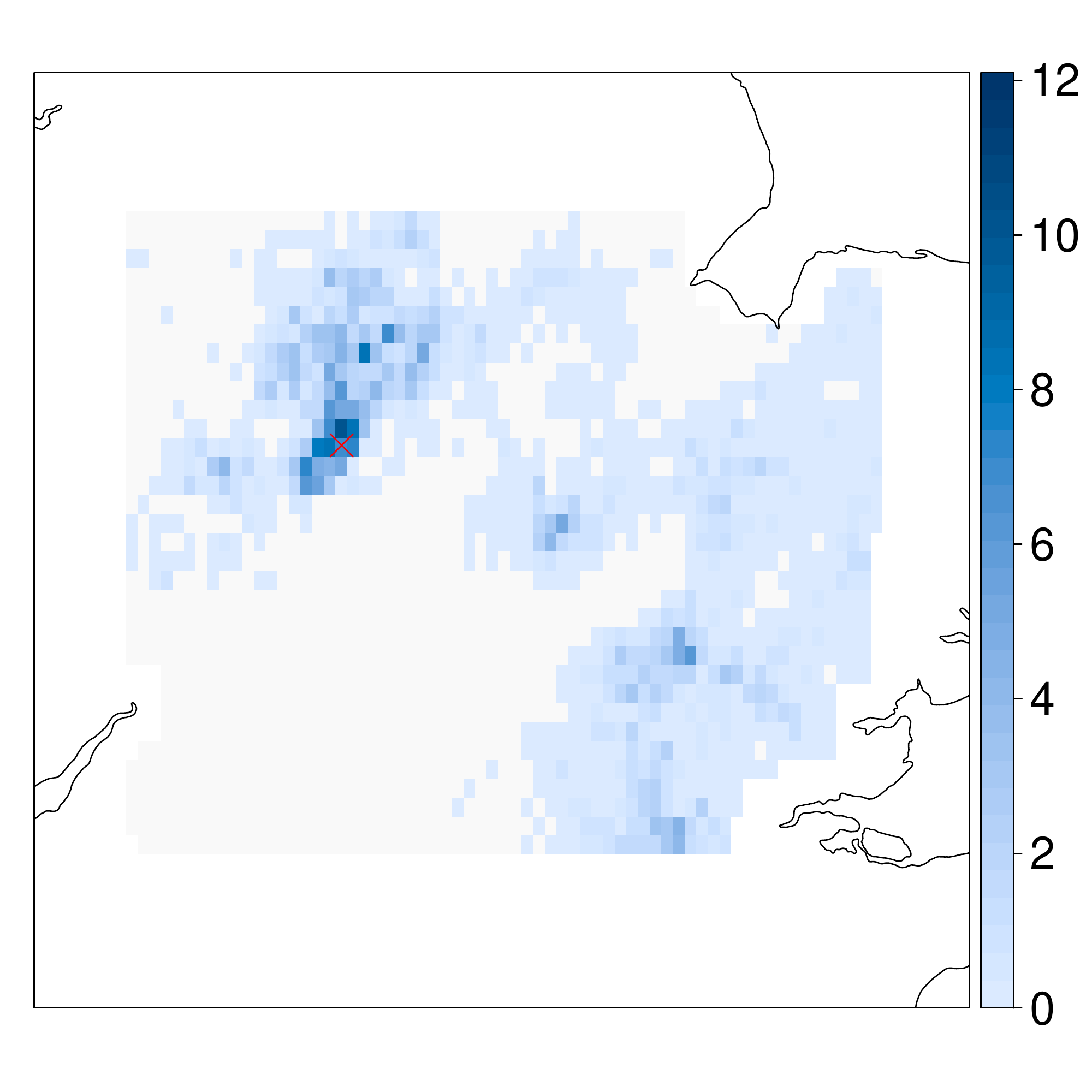} 
\end{minipage}
\begin{minipage}{0.32\linewidth}
\centering
\includegraphics[width=\linewidth]{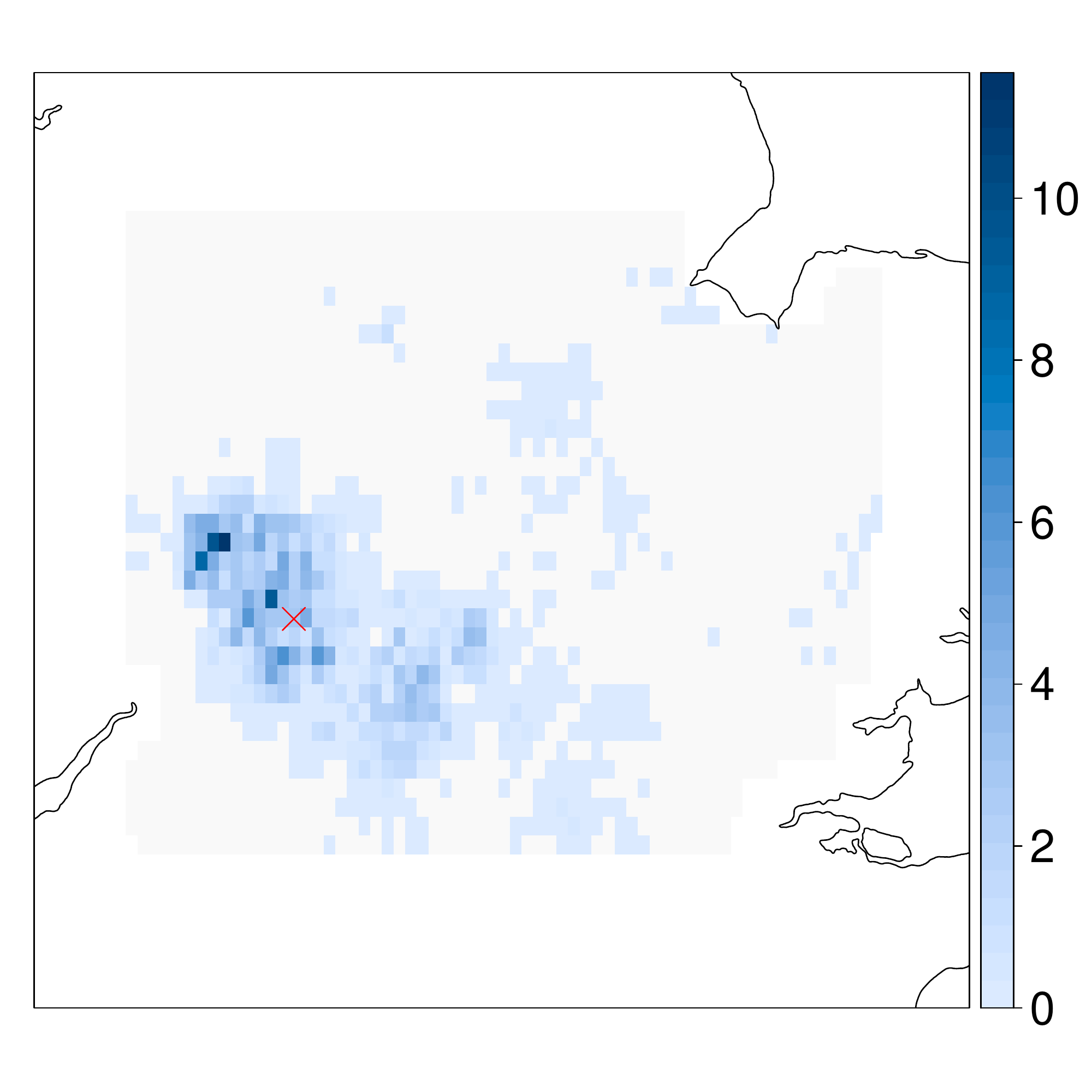} 
\end{minipage}
\begin{minipage}{0.32\linewidth}
\centering
\includegraphics[width=\linewidth]{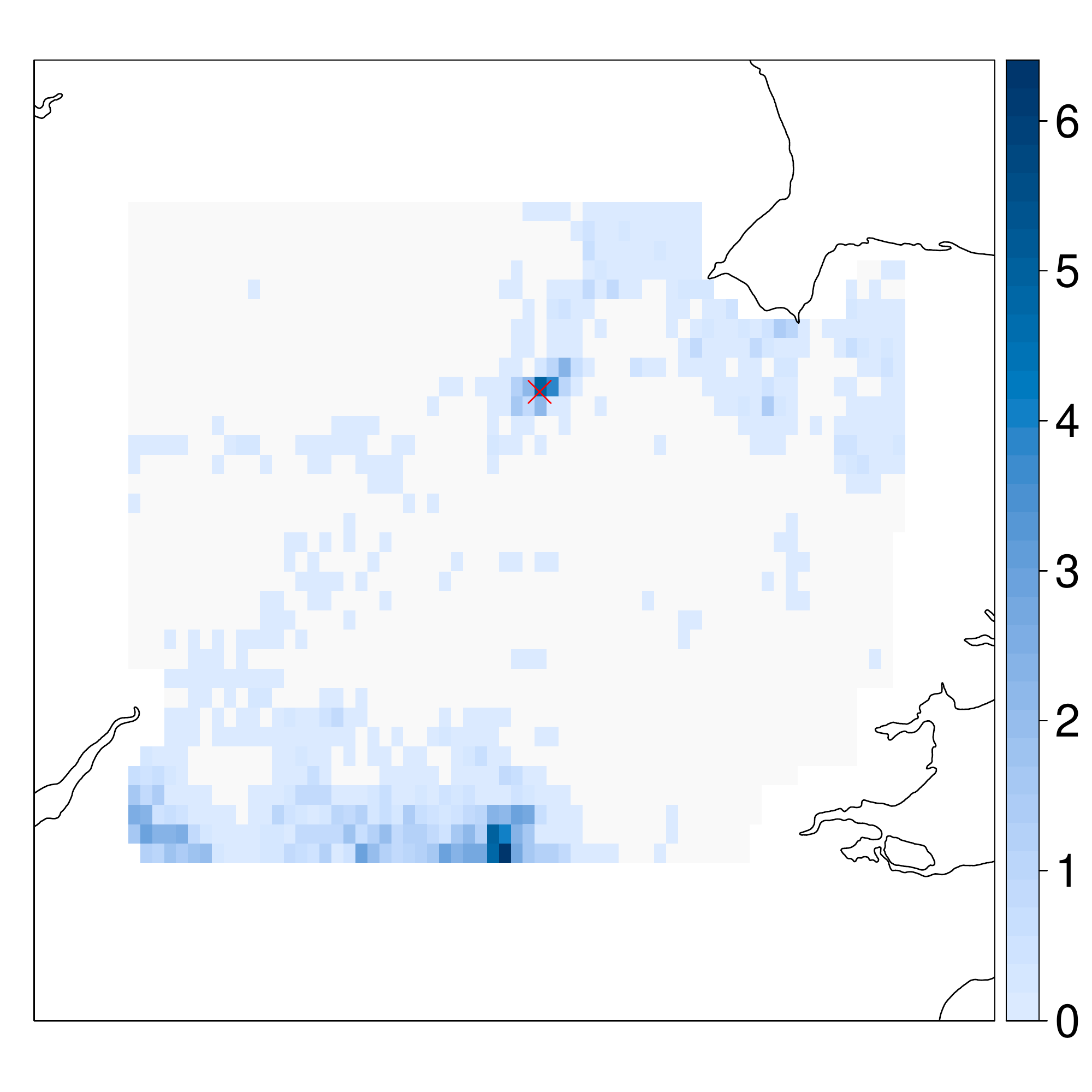} 
\end{minipage}
\caption{Extreme precipitation fields (mm/hr). Realisations from the
fitted model for all rainfall ($Y_\mathcal{E}$) for exceedances over the 0.99 quantile at conditioning site $s_O$, given by the red crosses. Scales differ in each panel.}
\label{figrealise_sup}
\end{figure}
\begin{figure}[h!]
\begin{minipage}{0.48\textwidth}
\centering
\includegraphics[page=1,width=\textwidth]{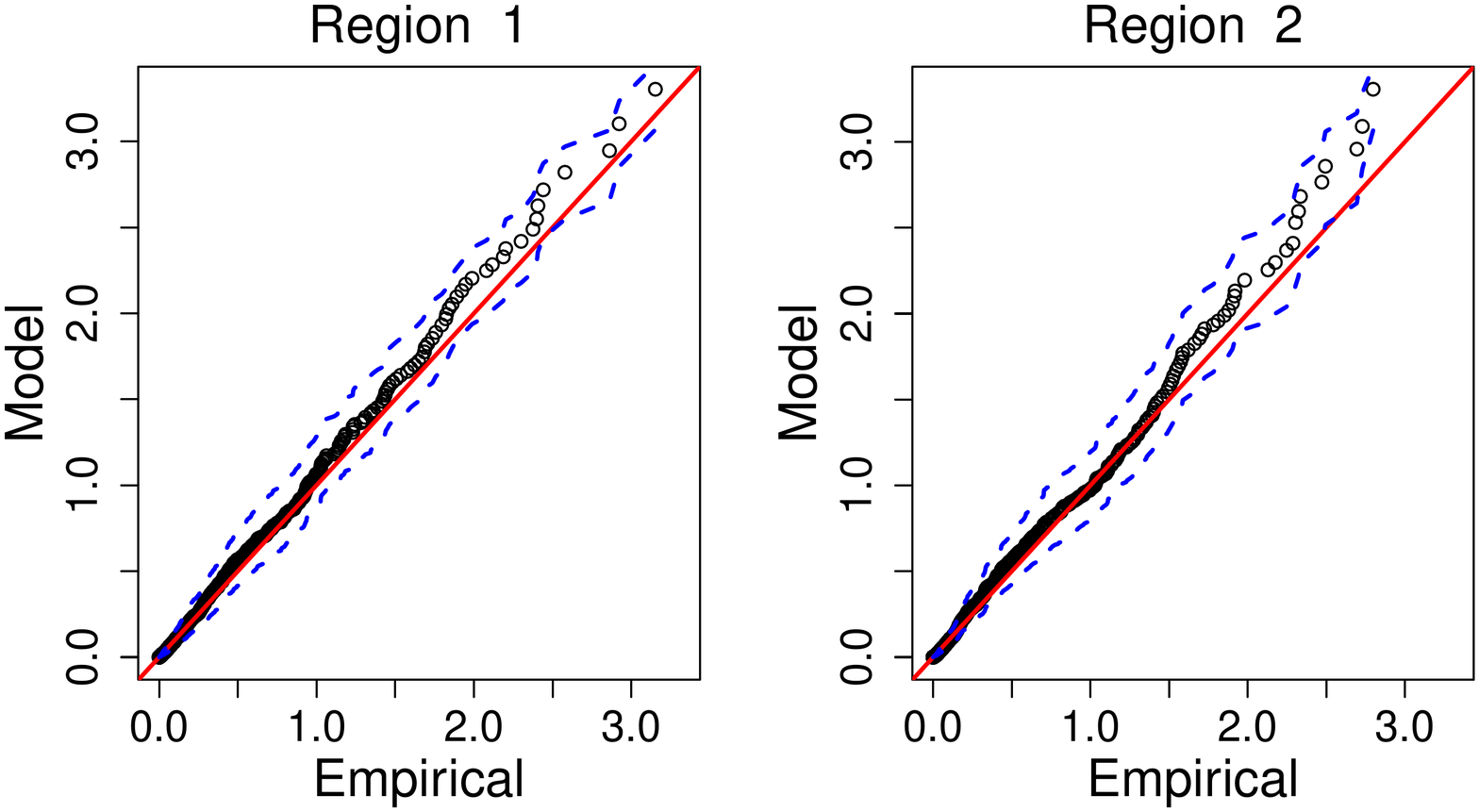} 
\end{minipage}
\hfill
\begin{minipage}{0.48\textwidth}
\centering
\includegraphics[page=2,width=\textwidth]{figures/Agg_Front_boot.pdf} 
\end{minipage}
\begin{minipage}{0.48\textwidth}
\centering
\includegraphics[page=1,width=\textwidth]{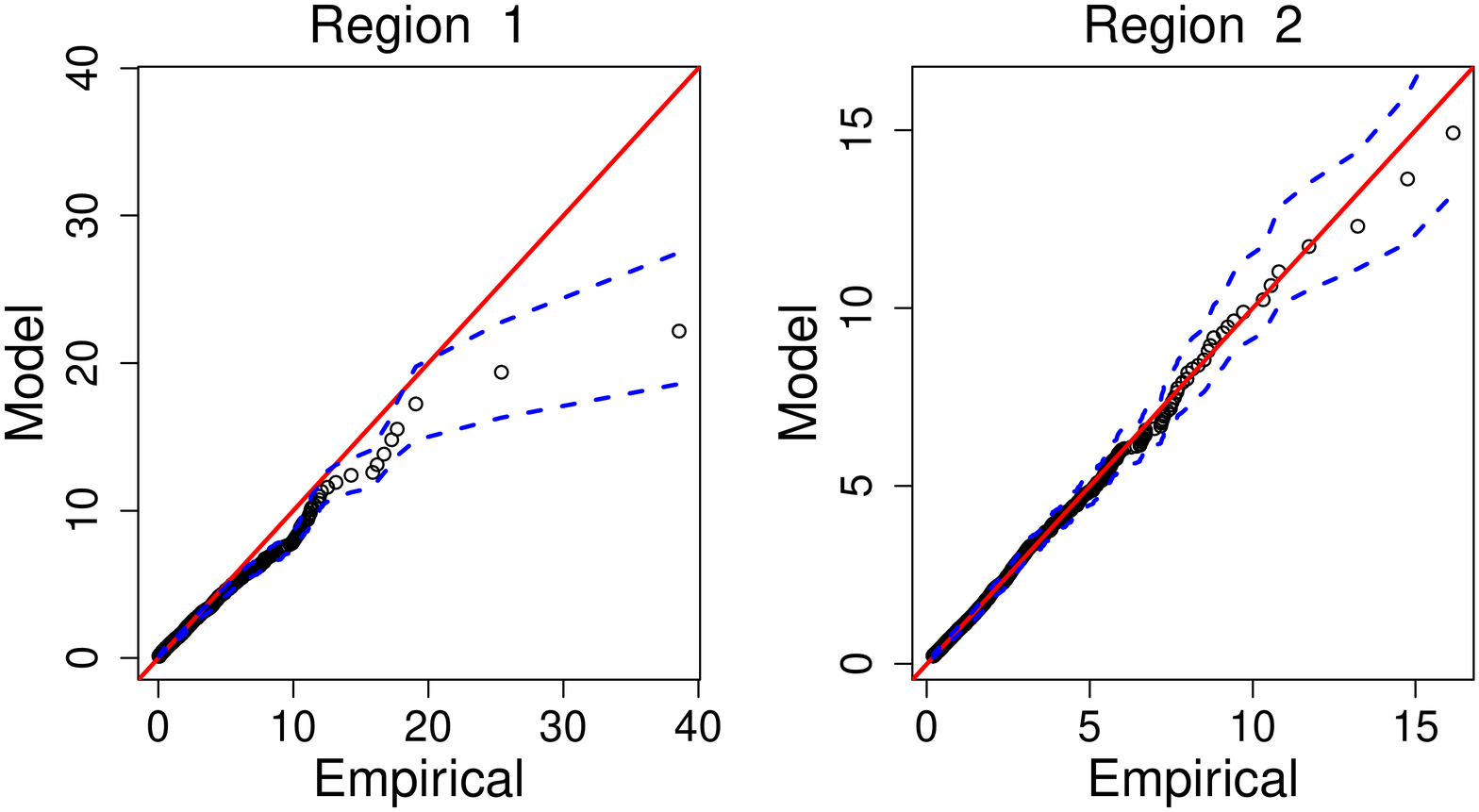} 
\end{minipage}
\hfill
\begin{minipage}{0.48\textwidth}
\centering
\includegraphics[page=2,width=\textwidth]{figures/Agg_Conv_boot.pdf} 
\end{minipage}

\caption{Q-Q plots for the model, and empirical, separate contributions of convective and frontal rainfall to the spatial aggregate $\bar{R}_{\mathcal{M},\mathcal{A}}$ for regions of increasing size. Top: non-convective contribution ($\bar{R}_{\mathcal{N},\mathcal{A}}$), bottom: convective contribution ($\bar{R}_{\mathcal{C},\mathcal{A}}$). The considered quantiles range from the 0.8-quantile to a value corresponding to the respective 20 year return level. Blue dashed lines denote point-wise $95\%$ quantile estimates. Regions 1--4 correspond to those illustrated in Figure~\ref{agg_diags_locs} of the main text.}
\label{agg_diags1}
\end{figure}

\end{appendix}

\end{document}